\documentclass[fleqn,usenatbib]{mnras}

\usepackage{newtxtext,newtxmath}
\usepackage[T1]{fontenc}

\DeclareRobustCommand{\VAN}[3]{#2}
\let\VANthebibliography\thebibliography
\def\thebibliography{\DeclareRobustCommand{\VAN}[3]{##3}\VANthebibliography}

\usepackage{graphicx}	% Including figure files
\usepackage{amsmath}	% Advanced maths commands
\title[Collision Between Molecular Clouds-III]{Collision Between Molecular Clouds-III: The effects of cloud initial density profile on head-on collisions}

\author[T.S Tanvir et al.]{
Tabassum S Tanvir,$^{1}$\thanks{E-mail: tabassum.tanvir@anu.edu.au(TST)}
James E Dale$^{2}$
\\
$^{1}$Research School of Astronomy and Astrophysics, The Australian National University,Cotter Rd, Weston Creek, ACT 2611, Australia\\
$^{2}$Centre for Astrophysics Research, University of Hertfordshire, College Lane,Hatfield, Hertfordshire AL10 9AB, UK\\
}

\date{Accepted XXX. Received YYY; in original form ZZZ}

\pubyear{2020}

\begin{document}
\label{firstpage}
\pagerange{\pageref{firstpage}--\pageref{lastpage}}
\maketitle

\begin{abstract}
In this third paper in our cloud-collision series, we present the results from simulations of head--on collisions with a strongly centrally--condensed initial density profile of $\rm \rho \propto R^{-2}$. We investigate the impact of these density profiles on the overall evolution of the simulations: the structures formed, their dynamical evolution, and their star formation activity. We consider clouds which are globally bound and globally unbound, leading to three different scenarios -- the collision of a bound cloud with a bound cloud, the collision of two unbound clouds, or the collision of one cloud of each type.\\
\indent In all the simulations dense star clusters form before the collisions occur, and we find that star formation remains confined to these systems and is little affected by the collisions. If the clouds' are both initially bound, the collision forms a filamentary structure, but otherwise this does not occur. We observe that rotating structures form around the clusters, but they also form in our non--colliding control simulations, so are not a consequence of the collisions. Dissipation of kinetic energy in these simulations is inefficient because of the substructure created in the clouds by turbulence before the collisions. As a result, although some gas is left bound in the COM frame, the star clusters formed in the simulations do not become bound to each other.
\end{abstract}


%%%%%%%%%%%%%%%%%%%%%%%%%%%%%%%%%%%%%%%%%%%%%%%%%%

%%%%%%%%%%%%%%%%% BODY OF PAPER %%%%%%%%%%%%%%%%%%

\section{Introduction}

Stars are predominantly formed in giant molecular clouds (GMCs) and understanding the regulation of star formation on galactic scales, and the overall evolution of GMCs is a long--standing problem in astrophysics. Both simulations \citep[e.g.][]{2009ApJ...700..358T,2015MNRAS.446.3608D} and observations \citep[e.g.][]{2009ApJ...696L.115F,2014ApJ...780...36F} suggest that encounters, ranging from tidal interactions to outright collisions, between clouds may play an important role in their evolution\\
\indent Galactic--scale simulations have established typical rates and timescales for collisions between clouds, but generally lack the resolution to examine in detail the effect of the encounter on the clouds themselves. Single interactions between pairs of clouds have been studied extensively using hydrodynamical simulations.\\
\indent A great deal of modelling effort has concentrated on smooth clouds or smooth shocked layers \citep[e.g][]{1994A&A...290..421W,2015MNRAS.453.2471B}. In such cases, it is essentially inevitable that the clouds will become bound, because the smooth contact layer is very efficient at radiating away kinetic energy. This produces an almost stationary structure, which is highly likely to become gravitationally unstable, so that the collision almost inevitably triggers star formation.\\
\indent Several other studies have used Bonner--Ebert spheres as initial conditions \citep[e.g.][]{1998MNRAS.297..435B,2014ApJ...792...63T} in order to study the influence of collisions on clumps whose masses would allow them to form small multiple systems. These models produce broadly similar results to those involving uniform clouds; the collision disturbs the equilibrium of the clouds, and gravitational instabilities occur at a limited number of locations in the shocked layers.\\
\indent Since molecular clouds are known to be turbulent, other studies have taken clouds seeded with turbulent velocity fields as their starting point. \citet{2014ApJ...792...63T} repeated their models of colliding Bonner--Ebert spheres seeded with turbulence, and found a much richer phenomenology. Their models produced complex filamentary structures and multiple sites of star formation. \cite{2018PASJ...70S..54S} also modelled the collision of turbulent Bonner--Ebert spheres, and found that the collisions were responsible for forming more stars and more stellar mass than in a control run in which the clouds remained isolated. They also found that higher collision velocities resulted in more stars.\\
\indent Several studies have reported that the shocked layers produced by collisions are efficient at producing \textit{massive} stars in particular. Observationally, a potential relation between cloud collision and massive star formation has been found by, e.g.,  \citet{2009ApJ...696L.115F, 2014ApJ...780...36F, 2016ApJ...820...26F,2010ApJ...709..975O, 2011ApJ...738...46T, 2015ApJ...806....7T, 2017ApJ...835..142T}. There is also numerical support for this idea. \citet{1994MNRAS.268..291W} showed analytically that shocked gas layers, possibly produced by collisions, preferentially form massive stars. \citet{2015MNRAS.453.2471B} and \citet{2017MNRAS.465.3483B} found that collisions with lower relative velocity produce more massive stars.\\
\indent Most of the work published on this subject takes the clouds to be initially in virial equilibrium. In \citet{2020MNRAS.494..246T} (hereafter, Paper I) we investigated the effect of the clouds' virial ratios on head-on collisions between two clouds of same mass and radius with an initially uniform density profile. We found that colliding  clouds that are both initially bound will produced about twice as much stellar mass as a collision where one cloud is bound and one is unbound. Simulations where both clouds are initially unbound produced even less stellar mass. Comparing to control runs (where the clouds evolve in isolation) for each of the scenarios, we found that collisions do increase the star formation efficiency modestly in both the bound--bound and bound--unbound cases. However, the collision seems to have no effect in the star formation in the unbound-unbound case.\\
\indent The initial conditions from which simulations are started, such as initial density profile and level of turbulent support play an important role in determining the outcome of models. In their study \citet{2011MNRAS.413.2741G} investigated four initial density profiles to understand their influence on the evolution of their simulated clouds and in particular on the stellar mass distribution. The conclusion drawn from the paper was that flat density distributions tend to produce more lower mass stars whereas concentrated density profiles tend to produce more massive stars, usually at the core of the clouds with lower mass stars further out.\\
\indent In this paper we knit together several of the issues discussed above. We investigate the role of the initial density profile on the evolution of collisions between turbulent molecular clouds. We choose an initial density profile of $\rm \rho \propto R^{-2}$, appropriate for a hydrostatically--supported isothermal sphere. Such an object will undergo inside--out collapse, producing an embedded star cluster inside the remaining infalling molecular gas \citep{1977ApJ...214..488S}.\\
\indent We use the smooth particle hydrodynamics (SPH) code {\sc gandalf} \citep{2018MNRAS.473.1603H}. The aim of the paper is to explore, by comparison with Paper I, the role the initial density profile plays on the evolution of the collision structure and on the overall star formation efficiencies.\\
\indent In Section 2, we briefly discuss our numerical methods, model assumptions and initial conditions. Section 3 contains the results of our calculations, and our discussion and conclusions are presented in Section 4.\\
%Re hashing the intro from paper I --- add some bits here and there and also find studies that did simulations on  
\section{Numerical Methods and Initial Conditions}
Like paper-I all of our simulations here are performed using the smooth particle hydrodynamics code {\sc gandalf}\citep{2018MNRAS.473.1603H}. {\sc gandalf} uses the  gradh-SPH formalism \citep[][]{2002MNRAS.333..649S,2004MNRAS.348..139P} and a leapfrog  kick--drift--kick (KDK) integrator in order to solve the fluid equations. An octal tree is used to compute self gravitational forces. Artificial viscosity forces are calculated using the \cite{1997JCoPh.136..298M} scheme with $\rm \alpha$ = 1 and $\rm \beta$ = 2. Sink particles replace gravitationally--unstable groups of gas particles, provided they are collapsing. The sink particle density threshold in our simulations are 10$^{-17}$\,g\,cm$^{-3}$. No background medium is used in these simulations. We repeated some simulations with $10^{5}$, $3\times10^{5}$ particles and found that convergence (in terms of kinetic energy dissipated in the collision, and the rate and efficiency of star formation) was excellent (the convergence tests will be published in a subsequent paper).\\
\indent A barotropic equation of state is used to model the thermodynamics of the gas (e.g:\citep[e.g.][]{2011A&A...529A..27H}) with a critical density of 10$^{-16}$\,g\,cm$^{-3}$ above which the gas behaves adiabatically. Given the sink density formation threshold, this essentially makes the state of the gas isothermal with a temperature of 30K. The reasoning behind this initial temperature selection is that this is an intermediate temperature between the coldest and densest part of the gas in a typical molecular cloud, where the coldest densest regions have temperatures of $\sim$10 K and the outer regions where the temperature can be $\sim$100 K. We take the mean molecular weight to be 2.35.\\ 
\indent Each cloud is modelled with $\rm 10^{6}$ particles which gives a mean mass resolution of 0.5-1.0 $\rm M_{\odot}$ (since the number of neighbours for particles is 50). Both clouds are provided with a divergence--free turbulent velocity field using a power spectrum  $P(k)\propto k^{-4}$. This power spectrum is appropriate for supersonic turbulence. The turbulence is not artificially driven but is purely initial. The clouds are seeded with the turbulent velocity fields where the root--mean--square velocities are scaled in order to generate their virial ratios.\\
\indent The clouds start with the same mass and same radii with their centres of mass are placed at d=40 pc from each other in the x-direction. The bulk velocity in these models in the x-axis are $\rm \pm v_{0}/2$ where $\rm v_{0}$ is 10 \,km\,s$^{-1}$. The simulations here are head--on collisions therefore the impact parameter, b = 0. Table 1 shows the initial conditions of the simulations used in this study. \\
\begin{table*}
\centering
\begin{tabular}{ccccccccccccccccc}
\hline \hline
Run  & Mass & Radius &Temperature&$\rm b_{0}$&$\rm v_{0}$&$\rm \alpha$&$\rm t_{ff}$&$\rm t_{cross}$& $\rm \rho$ \\ 
&$\rm M_{\odot}$&pc&K&&\,km\,s$^{-1}$&&Myr&Myr& g$cm^{-3}$\\
\hline \hline
 BB\_rho\_r-2 & 10000&10&30&0&10&1.0&5.2&8.61& $\rm 1.62\times 10^{-22}$\\
  BU\_rho\_r-2 & 10000&10&30&0&10&1.0, 5.0&5.2&8.61,3.85& $\rm 1.62\times 10^{-22}$\\
  % C & 10000&10&30&5&10&2.5,2.5&5.2&5.44\\
UU\_rho\_r-2 & 10000&10&30&0&10&5.0&5.2&3.85& $\rm 1.62\times 10^{-22}$\\
\hline \hline
\end{tabular}
\caption{Simulation parameters: Run name, masses of individual clouds, radius of clouds, gas temperature, impact parameter, initial relative velocity, global initial virial parameter(s), global initial freefall time and initial crossing time and initial density}
\end{table*}
\indent As in paper I there is an enormous amount of parameter space to explore. Therefore, like paper I we chose to explore the effect of virial ratio of the clouds along with the non uniform densities of the cloud. The aim of this paper is to investigate whether the collision result in greater star formation than it would have if the clouds were to evolve in isolation and whether the collision will result in the clouds merging together to create singular object.\\
%{\bf I'm not sure if it makes sense to talk about a single virial ratio for these condensed clouds. The main thing we are looking at here is what happens if one collides condensed rather than uniform clouds.}

\begin{figure}
	\includegraphics[width=1.0\columnwidth]{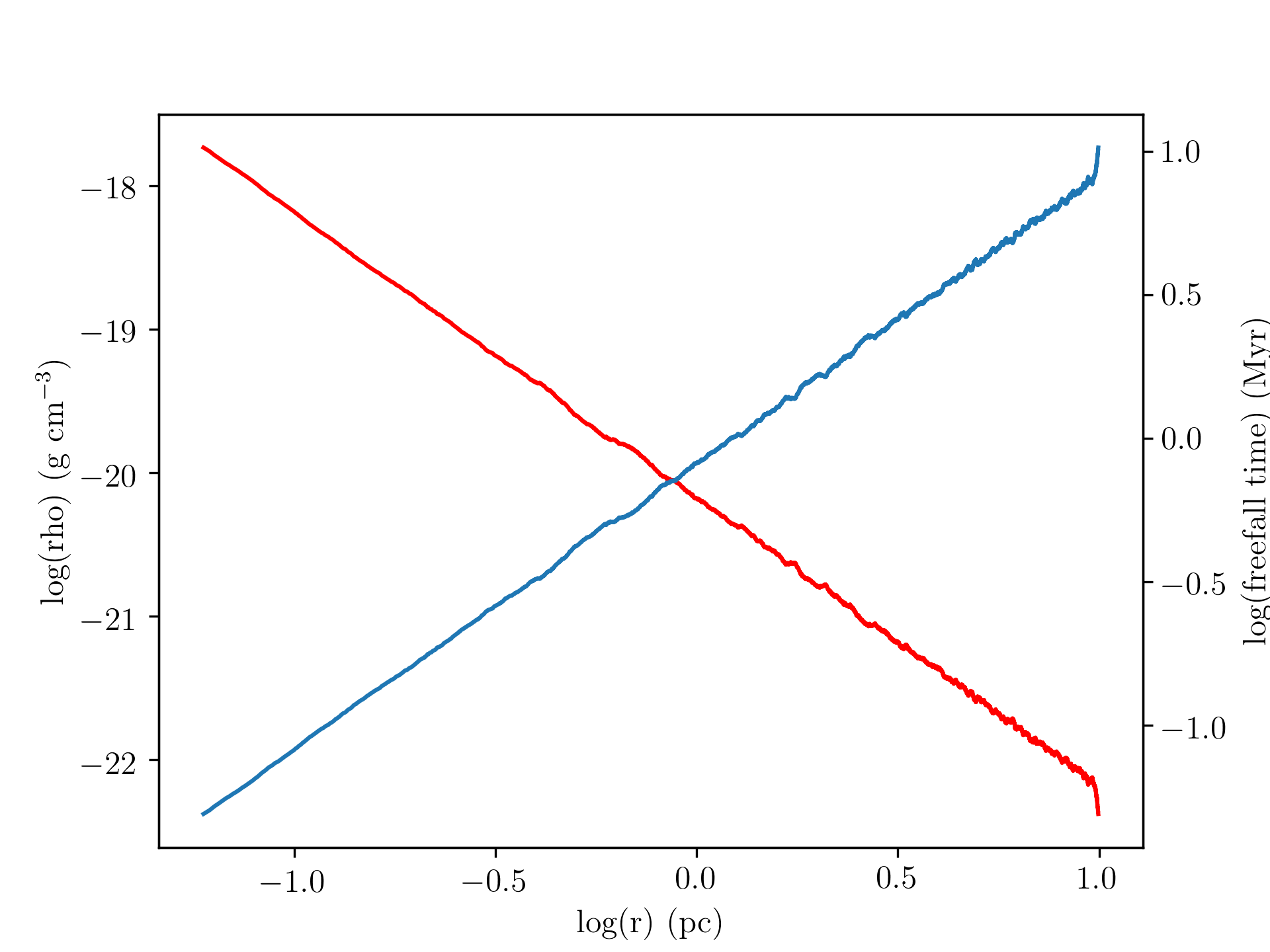}
    \caption{Red line, left axis: initial density as a function of radius. Blue line, right axis: initial freefall time, computed simple as a function of density.}
    \label{fig:density_tff}
\end{figure}

\begin{figure}
	\includegraphics[width=1.0\columnwidth]{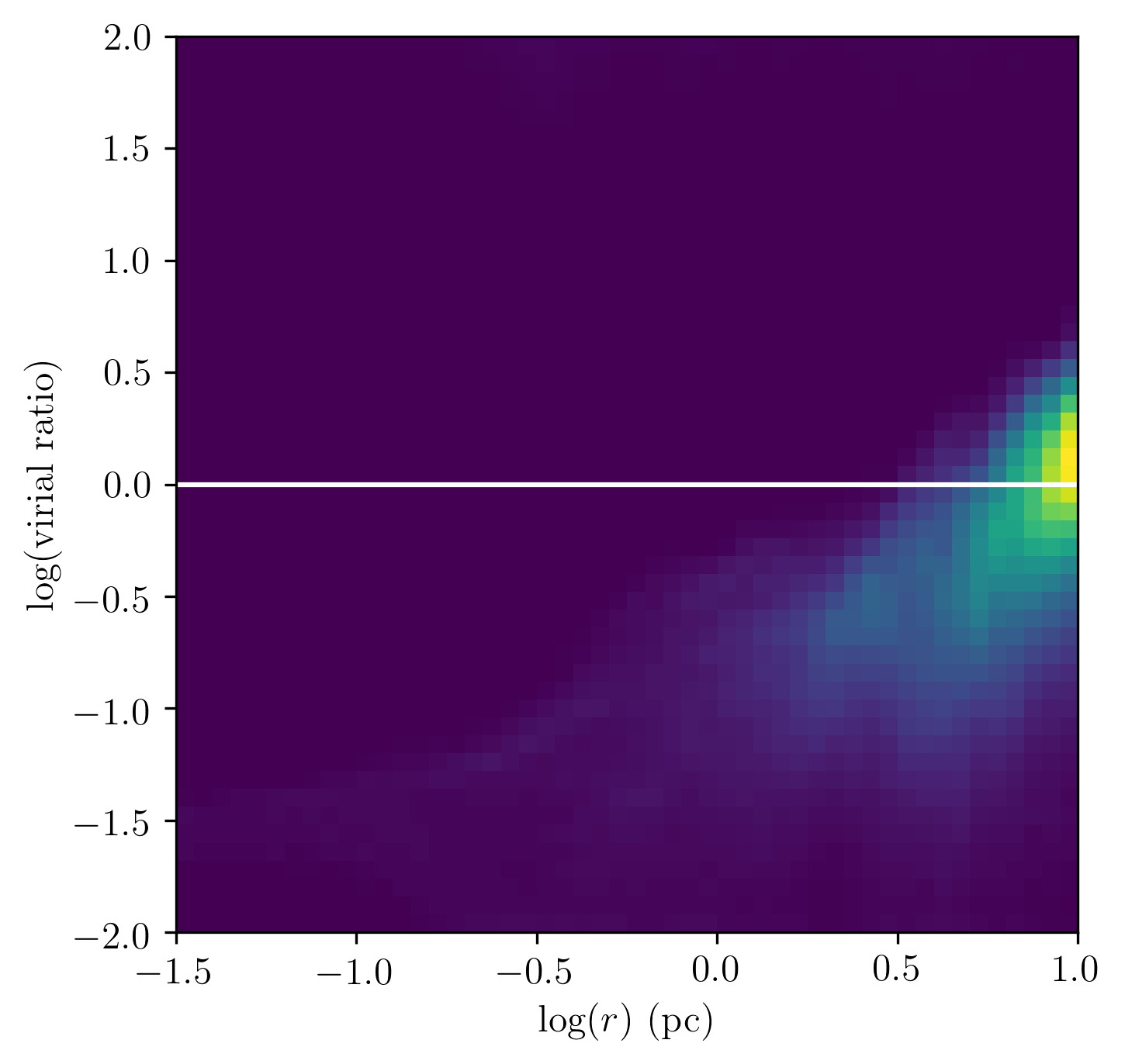}
    \caption{Two--dimensional histogram in the log($r$)--log(virial ratio) plane for the initial conditions in the bound cloud with a global virial ratio of 1.0. The white line represents a virial ratio of unity.}
    \label{fig:alpha_vs_r_bound}
\end{figure}

\begin{figure}
	\includegraphics[width=1.0\columnwidth]{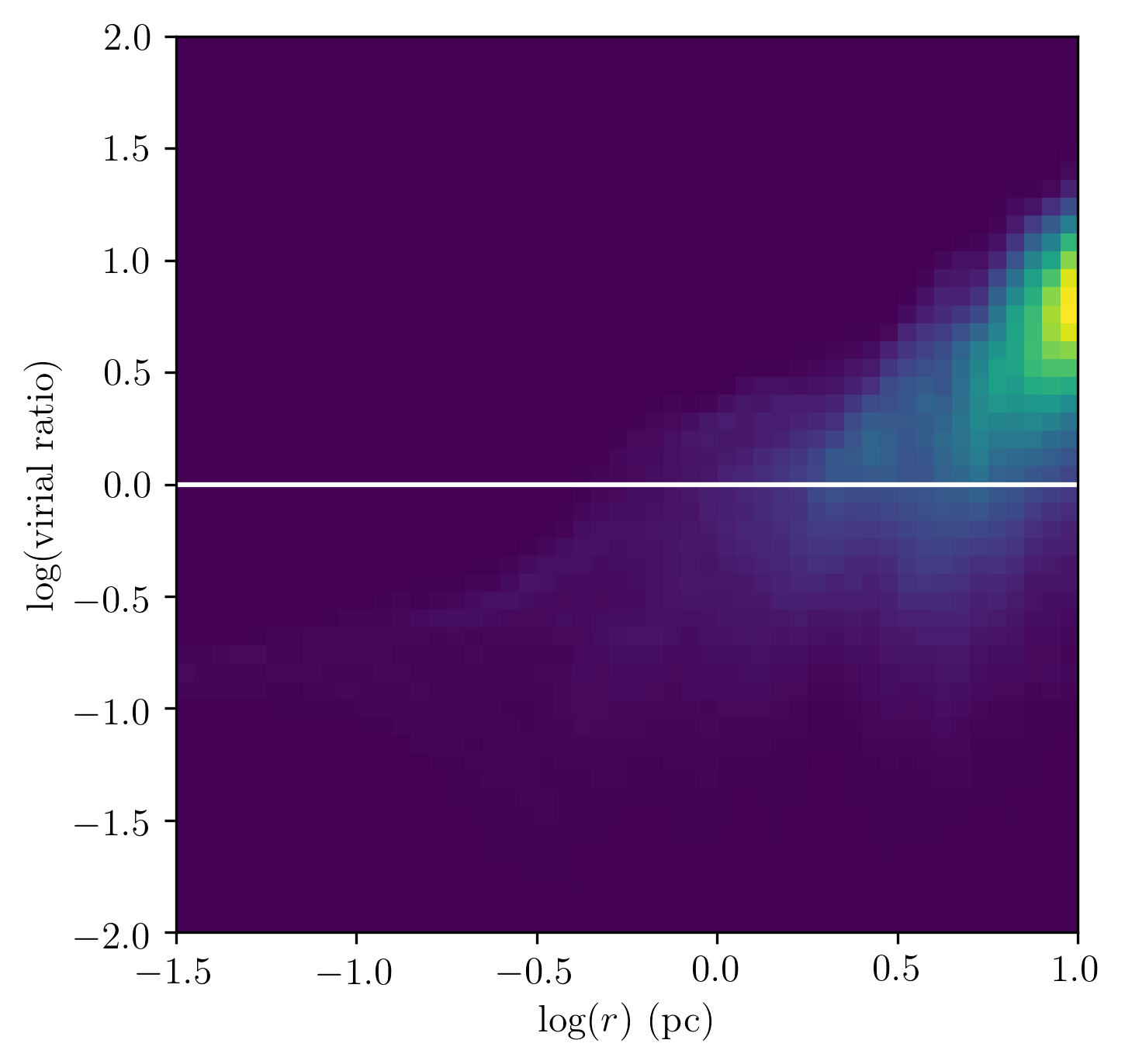}
    \caption{Two--dimensional histogram in the log($r$)--log(virial ratio) plane for the initial conditions in the unbound cloud with a global virial ratio of 5.0. The white line represents a virial ratio of unity.}
    \label{fig:alpha_vs_r_unbound}
\end{figure}

Like paper-I there are a few timescales in which some are constant while others are variable. \\
\indent Individual clouds are characterised by their masses, radii and imposed turbulent velocity dispersion. The clouds are initially modelled with a density distribution of $\rm \rho \propto R^{-2}$. The mass and radius determine the mean density $\rm 1.62\times 10^{-22}$ g$\rm cm^{-3}$ and set the \textit{global} freefall time from\\
\begin{equation}
t_{\rm ff}= \sqrt{\frac{3\pi}{32G\rho}} \approx 5.2 {\rm\,Myr}
\end{equation}
\indent However, since the density distribution is always strongly non--uniform, the global freefall time does not adequately capture the clouds' likely behaviour. The clouds are modelled with the density distribution of $\rm \rho \propto R^{-2}$. Therefore, the freefall time can be thought of as varying with radius as
\begin{equation}
t_{\rm ff} \propto R
\end{equation}
\indent Figure \ref{fig:density_tff} shows the relation between freefall time, density and radius. It shows that the freefall time in the innermost part of the cloud is $\leq10^{5}$\,yr, so the inner regions will collapse very quickly compared to the overall/global collapse of the cloud, which will occur on timescales of several Myr. \\
\indent The virial ratio of the clouds can be calculated from the following equation
\begin{equation}
\alpha = \frac{E_{\rm turb,0}}{\mid E_{\rm self,0} \mid}
\end{equation}
where the $E_{\rm turb,0}$ represents  the initial turbulent kinetic energy and $E_{\rm self,0}$ represents the self gravitational potential energy of the clouds.\\
\indent The kinetic energy can be straightforwardly computed from
\begin{equation}
E_{\rm turb,0}=\frac{M_{\rm cloud}\sigma ^{2}}{2}
\end{equation}
\indent To compute the gravitational potential energy, we compute the infinitessimal quantity of energy required to remove a spherical shell of thickness d$r$ and density $\rho(r)$ from the surface of a sphere of radius $r$ and mass $M(r)$, given by
\begin{equation}
{\rm d}E=\frac{G}{r}4\pi\rho(r)r^{2}M(r){\rm d}r.
\end{equation}
Writing $\rho(r)=\rho_{0}r_{0}^{2}r^{-2}$, 
\begin{equation}
M(r)=\int_{0}^{r}4\pi\rho_{0}r_{0}^{2}{\rm d}r=4\pi\rho_{0}^{2}r
\end{equation}
so that 
\begin{equation}
{\rm d}E=16\pi^{2}G\rho_{0}^{2}r_{0}^{4}{\rm d}r.
\end{equation}
giving the gravitational binding energy as
\begin{equation}
\mid E_{\rm self,0} \mid = \frac{GM_{\rm cloud}^{2}}{R_{\rm cloud}}
\end{equation}
\indent As with the global freefall time, the global virial ratio may give a misleading impression of the dynamical states of the clouds. Our two clouds have virial ratios of 1.0 and 5.0, and it is instructive to compute, using \textsc{gandalf}'s tree, the gravitational potential energy of every particle and compare it to that particle's turbulent kinetic energy, to obtain a `local' virial ratio. We show the result of this exercise in Figures \ref{fig:alpha_vs_r_bound} and \ref{fig:alpha_vs_r_unbound}.\\
\indent Evidently, a considerable fraction of the gas in the outer regions of the `bound' cloud is in fact unbound, while the inner regions are very strongly bound. Conversely, although most of the gas in the `unbound' cloud really is, non--negligible quantities of material at small to intermediate radii are actually bound. As will be seen shortly, these observations have important consequences for the star formation behaviour of the clouds.\\
\indent The turbulent crossing time/dissipation time is set by the velocity dispersion.\\
\begin{equation}
t_{\rm diss} \approx \frac{2R_{cloud}}{\sigma}
 \end{equation}
\indent Turning to the actual collisions, these are characterised by three parameters; the impact parameter $b$; the relative velocity at infinite separation $\rm v_{0}$; the initial cloud separation $d$. If $\rm d >> b$, the elapsed time before the clouds strike each other -- the collision time -- can be determined by
\begin{equation}
t_{\rm coll} \approx \frac{d}{v_{0}}
 \end{equation}
\indent The cloud crushing time is defined as the time between when the clouds first touch and the time when as much of the cloud as is going to has entered the shocked collision region. The cloud crushing time, $\rm t_{crush}$ is,\\
\begin{equation}
t_{\rm crush} \approx \frac{2R_{\rm cloud}}{v_{0}} \approx {\rm2.0\,Myr}
\end{equation}
\indent These timescales are expected to have a significant role in the outcome of the simulations. If the collision time is shorter than the turbulent crossing time, the window is small for structure to develop before the collision or for the gas to enter the shocked region. The clouds would then be relatively smooth when they collide.\\ 
\indent The unbound clouds are expected to expand. If the collision time is short compared with the turbulent crossing time, the degree of expansion will be small.\\
\indent A key difference between the simulations presented here and those in Paper I is that the clouds studied in this paper are centrally condensed and the freefall time hence varies substantially with radius, as shown in Figure \ref{fig:density_tff}. The central freefall times of the clouds are shorter than any of the above timescales. Additionally, both clouds have bound material at their centres. We therefore expect both clouds to initiate star formation in their core regions long before they collide. In a companion paper (Tanvir et al. 2020,in prep.hereafter Paper IV), we explore the influence of the impact parameter and relative velocity along with the initial density profile in cloud-cloud collision.\\
\indent The speed of sound $c_{s}$ at 30K for a mean molecular weight of 2.36 amu is 324 m/s. The Mach number of the \textit{collision} is $v_{0}/c_{s}$=30.9 for all simulations.  The Mach number of the \textit{turbulence} is given by $\sigma/c_{s}$, which is 5.84 and 13.1 for the bound and unbound clouds respectively. No background medium is present in these simulations.\\
\indent One of the main targets of this paper is to understand the effect of collision on star formation efficiencies and rates. To understand the effect of the collision better we have also run control simulations where the clouds evolve in isolation.\\
%Normally the next section describes the techniques the authors used.
%It is frequently split into subsections, such as Section~\ref{sec:maths} below.

%\subsection{Maths}
%\label{sec:maths} % used for referring to this section from elsewhere

%Simple mathematics can be inserted into the flow of the text e.g. $2\times3=6$
%or $v=220$\,km\,s$^{-1}$, but more complicated expressions should be entered
%as a numbered equation:

%\begin{equation}
 %   x=\frac{-b\pm\sqrt{b^2-4ac}}{2a}.
%	\label{eq:quadratic}
%\end{equation}

%Refer back to them as e.g. equation~(\ref{eq:quadratic}).

\section{Results}
This section is divided into four subsections. The first discusses the morphology and gross evolution of the models, the second explores the star formation efficiencies and rates, the third examines the angular momentum distribution within the collision products, and the fourth discusses the formation of bound structures.\\ 
\subsection{Morphology of Collision Simulations}
\subsubsection{Run BB\_rho\_r-2}
\begin{figure}
	\includegraphics[width=1.0\columnwidth]{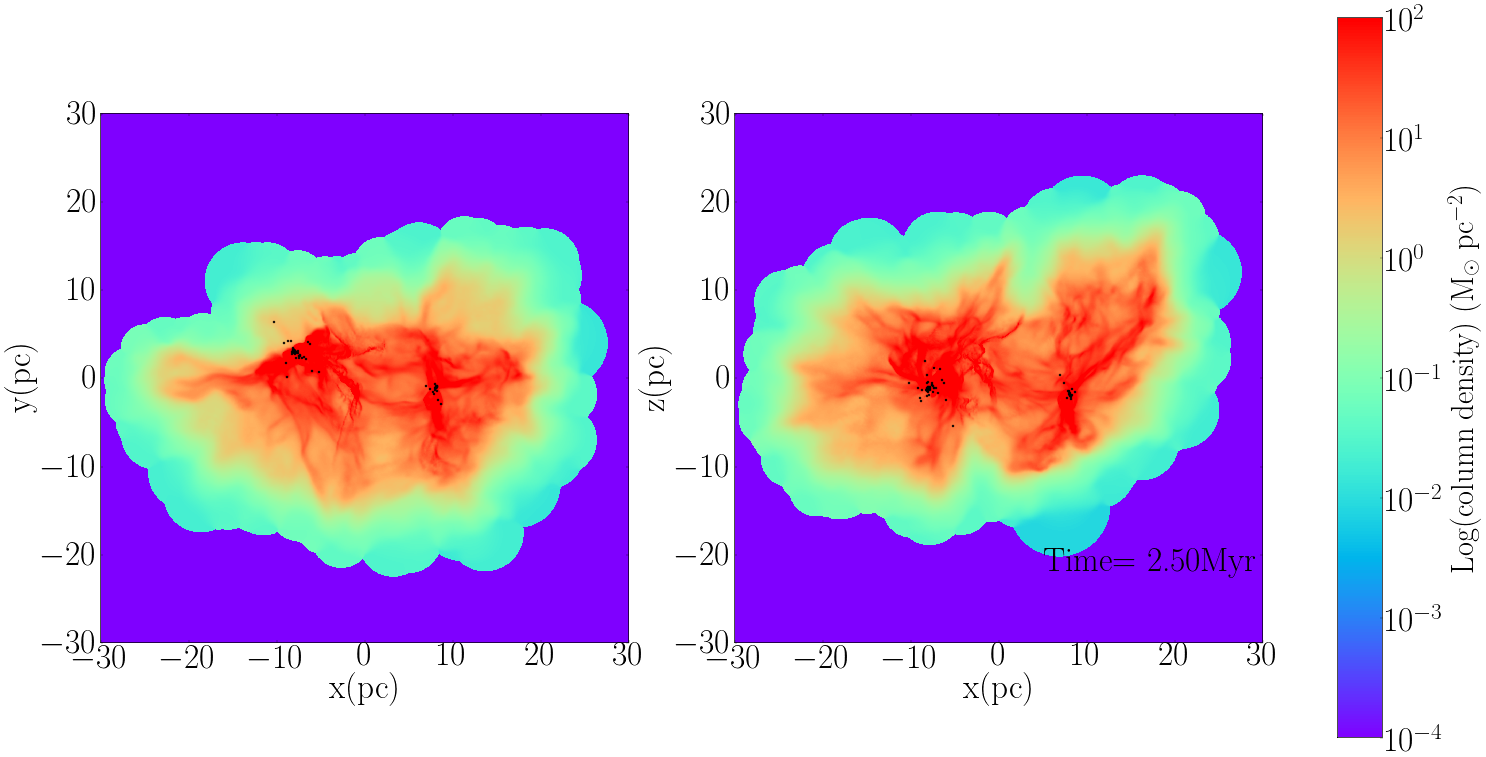}
		\includegraphics[width=\columnwidth]{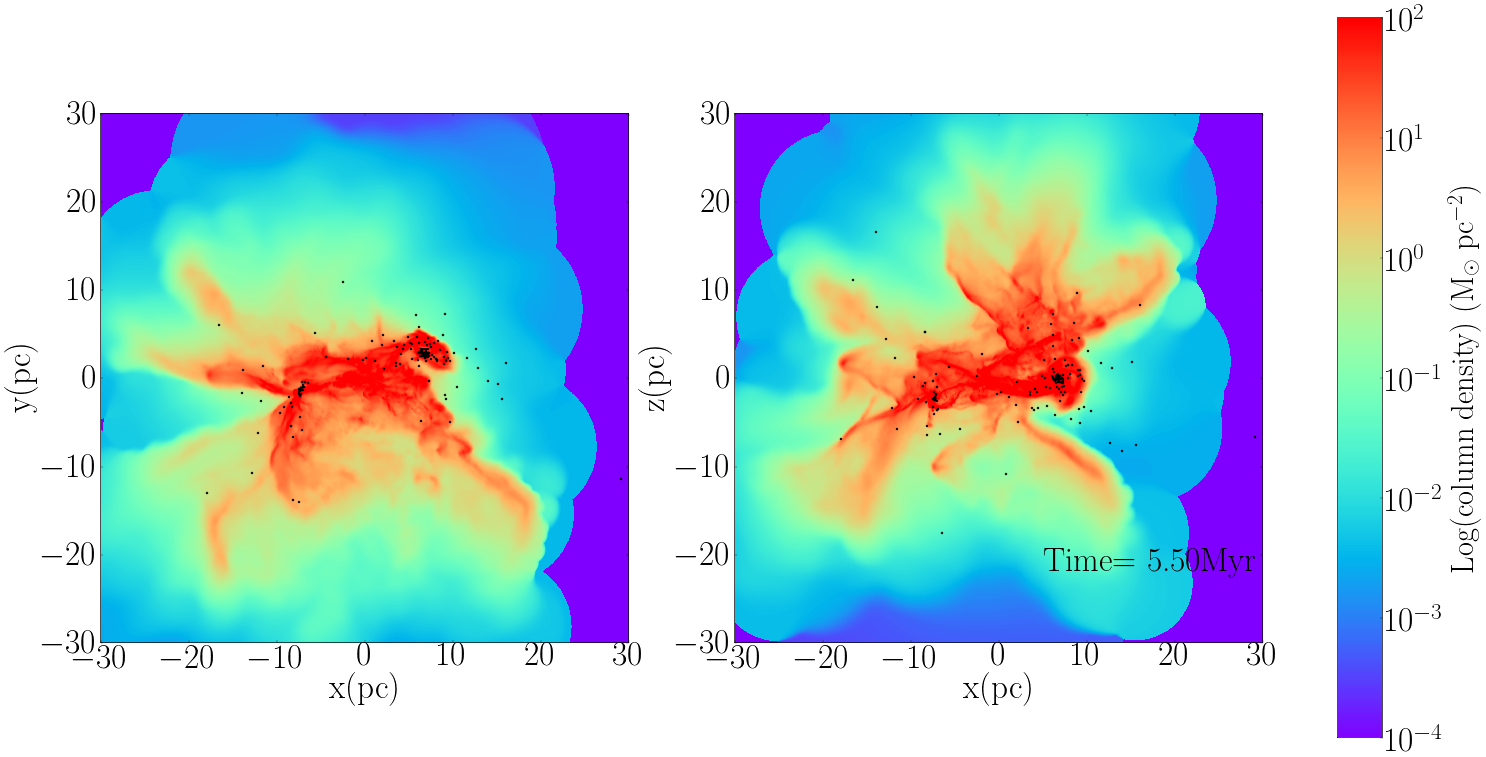}
			\includegraphics[width=\columnwidth]{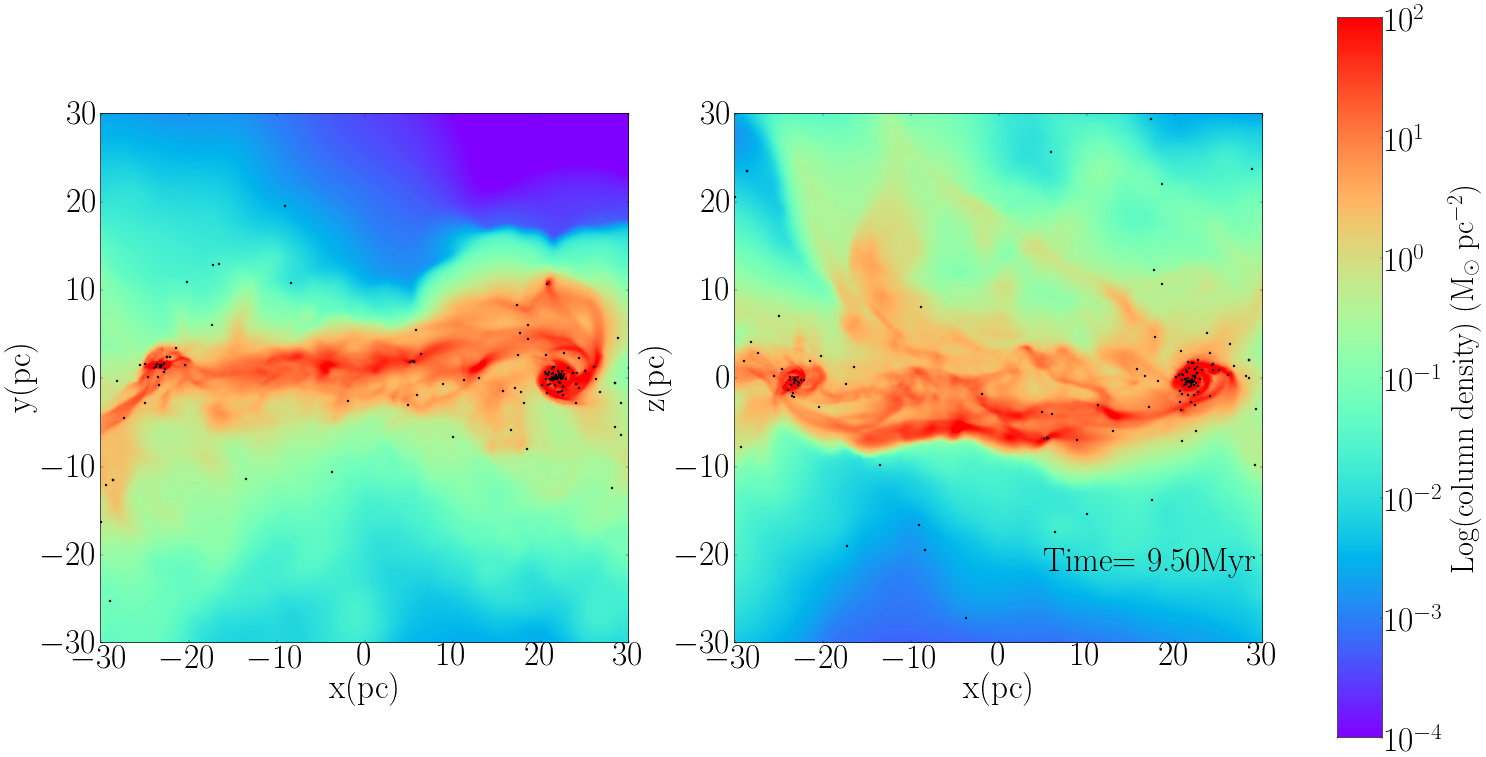}
    \caption{Column-density plots from Run BB simulation. The first plot is shortly after the cloud crushing time at 2.5 Myr. The second plot is an intermediate plot at 5.5 Myr. At this time it is harder to separate the two clouds from one another. The third plot shows the state of the clouds at 9.5 Myr. At this time the remainder of the clouds are clearly separated.}
    \label{fig:runA}
\end{figure}

\begin{figure}
	\includegraphics[width=1.0\columnwidth]{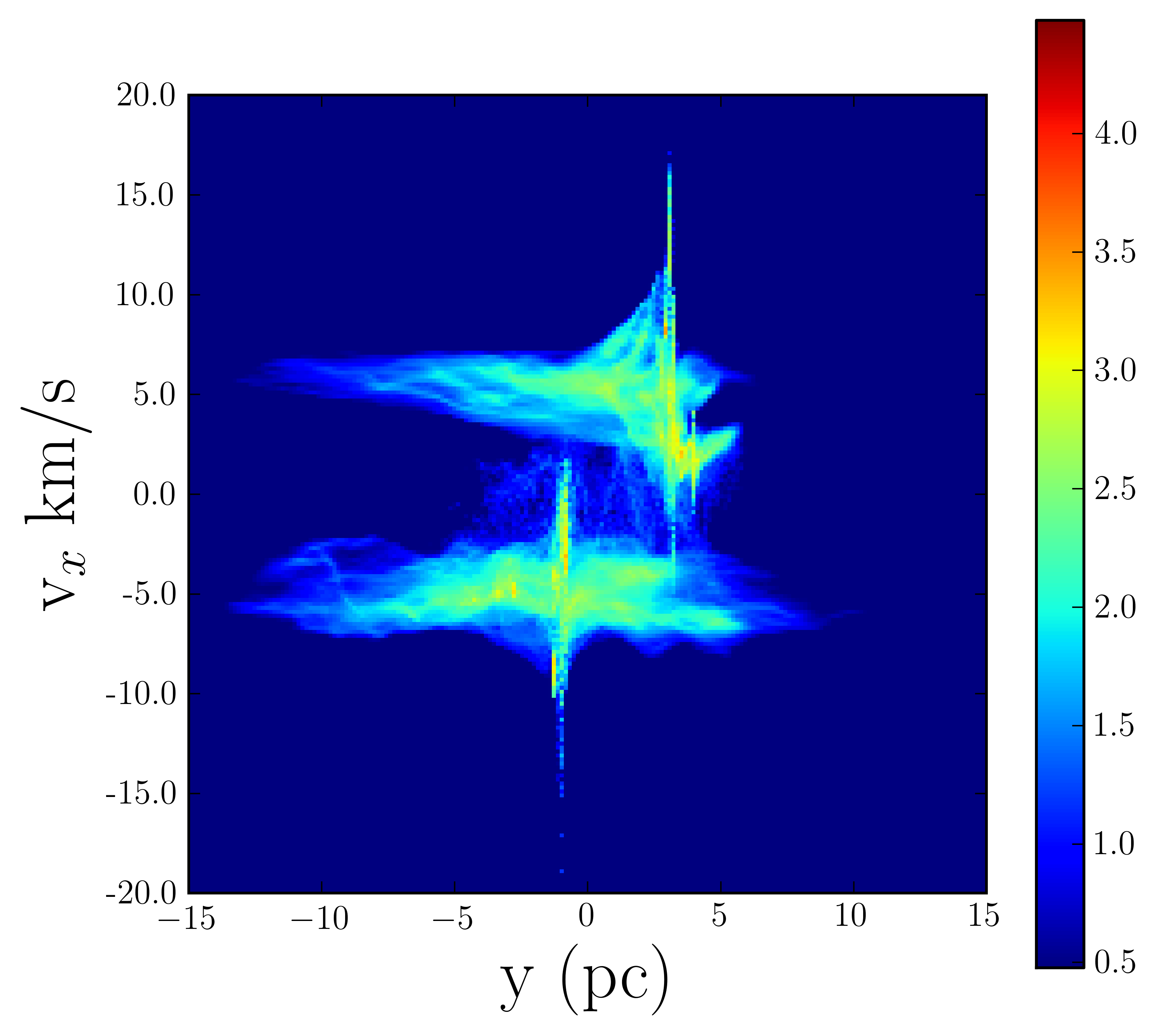}
		\includegraphics[width=\columnwidth]{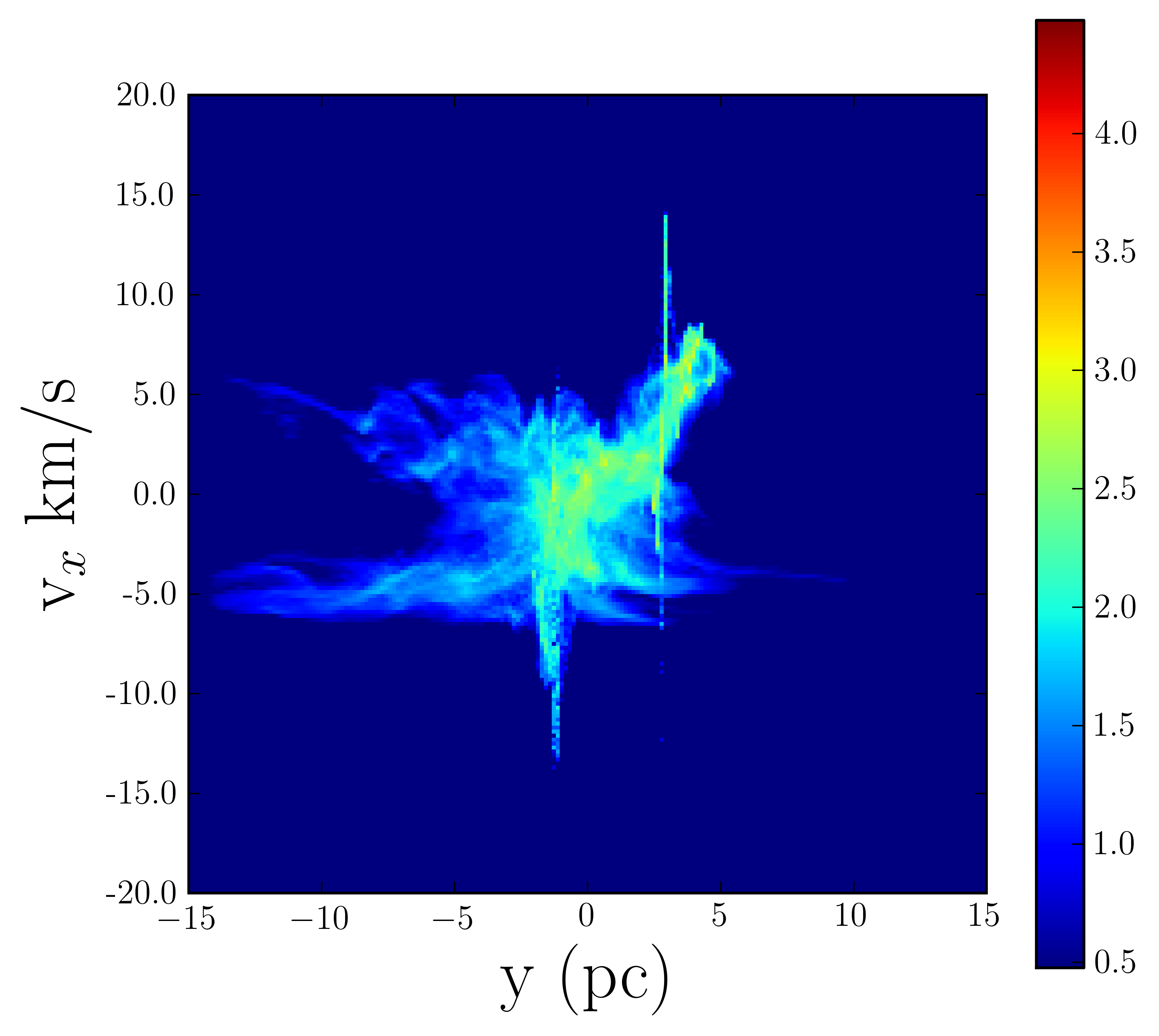}
			\includegraphics[width=\columnwidth]{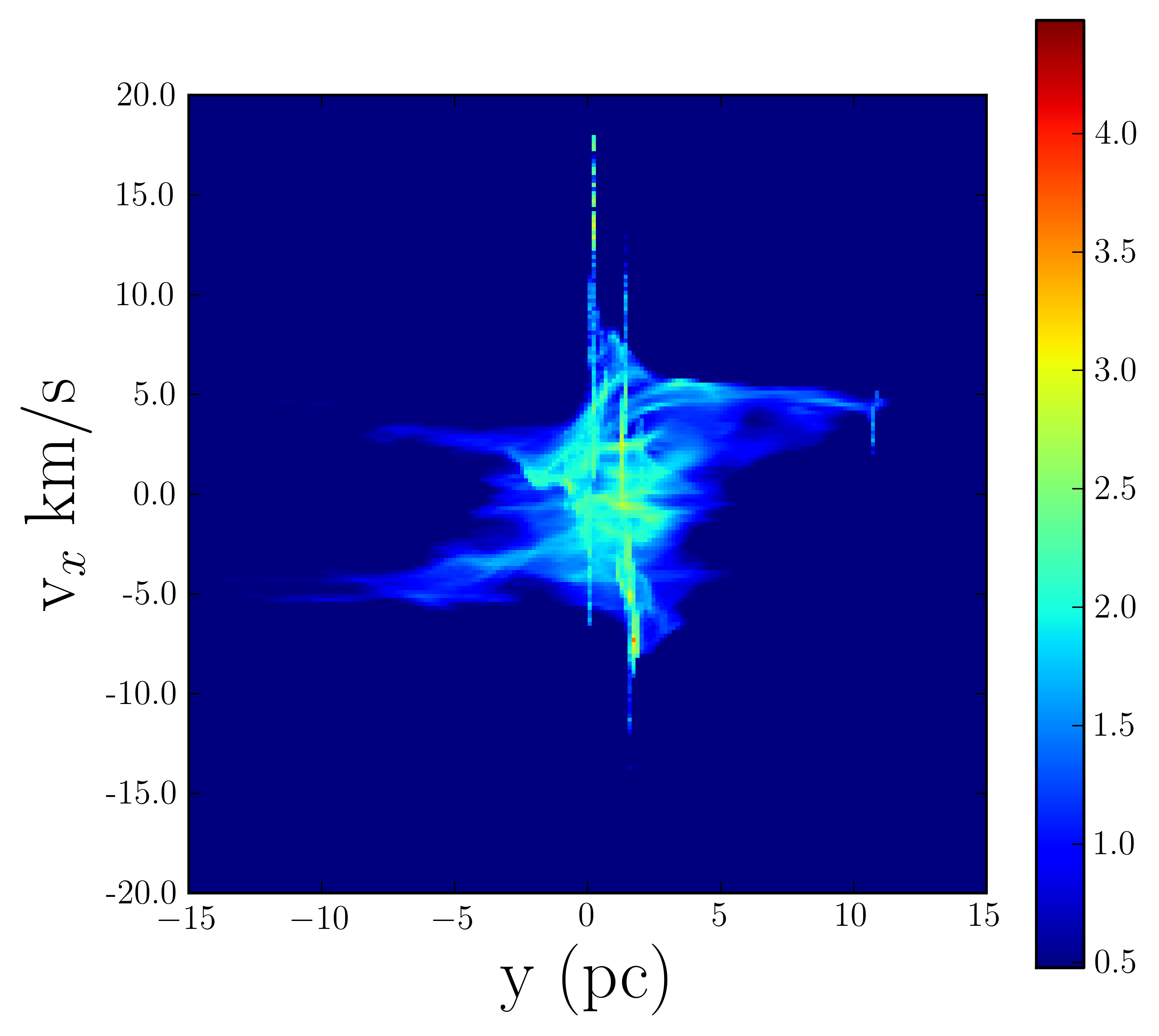}
    \caption{Position-velocity plots from Run BB simulation.The first plot is taken prior to the collision where the clouds can be separated by their pre collisional relative velocity. The second plot shows the state of the clouds during the collision. The third plot is taken after the clouds have collided. The timeline for these plots are the same timeline as the column-density plots.}
    \label{fig:runA_PV}
\end{figure}

\begin{figure}
	\includegraphics[width=1.0\columnwidth]{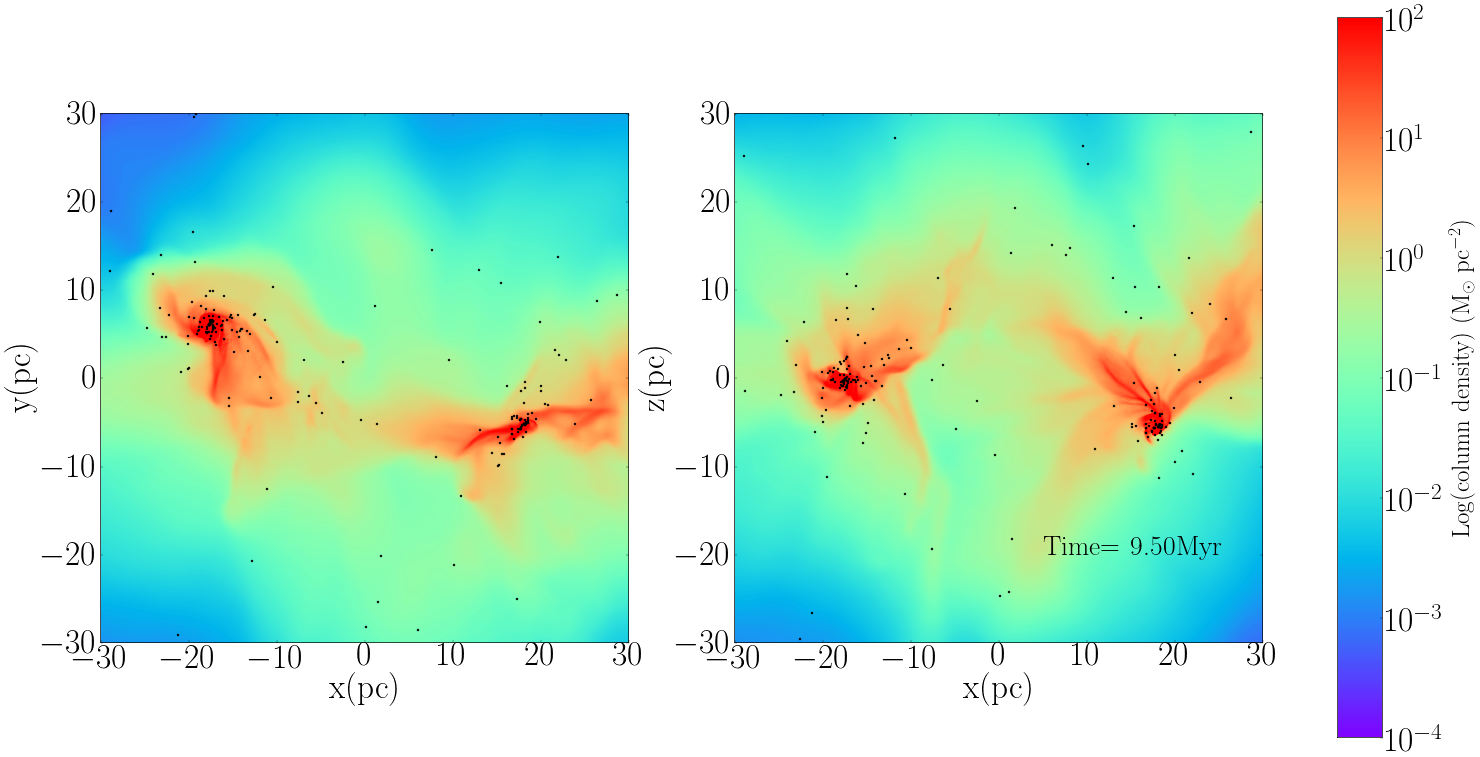}
		\includegraphics[width=\columnwidth]{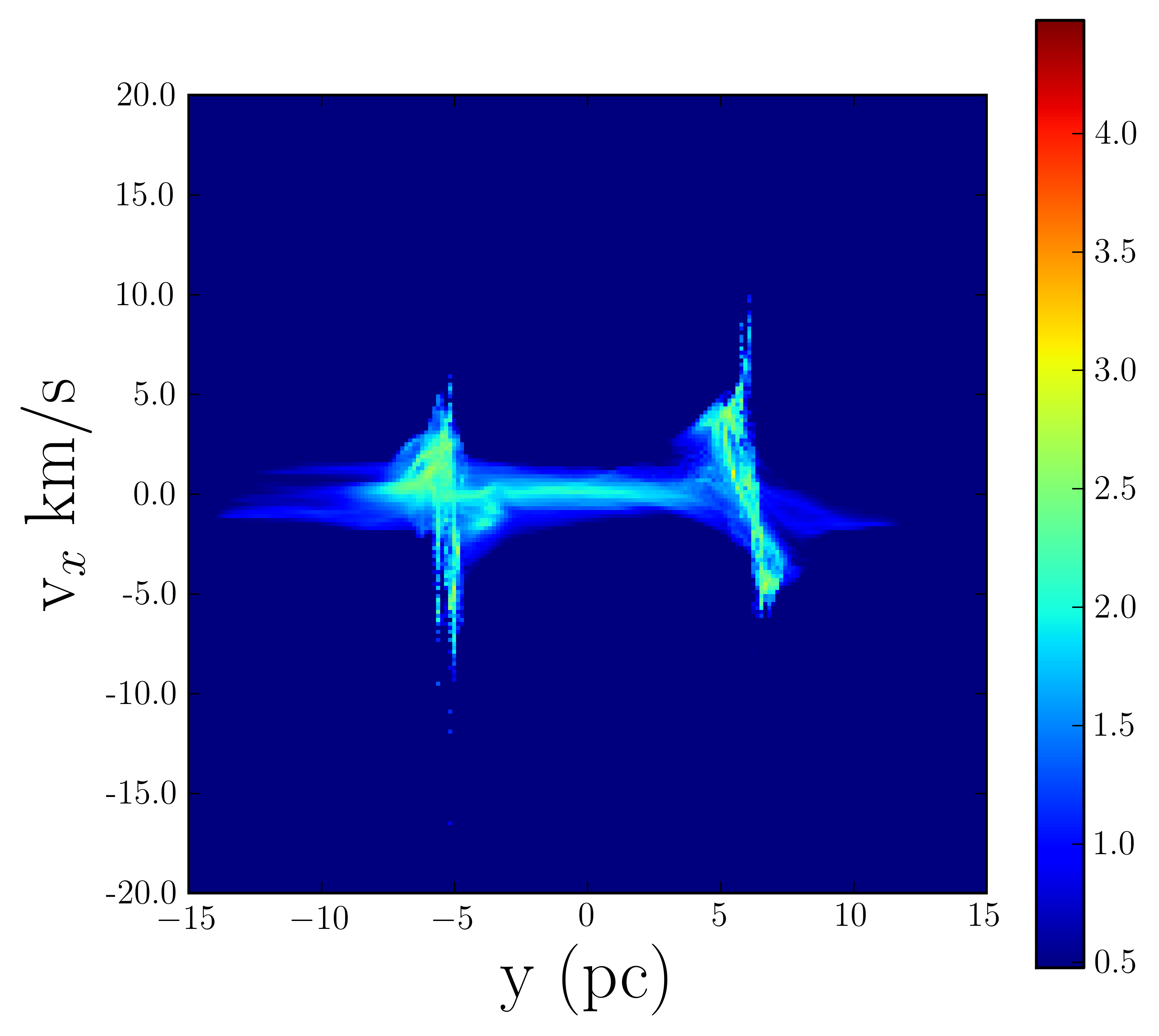}
		\caption{Column-density and position-velocity plots from the controlled runs of run BB simulation. Both these plots are taken at 9.5 Myr of the simulation. From the position-velocity diagram it is hard to distinguish the clouds since they both occupy the same width along the velocity axis.}
    \label{fig:runAcontrol}
\end{figure}

In this run, both clouds are initially gravitationally bound with a global virial ratio of 1. The collision time here is 1.96 Myr.\\
\indent We first examine the morphology of the collision using column density position--position plots at three epochs imaged along the two directions orthogonal to the collision axis, shown in Figure \ref{fig:runA}. The first plots, shown in the upper panels of the figure, are from 2.50 Myr, $\approx$ 0.6 Myr after the collision time (which we define as the moment when the clouds first touch).\\
\indent The collision time is shorter by a factor of about two than the turbulent crossing time, but the turbulence has nevertheless generated significant substructure in both clouds. The collision time is also shorter than the clouds \textit{global} freefall times, and there is thus no evidence of global collapse or shrinking. However, as discussed in Section 2 above, the steep density profiles of clouds leads to very short \textit{local} freefall times in the clouds' centres, and Figure \ref{fig:runA} shows that star formation has already initiated there, forming two very compact clusters. presence of two star clusters at the centres of the clouds. This is in stark contrast to the simulations presented in Paper I, in which star formation activity was delayed until  $\approx$ 3 Myr, after the onset of the collision, and was more randomly distributed through the cloud volumes.\\
\indent The second pair of position-position plots, shown in the middle panels of Figure \ref{fig:runA}, is taken around one global freefall time. The two clusters have now passed each other -- even though the impact parameter of these simulations is zero, the clusters formed by the condensed clouds are so small that even a minor component of the turbulent velocity field acting on the central gas perpendicular to the collision axis is sufficient to cause the clusters to miss each other.\\
\indent Both star clusters, but particularly that on the right hand side of these images, have grown significantly in numbers of stars and are sufficiently crowded that numerous stars have been ejected by dynamical encounters.\\
\indent Both clusters are still evidently accreting from local dense gas, and the structures in the gas are suggestive of rotation. The overall gas has started to acquire a more elongated distribution compared to the distribution in the first position-position plot.\\
\indent The third pair of panels in Figure \ref{fig:runA} are taken at $\approx$ 7.5 Myr after the clouds have collided (i.e. about 5.0 Myr after the images in the first panels). The remaining gas is now mainly part of a dense filamentary structure connecting the cores of the two star clusters, in a similar fashion to that described in Paper I. This is due once again to the densest parts of the clouds being largely unaffected by the collision, and being thus able to propagate themselves the farthest from the point of impact.\\
\indent Star formation is evidently not occurring uniformly along the filament, but is mostly confined to the densest material in the potential wells of the clusters. Rotation in this material is even more clearly visible, and will be discussed in detail below. Although star formation is concentrated around the clusters, significant numbers of stars have been ejected not only from the clusters, but from the filament as well, and these objects must therefore be starved of fresh material to accrete.\\
\indent Since position-position plots contain limited dynamical information, we turn to position-velocity plots, imaged \textit{along} the collision axis which is here the $x$--axis, and with the $y$--axis representing the space dimension of the plots.\\
\indent Figure \ref{fig:runA_PV} shows the position-velocity at the same stages of simulation BB as the position--position plots. Note that we take no account of optical depth effects in generating these images.\\
\indent In the first panel, much of the vertical thickness of the clouds is determined by their turbulent velocity dispersions and the clouds are separated by their original relative velocity, 10 \,km\,s$^{-1}$. However, it is also clear that the compact central regions of the clouds are collapsing, due to their short freefall times, leading to almost linear vertical structures in the plots, with the strongest features being due to high velocity accretion onto one of the sink particles. In reality, much of these structures would be optically thick and therefore potentially unobservable. We plan to investigate this issue using detailed radiation transport modelling to generate more realistic synthetic observations in a future paper.\\
\indent The clouds are also connected by a small amount of material which has been decelerated by the collision from the individual rest frames of the clouds to close to the zero--momentum frame of the whole simulation. This is the beginning of the broad-bridge feature characteristic of cloud--cloud collisions in PV diagrams  \citep{2015MNRAS.450...10H,2015MNRAS.454.1634H}. However, these features are still useful, since they point out where in the position-velocity diagram star formation activity is occurring. The first panel in the PV plot shows that star formation activity is present and is very strongly spatially concentrated near the clusters at the centres of the clouds.\\
\indent The second panel in Figure \ref{fig:runA_PV} shows the state of the simulation at 5.50 Myr. It is still just possible to identify the original clouds, still separated by their original relative velocity. It can be seen that most of the clouds' combined mass have now entered the shocked broad--bridge layer, near the collision axis in space and zero on the velocity axis. Around $y=0$ and $\rm v_{x}$= 5 \,km\,s$^{-1}$ it can be seen materials from the clouds travelling in the negative $x$-direction has been decelerated. The vertical spikes {\ seen in the PV plots indicate the presence of accretion flows onto individual sinks, which produce a wide range of velocities concentrated in a small spatial extent. In reality, these flows would be optically thick and almost certainly not observable. However, they serve a useful purpose in our plots in revealing active star formation.} In Figure \ref{fig:runA_PV}, they show that star formation activity is well underway and suggest that it is mostly concentrated around the broad bridge region, but there is a second structure offset from the collision axis by $\approx 3$\,pc, indicating significant star formation activity in one of the clusters which has drifted off the collision axis.\\
\indent The last panel of Figure \ref{fig:runA_PV} shows the state of the simulations around 7.5 Myr after the collision time. The densest part of the clouds have to some extent survived the collision and have retained their identity and much of their original relative velocity. Star forming regions are roughly aligned along the collision axis at $y=0$. It is now all but impossible to discern the system as two separate objects, however.\\
%control runs will be added in the end%
\indent As we did in Paper I, we have also performed control versions of each run, which are identical except that the clouds are given initial bulk velocities of zero, so that we can observe how they evolve in isolation. Figures \ref{fig:runAcontrol} and \ref{fig:runA_PV} show column--density and position--velocity plots of the the state of the control run BB at 9.5 Myr. No filamentary structure have been formed in this controlled run unlike its collisional counterpart. The two clouds have formed centrally--condensed star clusters which, as in the counterpart run, have ejected numerous low--mass members and, as shown by the accretion structures in the PV plot, are the exclusive sites of star formation. The position-velocity plot clearly shows signatures of infall onto the clusters. Interesting, there is a prominent diagonal feature in the PV plot centred on the right--hand cluster, indicating the presence of a rotating structure, which can perhaps also be discerned in the the $xz$ column--density image. This suggests that the rotating structures seen in the collisional run BB are not necessarily driven by the collision, but are at least in part intrinsic to the original model clouds. We will discuss this issue in more detail later in this section.\\
\subsubsection{Run BU\_rho\_r-2}
\begin{figure}
	\includegraphics[width=1.0\columnwidth]{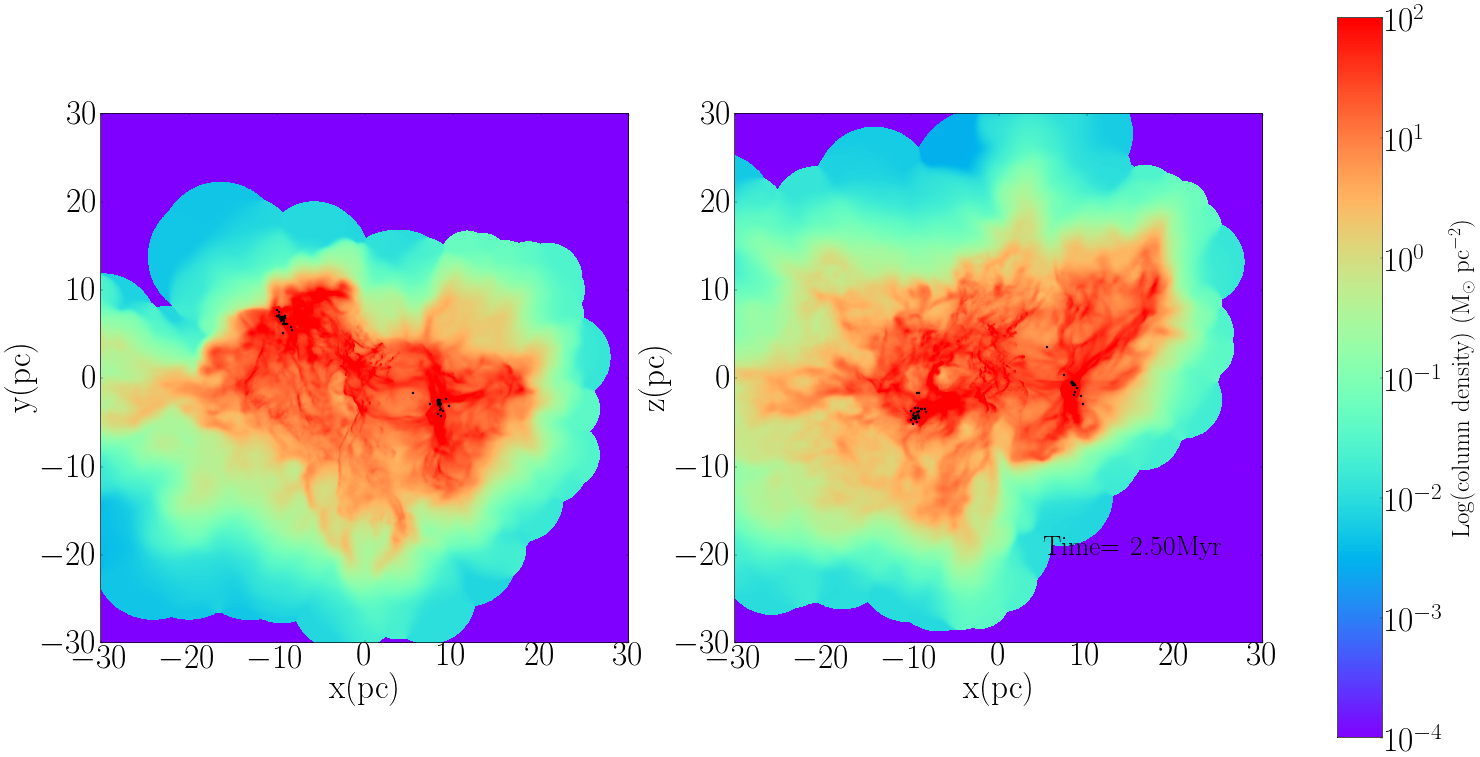}
		\includegraphics[width=\columnwidth]{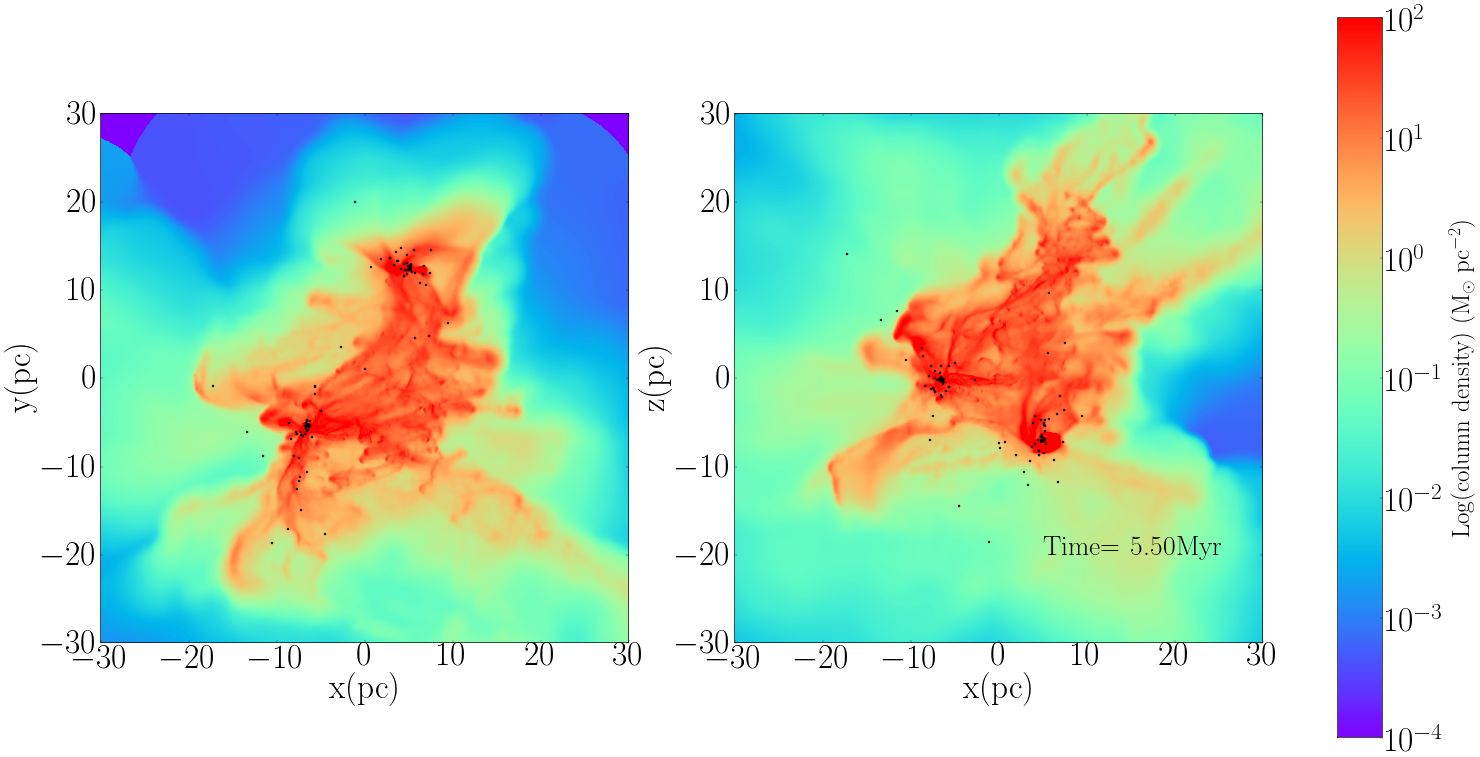}
			\includegraphics[width=\columnwidth]{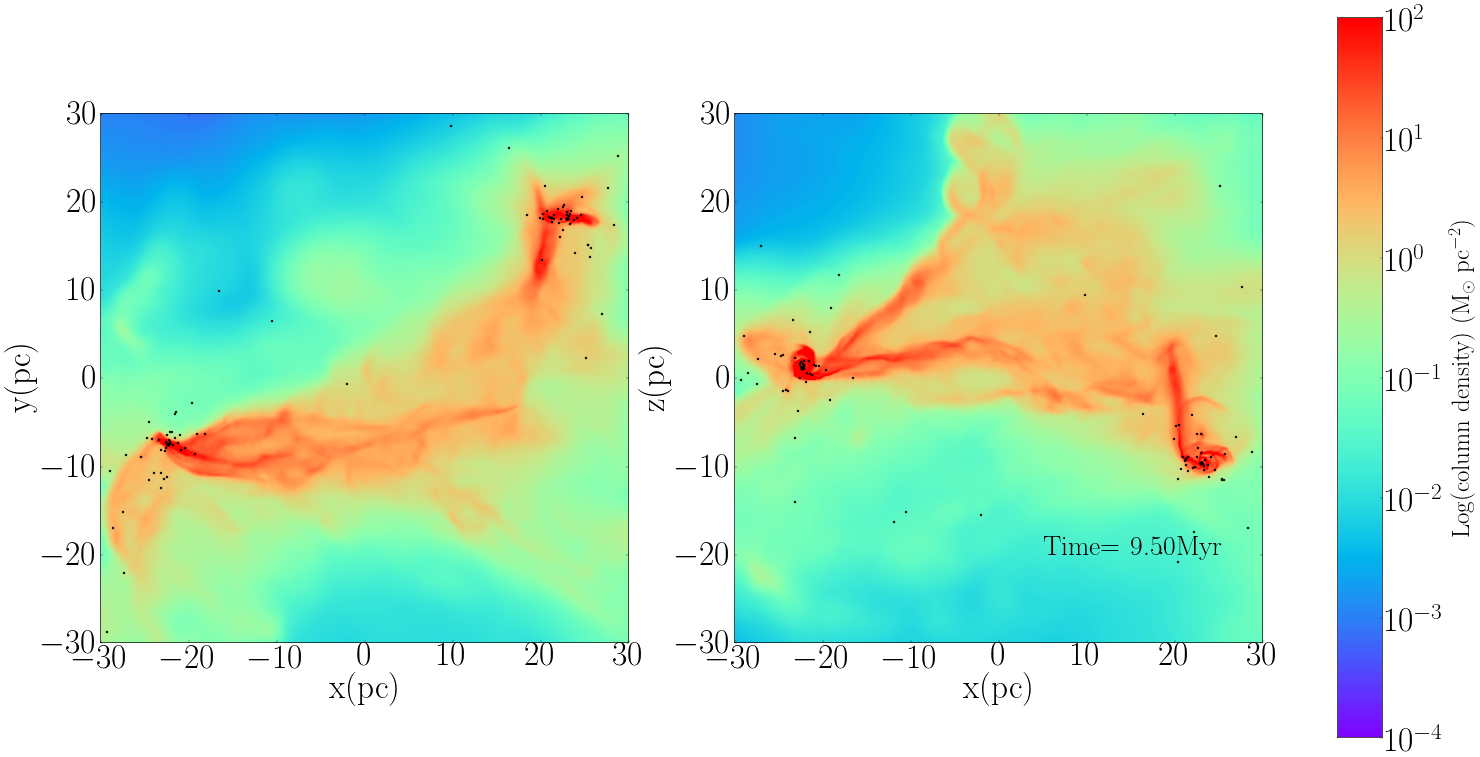}
    \caption{Column-density plots from Run BU simulation. The first plot is shortly after the cloud crushing time at 2.5 Myr. The second plot is an intermediate plot at 5.5 Myr. At this time it is harder to separate the two clouds from one another. The third plot shows the state of the clouds at 9.5 Myr. At this time the remainder of the clouds are clearly separated.}
    \label{fig:runB}
\end{figure}

\begin{figure}
	\includegraphics[width=1.0\columnwidth]{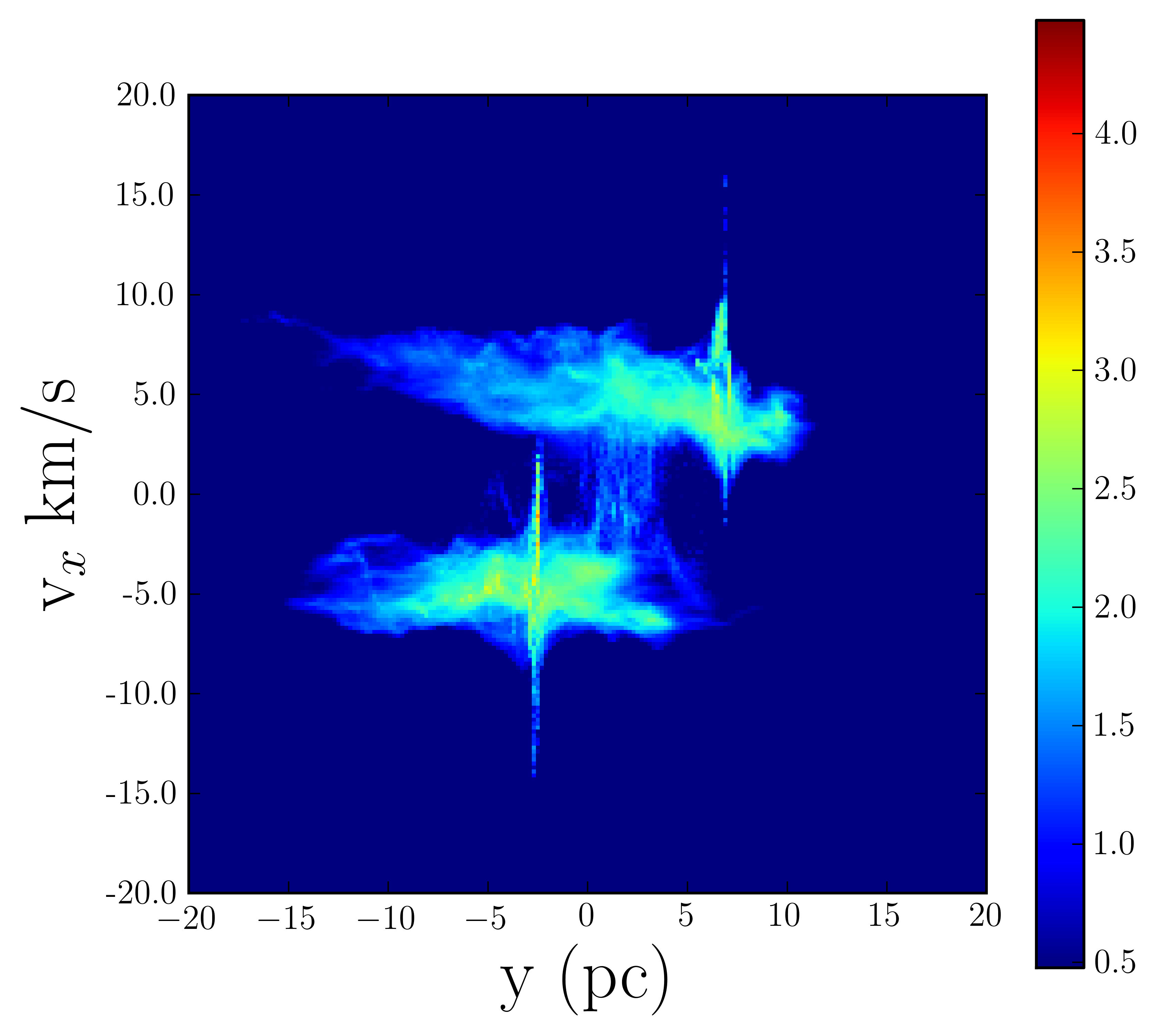}
		\includegraphics[width=\columnwidth]{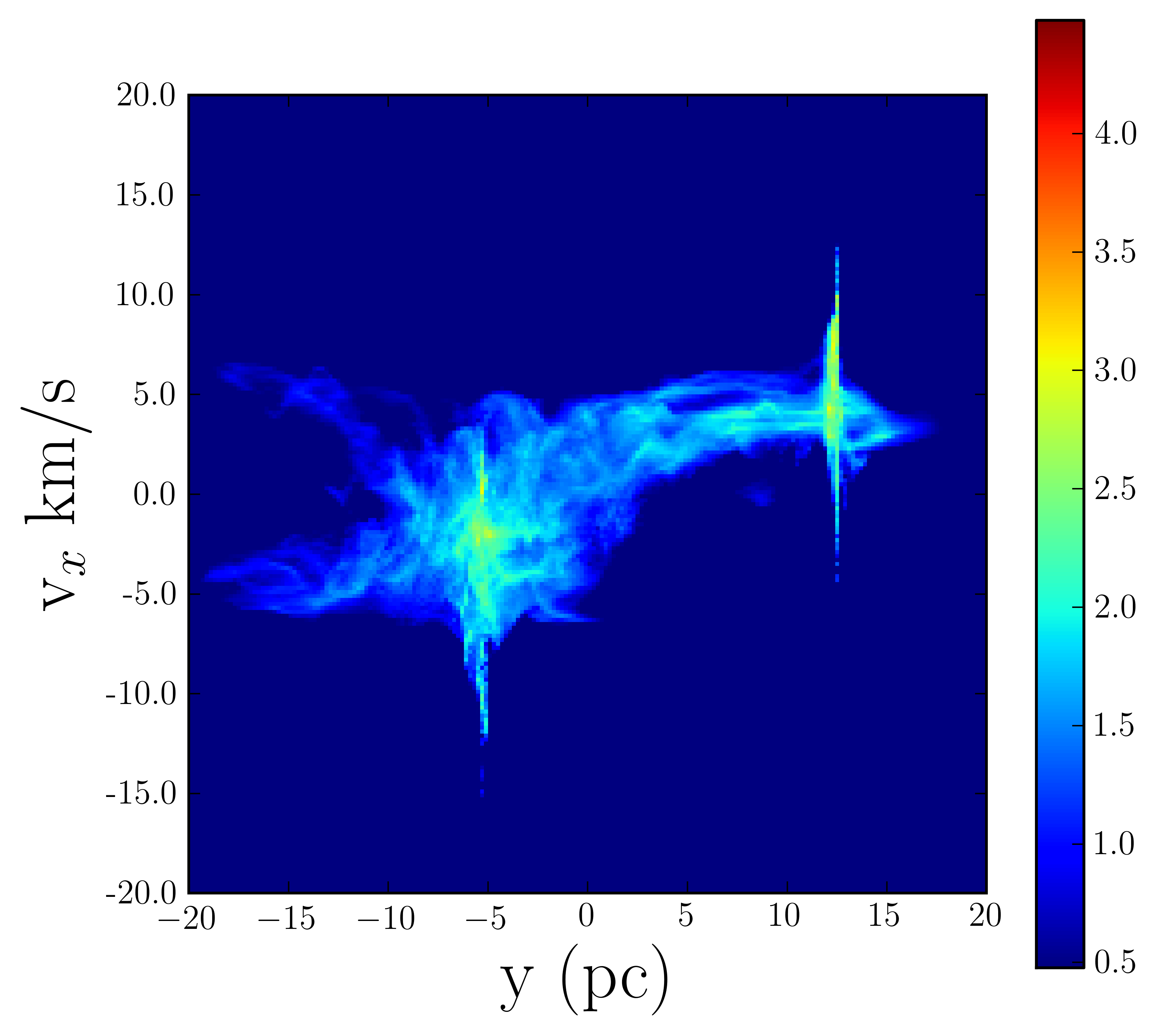}
			\includegraphics[width=\columnwidth]{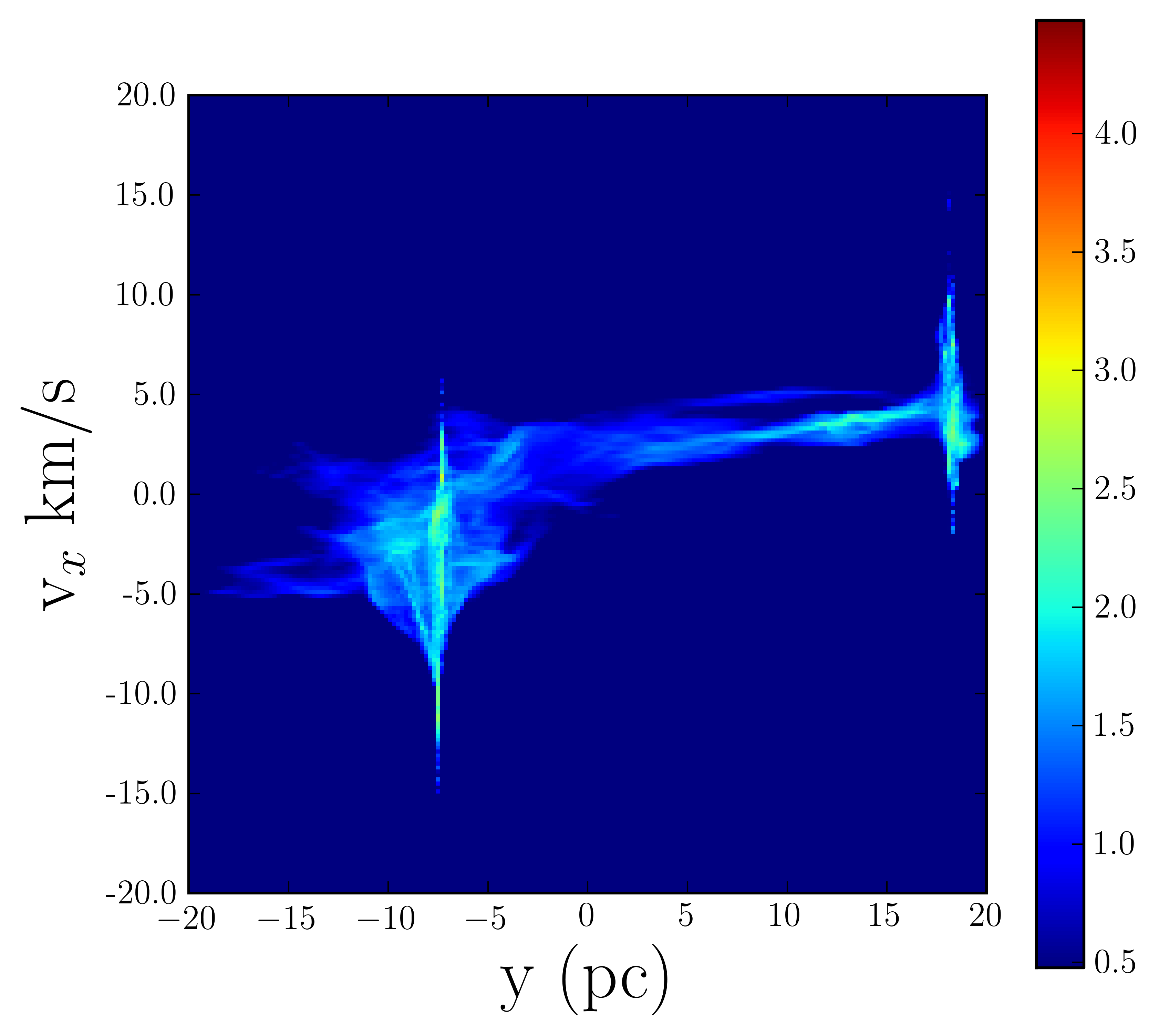}
    \caption{Position-velocity plots from Run BU simulation.The first plot is taken prior to the collision where the clouds can be separated by their pre collisional relative velocity. The second plot shows the state of the clouds during the collision. The third plot is taken after the clouds have collided. The timeline for these plots are the same timeline as the column-density plots.}
    \label{fig:runB_PV}
\end{figure}

\indent In this run the virial ratios of the clouds are 1.0 and 5.0 which means that one of the clouds is much more gravitationally unbound than the other cloud. In the case of uniform clouds, such a high virial ratio would be expected to strongly suppress star formation. However, it is not obvious that this should still be the case in such strongly condensed clouds.\\
\indent As in Run BB, in this simulation the clouds collide at 1.96 Myr. Figure \ref{fig:runB} shows column-density plots at the same three epochs of the simulation chosen in Figure \ref{fig:runA}, with the unbound cloud starting at a negative value of $x$, on the left of the images and moving from left to right. Comparing with the first panel of Figure \ref{fig:runA}, it is clear that the \textit{outer regions} of the unbound cloud have expanded around but that the denser regions are relatively unaffected. This observation is not surprising since, as shown in Figure \ref{fig:alpha_vs_r_unbound}, the outer regions of the unbound model cloud are indeed unbound, but the inner, denser regions, are at least somewhat bound.\\
\indent As in Run BB star formation is active in both clouds' centres/cores and two clusters have been formed, showing that the high global virial ratio of the unbound cloud has failed to prevent it forming stars. As we will show below, the expanding outer region of the unbound cloud does have the effect of \textit{slowing} the total star formation rate in this simulation, simply by reducing the rate at which the central cluster is able to accrete from the rest of the cloud.\\
\indent From the middle panels of Figure \ref{fig:runB} it is very difficult to tell the original clouds apart, and the two clusters have once again missed each other. The higher turbulent velocities in the unbound cloud (which is on the \textit{right} of the images in the middle and lower panels) has caused the clusters to miss each other by an even larger margin.\\
\indent The third row of Figure \ref{fig:runB}, taken at 9.50 Myr, shows the two star clusters widely separated. Unlike Run BB, and the corresponding bound--unbound uniform--cloud collision in Paper I, there is no strong central filament connecting the two clusters, and they appear to be completely independent objects.\\
\indent Once again, both clusters are at the centres of features which strongly suggest rotation. In particular, the left--hand cluster, which devolved from the \textit{bound} cloud, is evidently accreting from a disc--shaped structure several parsecs across, which is in turn being fed by two $\sim10$\,pc--scale filaments contacting the disc at a significant angular separation.\\
Figure\ref{fig:runB_PV} follows run BU via position-velocity diagrams. The first panel shows the clouds being separated by their pre-collisional velocity at 10 \,km\,s$^{-1}$. The unbound cloud can be easily recognised due to its thicker extent along the y axis. The vertical spikes reveal that star formation is once again confined to two very narrowly--defined regions, and a small broad bridge feature can also be seen.\\
\indent In the second diagram we see that, unlike the second panel of Figure 6 in Paper I, the clouds' core are widely spatially separated, appearing on either side of the diagram. This is due to the components of the turbulent velocity field perpendicular to the collision axis which the clouds' core regions were given at the start of the simulation . The broad bridge structure is by this stage rather ill--defined and is no longer obvious that one is looking at two initially--separate clouds, although the much thicker extent of the unbound cloud on the velocity axis might suggest this.\\
\indent The third panel shows the state of the simulation at 9.5  Myr, 7.5 Myr after the collision time. The densest region of the clouds have survived the collisions and remain widely separated in both space and velocity on either side of the plot. The remnants of the clouds are loosely connected by an almost horizontal structure, which is essentially what remains of the broad--bridge, since it lies at a velocity close to zero and is composed of material decelerated by the collision. Star formation is plainly still active in both the cores of the clouds. Unlike run B from Paper I star formation does not occur in the broad bridge structure. The outer region of the unbound cloud have spread out in velocity compared to its bound counterpart.\\
%control stuff coming up%
\begin{figure}
	\includegraphics[width=1.0\columnwidth]{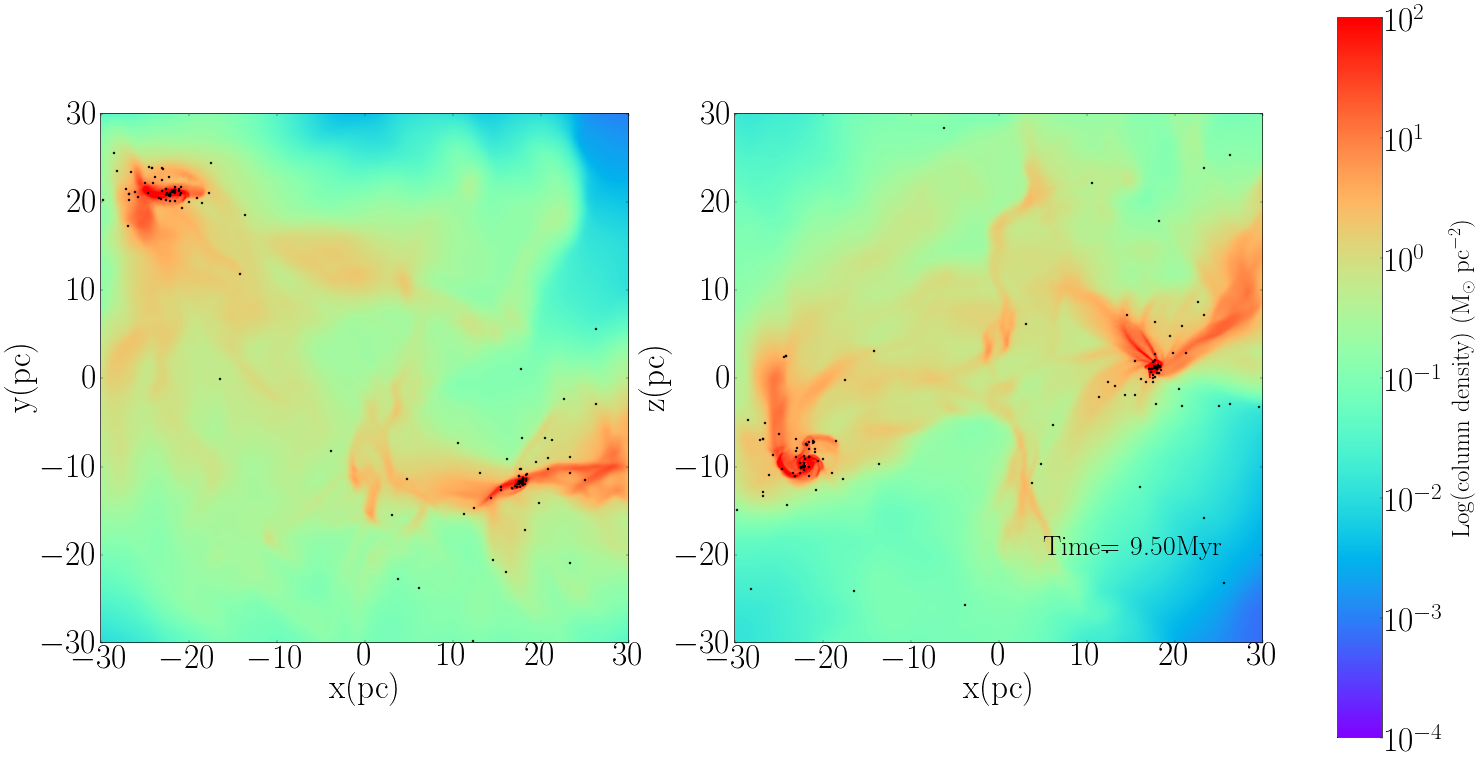}
		\includegraphics[width=\columnwidth]{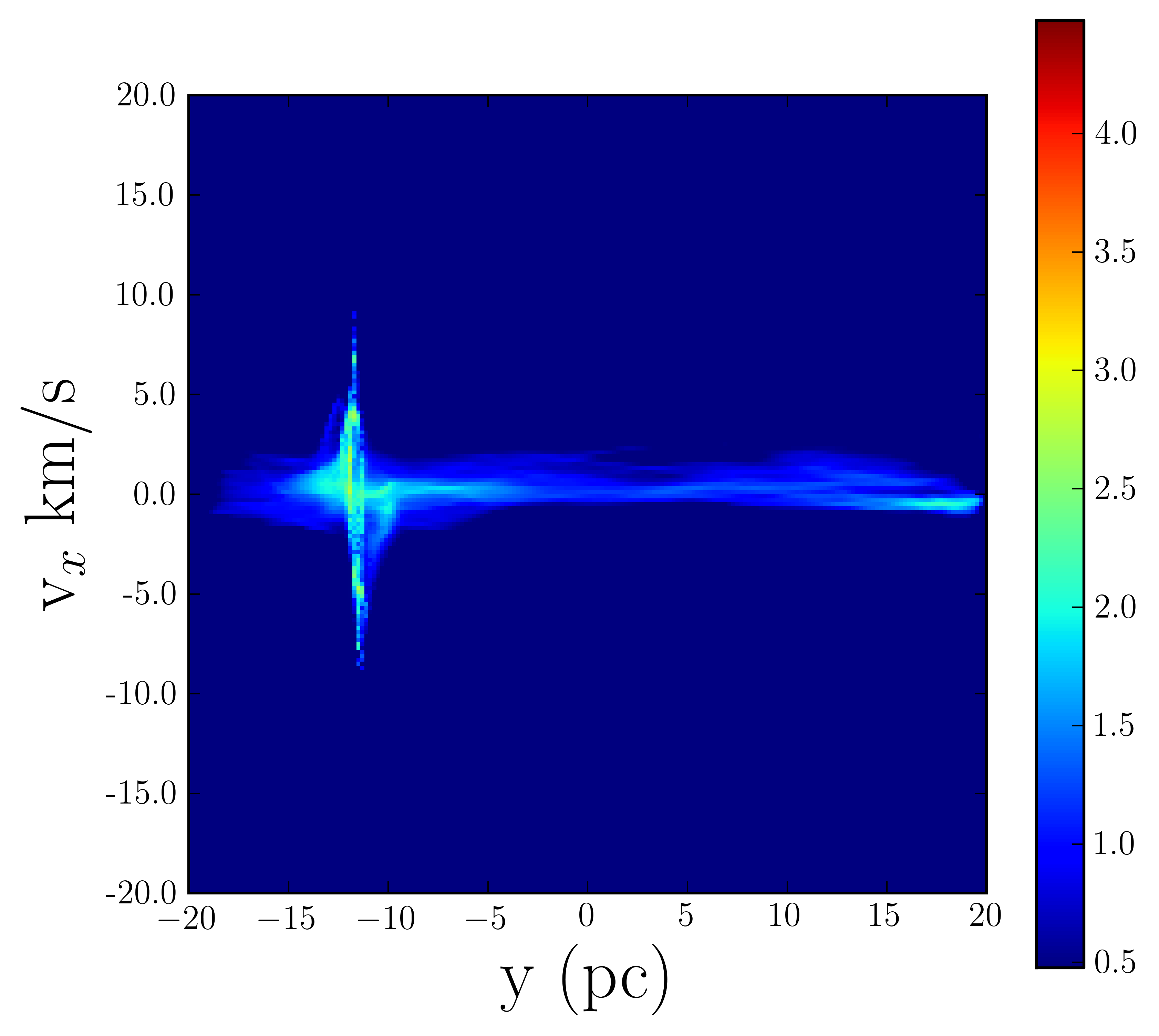}
		\caption{Column-density and position-velocity plots from the controlled runs of run BU simulation. Both these plots are taken at 9.5 Myr of the simulation. From the position-velocity diagram it is hard to distinguish the clouds since they both occupy the same width along the velocity axis.}
    \label{fig:runBcontrol}
\end{figure}
\indent Figure \ref{fig:runBcontrol} shows the state at 9.5 Myr of the controlled runs of the same cloud, with the unbound cloud on the left--hand side of the plot. From the column density plot it is clear that the controlled runs like its collision counter part does not form any filamentary structures. Two widely--separated star clusters have formed and there is again in both the PV and column--density plots evidence of rotating gaseous structures around the clusters.\\
\subsubsection{Run UU\_rho\_r-2}
\begin{figure}
	\includegraphics[width=1.0\columnwidth]{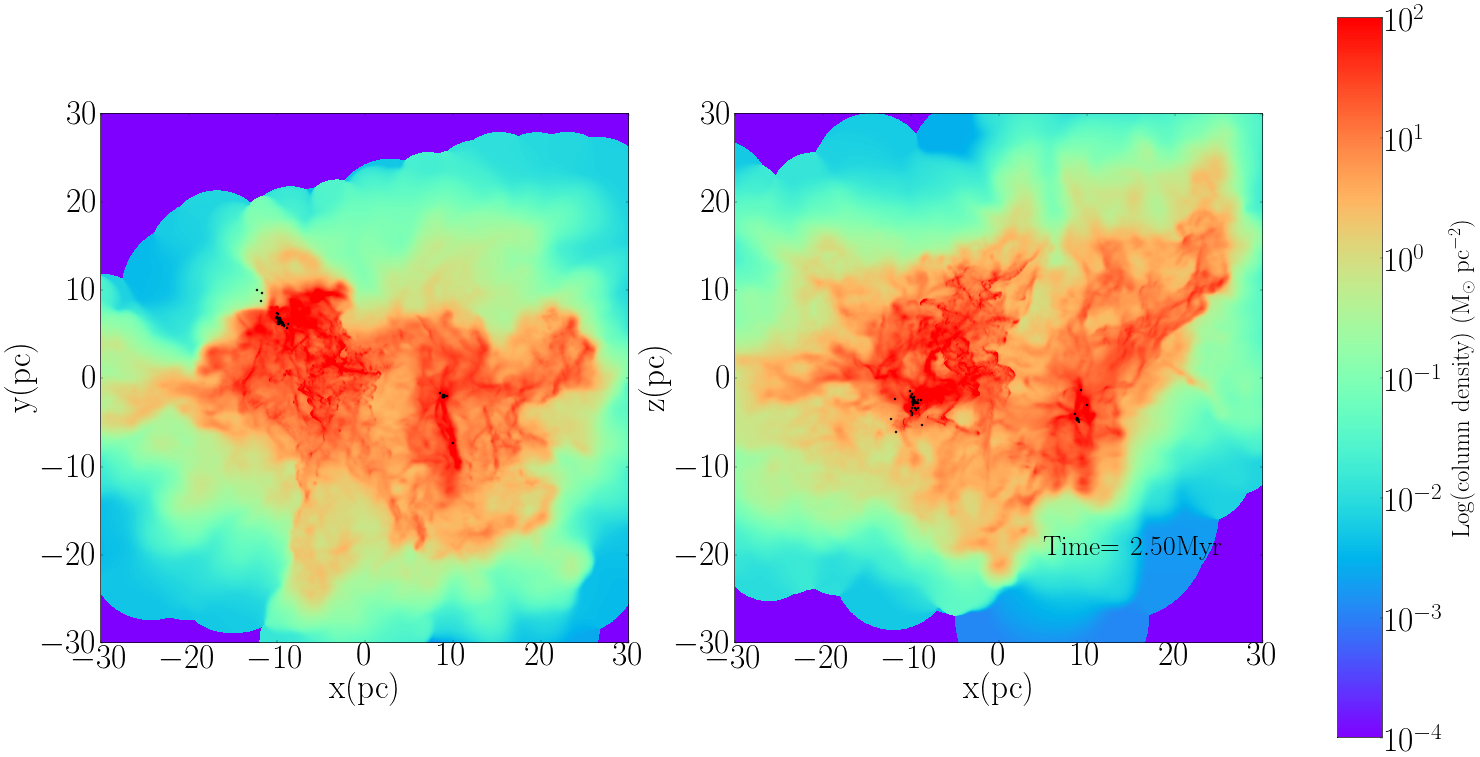}
		\includegraphics[width=\columnwidth]{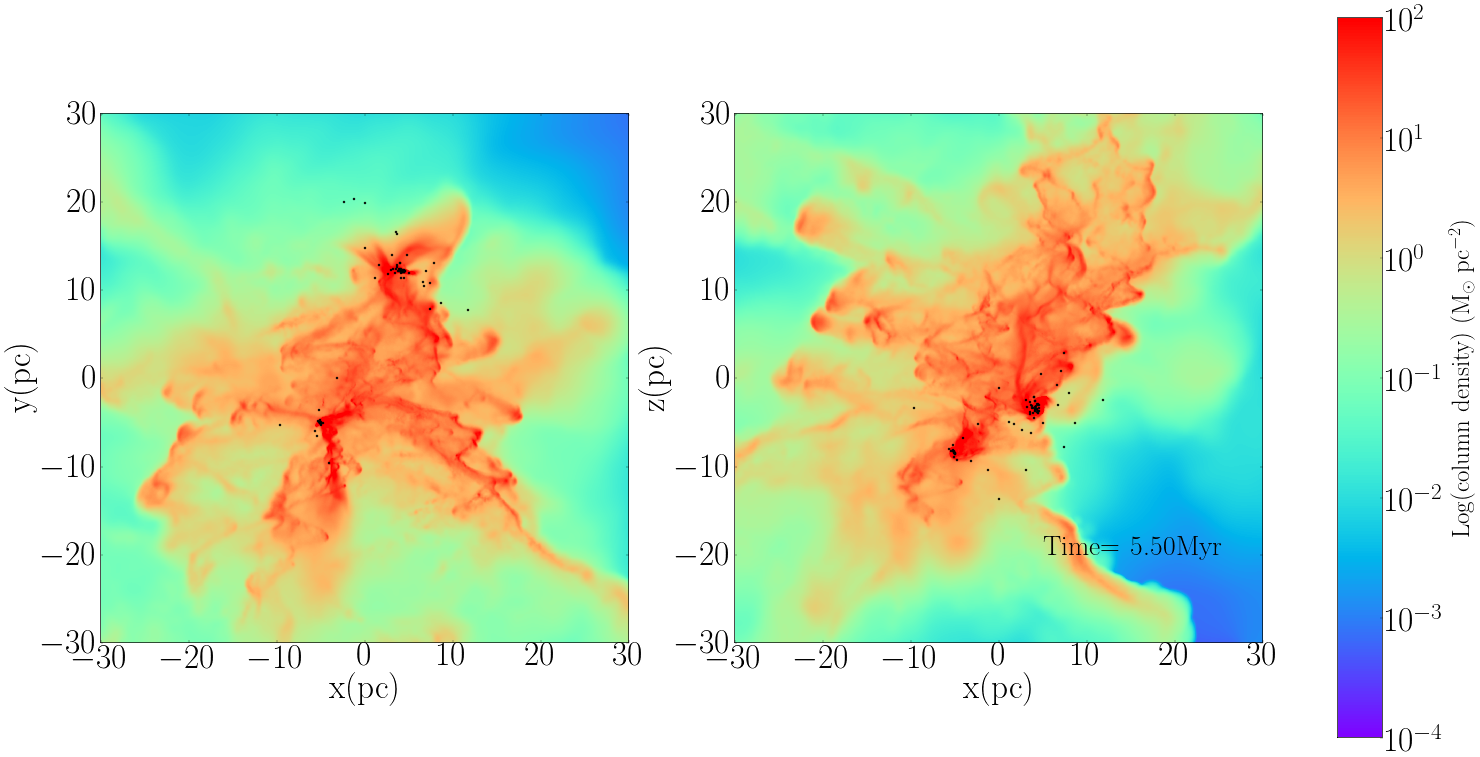}
		\includegraphics[width=\columnwidth]{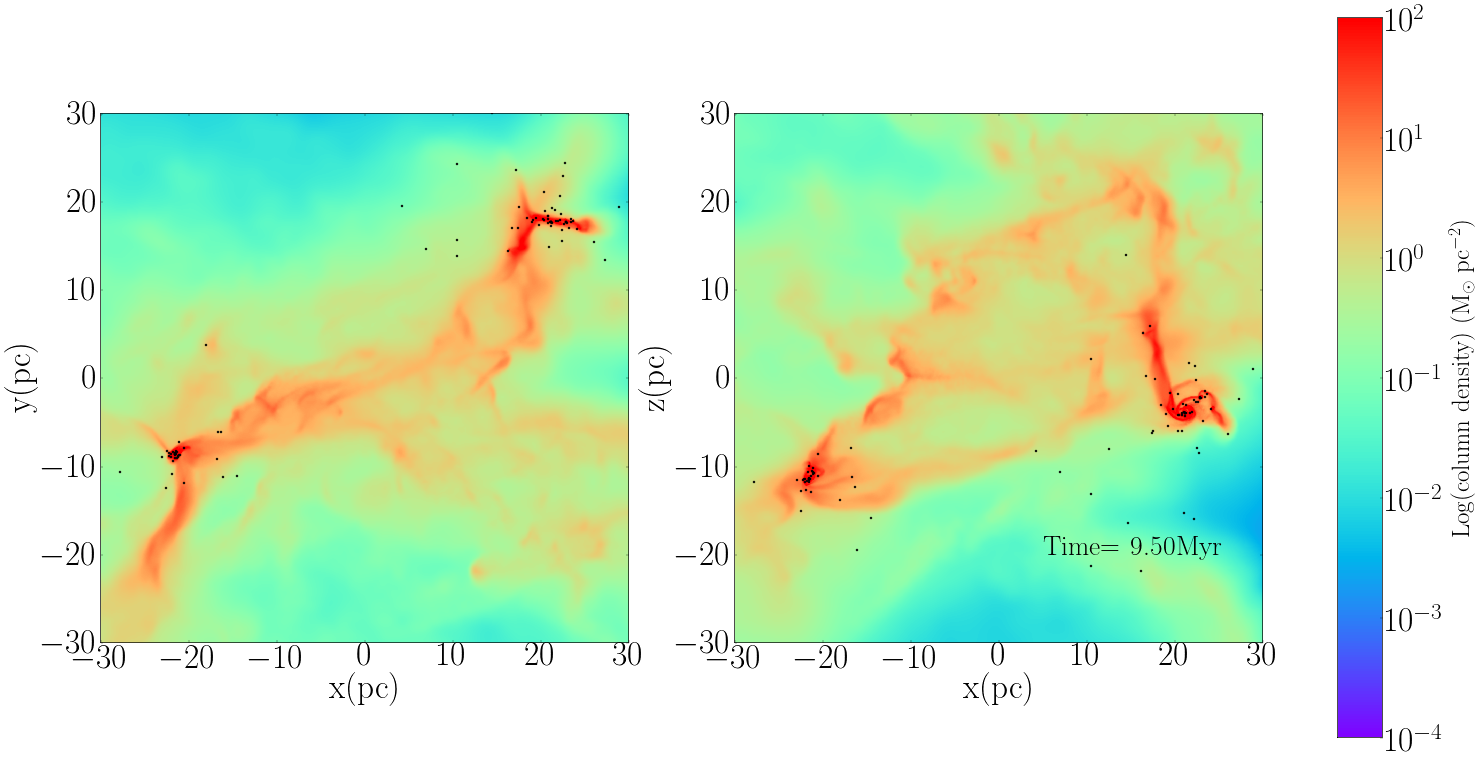}
    \caption{Column-density plots from Run UU simulation. The first plot is shortly after the cloud crushing time at 2.5 Myr. The second plot is an intermediate plot at 5.5 Myr. At this time it is harder to separate the two clouds from one another. The third plot shows the state of the clouds at 9.5 Myr. At this time the remainder of the clouds are clearly separated.}
    \label{fig:runC}
\end{figure}

\begin{figure}
	\includegraphics[width=1.0\columnwidth]{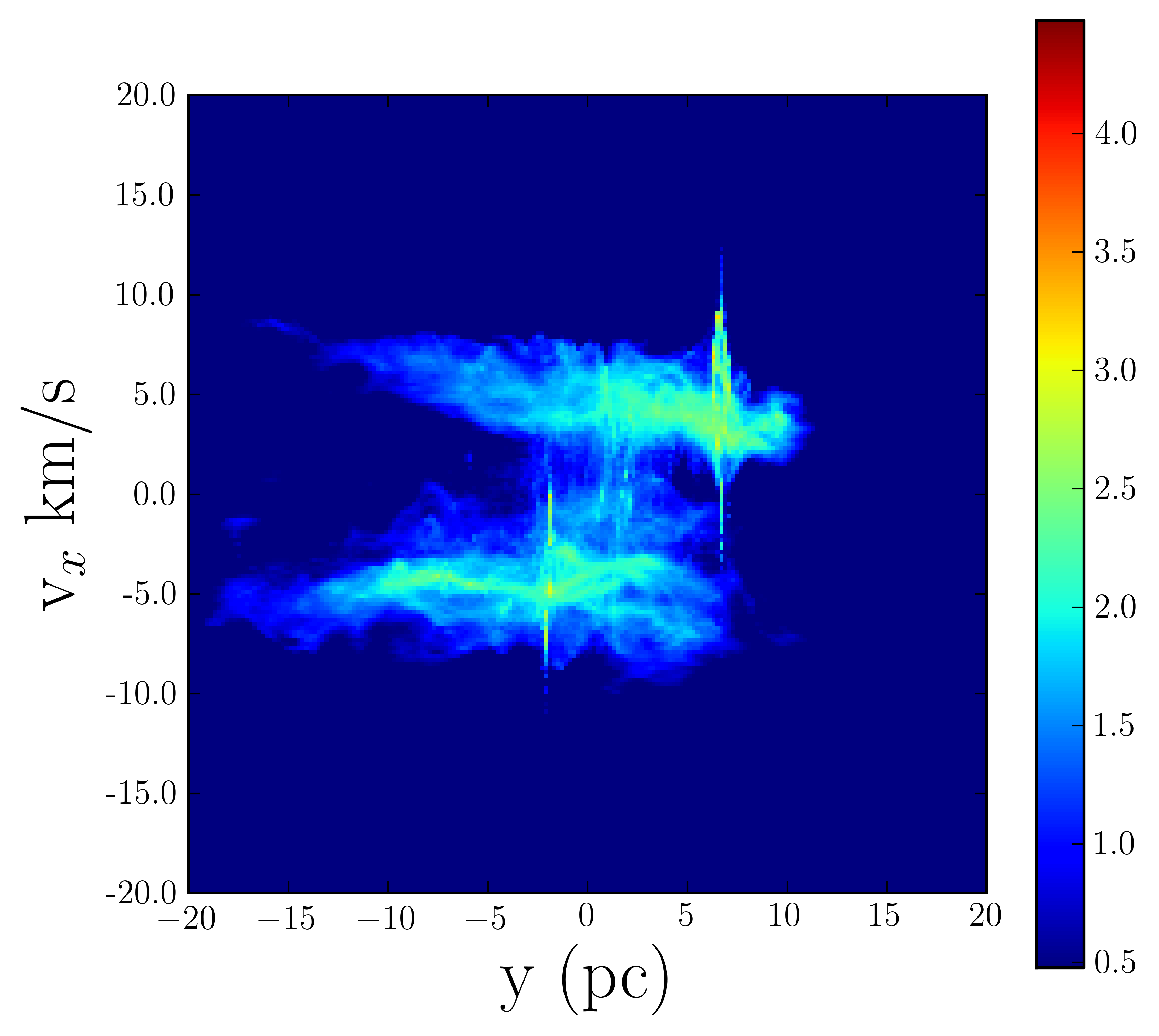}
		\includegraphics[width=\columnwidth]{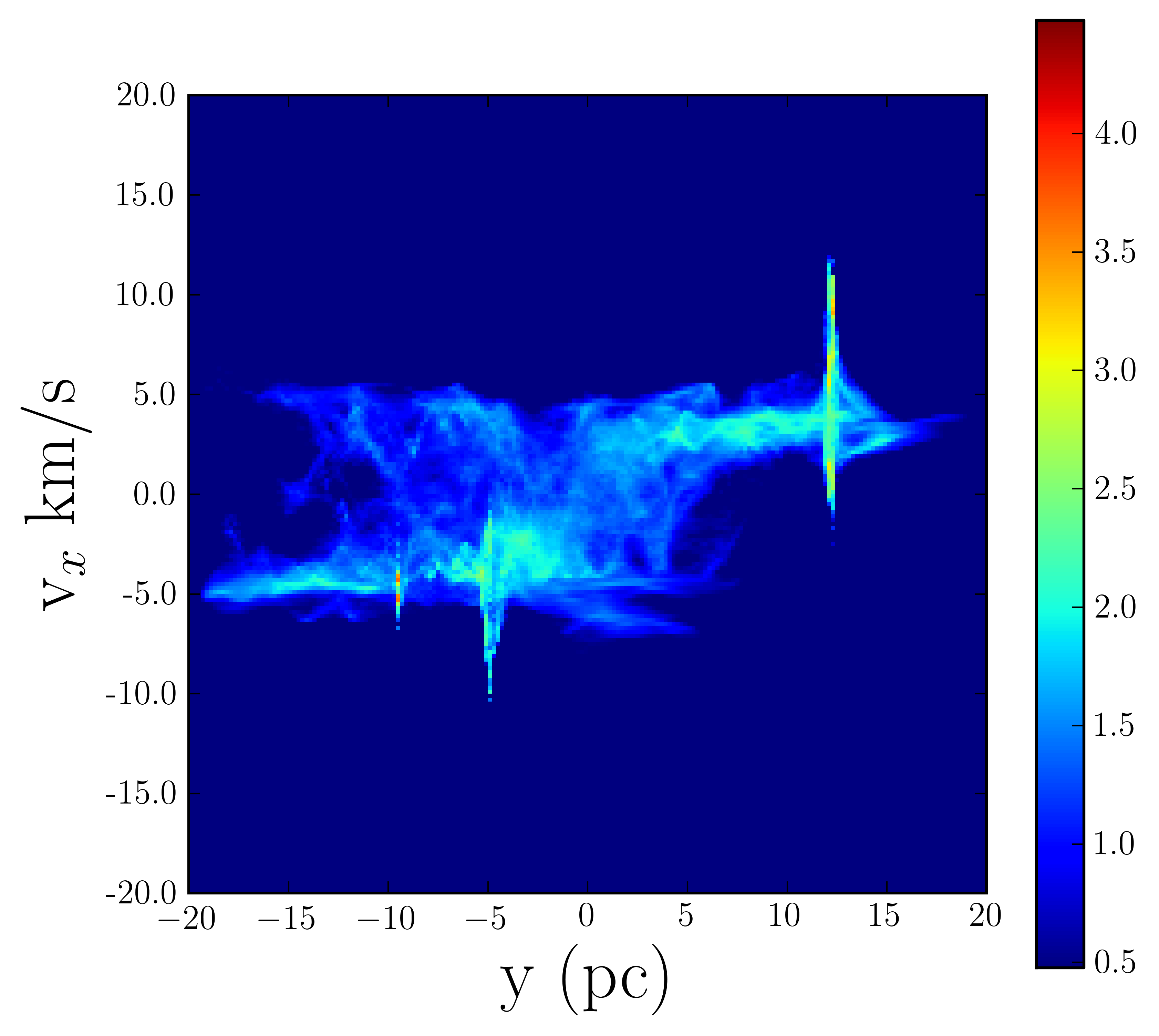}
		\includegraphics[width=\columnwidth]{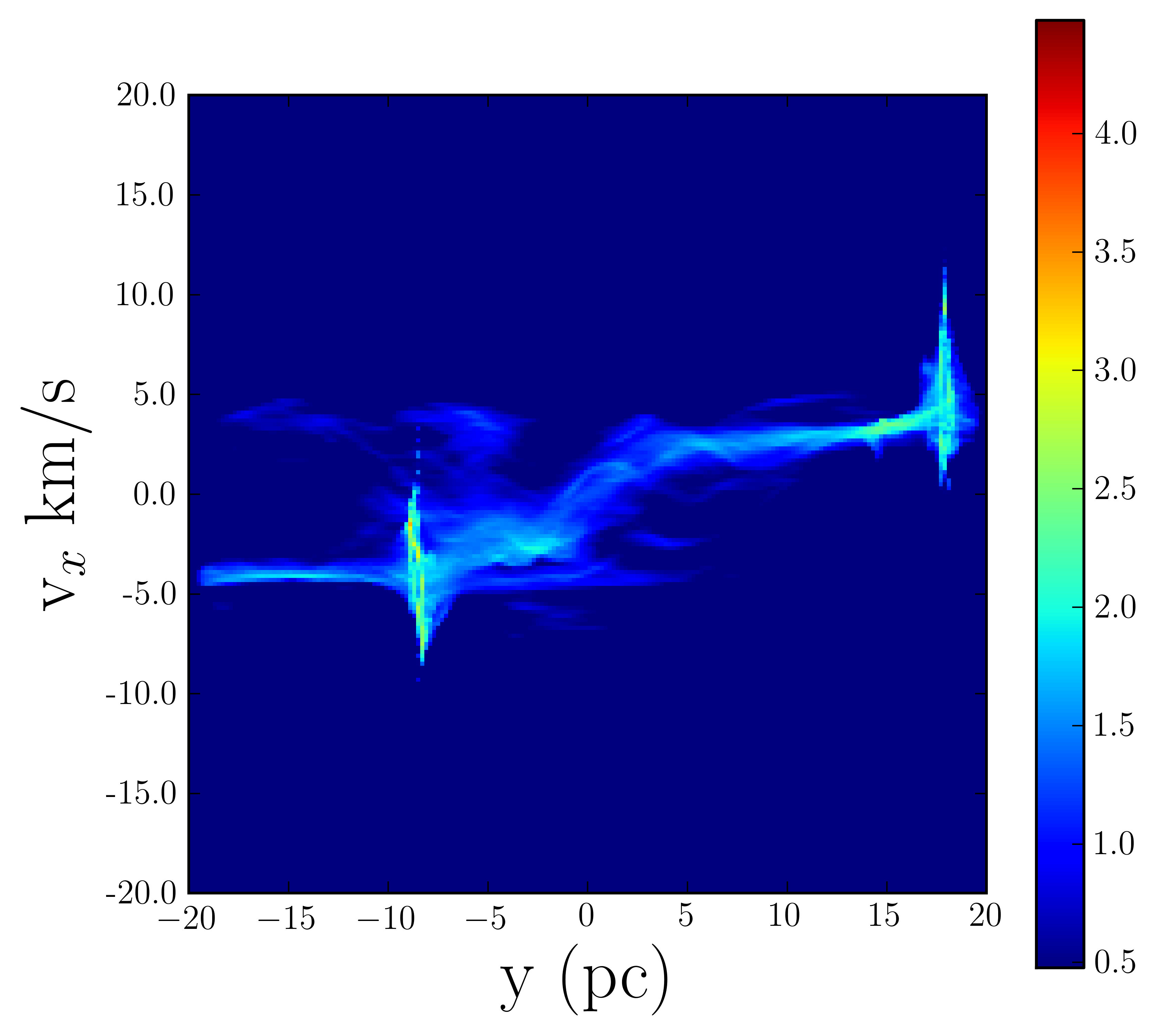}
    \caption{Position-velocity plots from Run UU simulation.The first plot is taken prior to the collision where the clouds can be separated by their pre collisional relative velocity. The second plot shows the state of the clouds during the collision. The third plot is taken after the clouds have collided. The timeline for these plots are the same timeline as the column-density plots.}
    \label{fig:runC_PV}
\end{figure}

In this run, both the clouds have an initial virial ratio of 5.0 and are thus globally gravitationally unbound and it is expected therefore that star formation activity will not be as vigorous as the previous two runs.\\
\indent Figure \ref{fig:runC} shows the column-density plots at different stages of the simulation. The collision time in this run is same as the last two runs at 1.96 Myr. The first panel is taken at 2.5 Myr. The excess turbulent kinetic energy has already caused both clouds' outer regions to expand significantly. However, once again, star formation activity is well underway at the densest regions of both clouds.\\
\indent The second panel in the plot the simulation at 5.5 Myr. Although there are some shocked structures present as a result of the collision, most of the outer regions of the cloud have continued dispersed due to rather large initial kinetic turbulent energy being injected into the clouds. This is an indication that little has changed as a result of the collision. Whatever star formation activity is present remains confined to the initial densest region of the clouds.\\
\indent The final diagram of Figure \ref{fig:runC} shows the state of the simulation 7.5 after the collision of the clouds. Much of the clouds have now dispersed, with the cores of the clouds having survived and being widely separated, hosting a star cluster each. There is clearly relatively little dense gas except the core regions. There is no filamentary structure present connecting the two cloud remnants, which we also observed to be the case in run C of Paper I. However, there is a structure resembling a filament present at the location of the right--hand cluster, almost at right angles the collision axis. This is evidently an accretion flow, and it is feeding a disc--like structure centred on the cluster.\\
\indent The corresponding position--velocity diagrams are shown in Figure \ref{fig:runC_PV}. The first panel shows that both the clouds' outer region have been expanded along the velocity axis due to the initial virial ratio and the broad bridge feature can be found like the other collisional simulations in this study. In common with the other runs presented here is the presence of star formation activity at the densest region of the clouds.\\
\indent The second panel shows that quite a considerable amount of material, particularly at negative $y$ locations, belonging to the cloud with an initial negative $x$--velocity has been decelerated by the collision, producing a very broad bridge feature.\\
\indent The third panel shows the state of the run at 9.5 Myr. The cores of the clouds have survived the collision and much of the remaining gas has dispersed. The broad bridge feature remains between the cores, which now reside at opposite ends of the plot, and accretion flows indicate continued star formation activity. 
\begin{figure}
	\includegraphics[width=1.0\columnwidth]{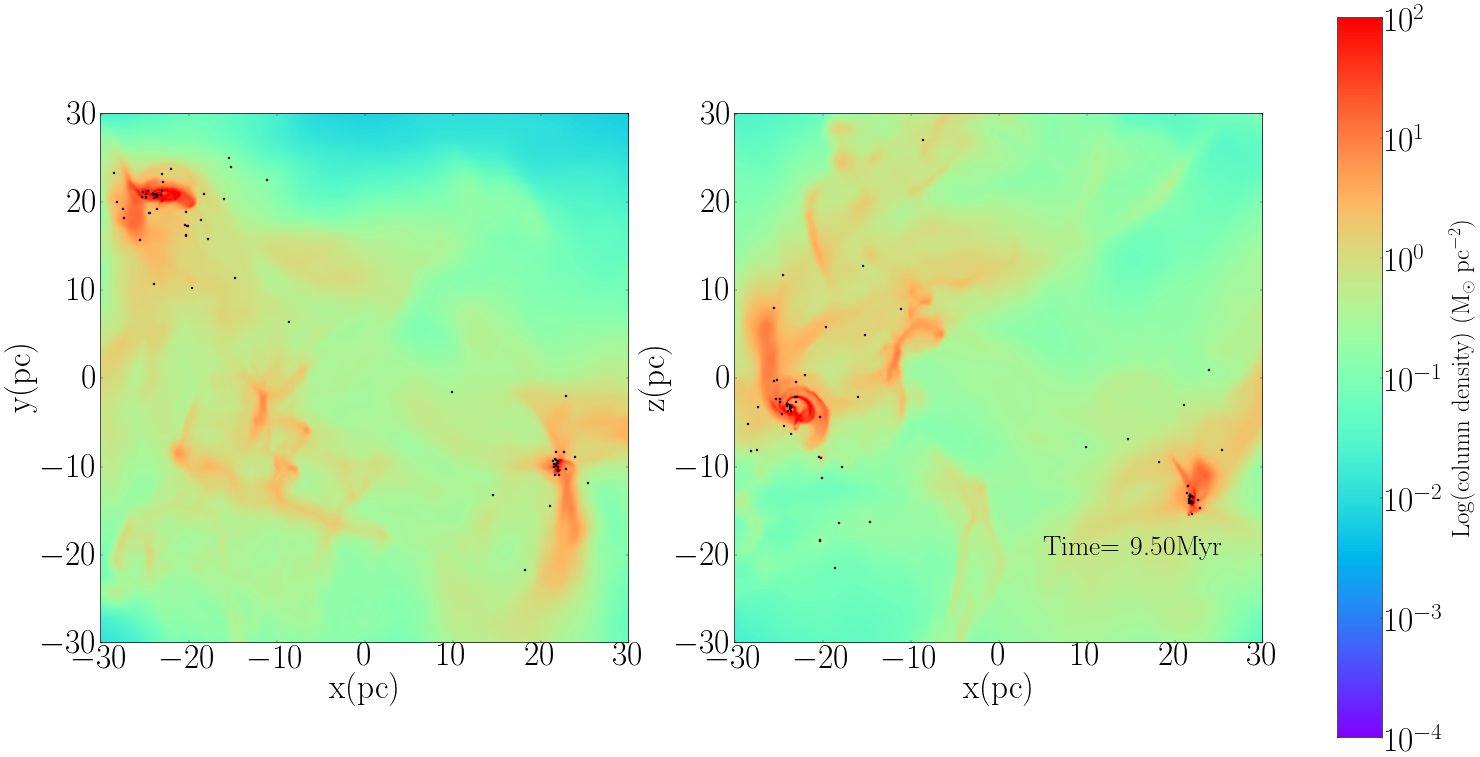}
		\includegraphics[width=\columnwidth]{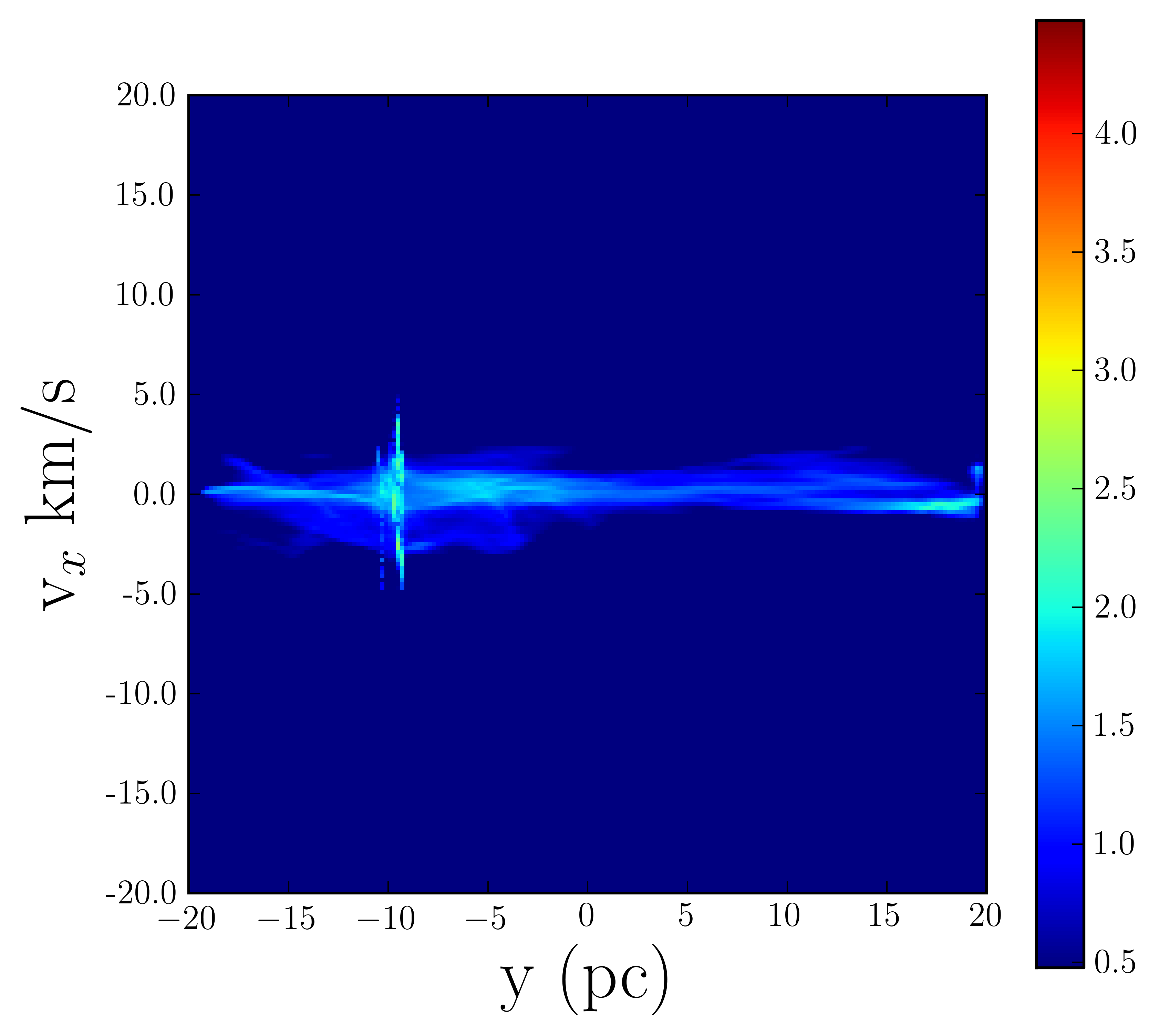}
		\caption{Column-density and position-velocity plots from the controlled runs of run UU simulation. Both these plots are taken at 9.5 Myr of the simulation. From the position-velocity diagram it is hard to distinguish the clouds since they both occupy the same width along the velocity axis.}
    \label{fig:runCcontrol}
\end{figure}
\indent Figure \ref{fig:runCcontrol} shows the state of the control run at 9.50 Myr. From the column-density plot it is clear that not a lot of dense gas is left. Like other controlled runs and like bound-unbound and unbound-unbound collision there is a clear absence of filamentary structures in this control run. Two star clusters formed and are situated at either ends of the cloud and rotating gaseous structure is present around the cluster. The position-velocity plot tells even less since both the clouds occupy the same width along the velocity axis. The vertical spikes are an indicator of whatever little star formation activity is present are mostly confined to the dense core regions of the clouds since most of the gas dispersed or are dispersing. \\
\subsection{Star formation efficiencies and rates}
\begin{figure}
	\includegraphics[width=1.0\columnwidth]{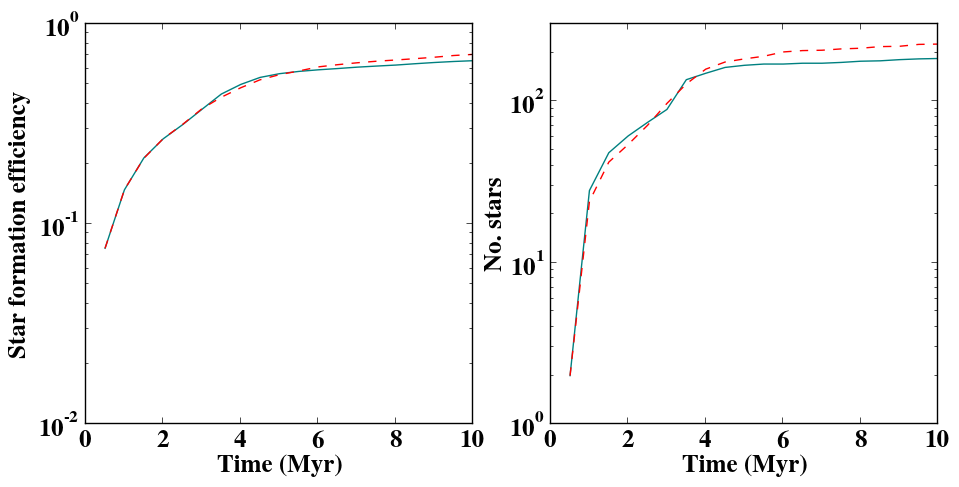}
		\includegraphics[width=\columnwidth]{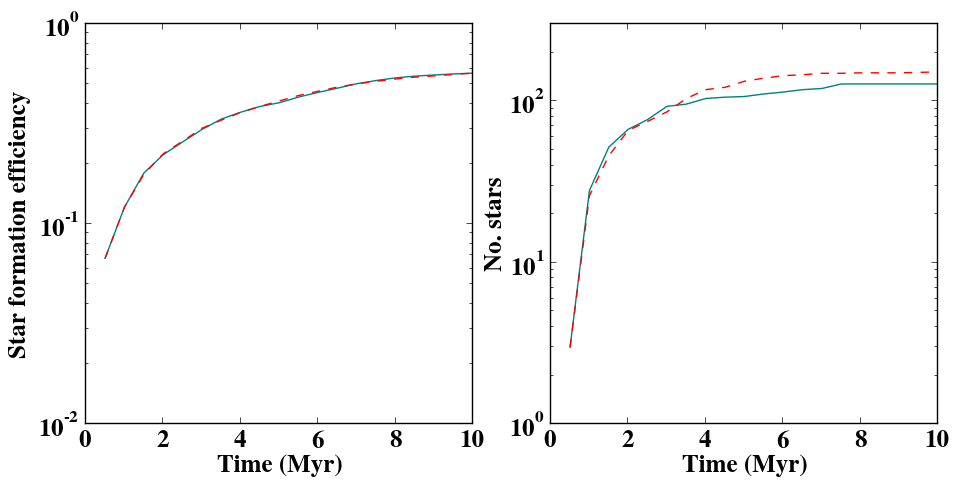}
		\includegraphics[width=\columnwidth]{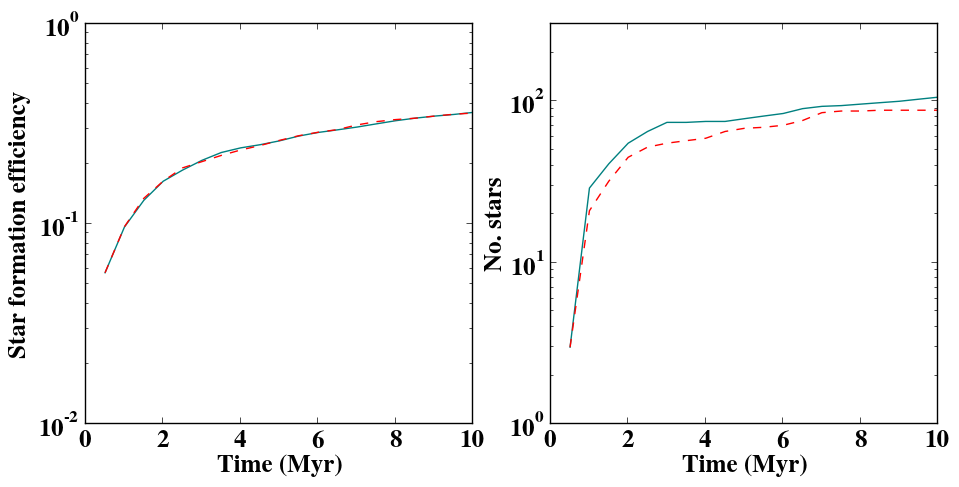}
    \caption{Star formation efficiency and number of stars formed as a function of time. The first row of plot is of run BB (bound-bound) simulation. The second row of plots is of run BU (bound-unbound) and the third row of plots is of run UU (unbound-unbound) simulations. The green lines represent the collision runs while the red lines represent the star formation efficiency and number of stars formed in the controlled runs.}
    \label{fig:sfe}
\end{figure}

\begin{figure*}
	\includegraphics[width=0.33\textwidth]{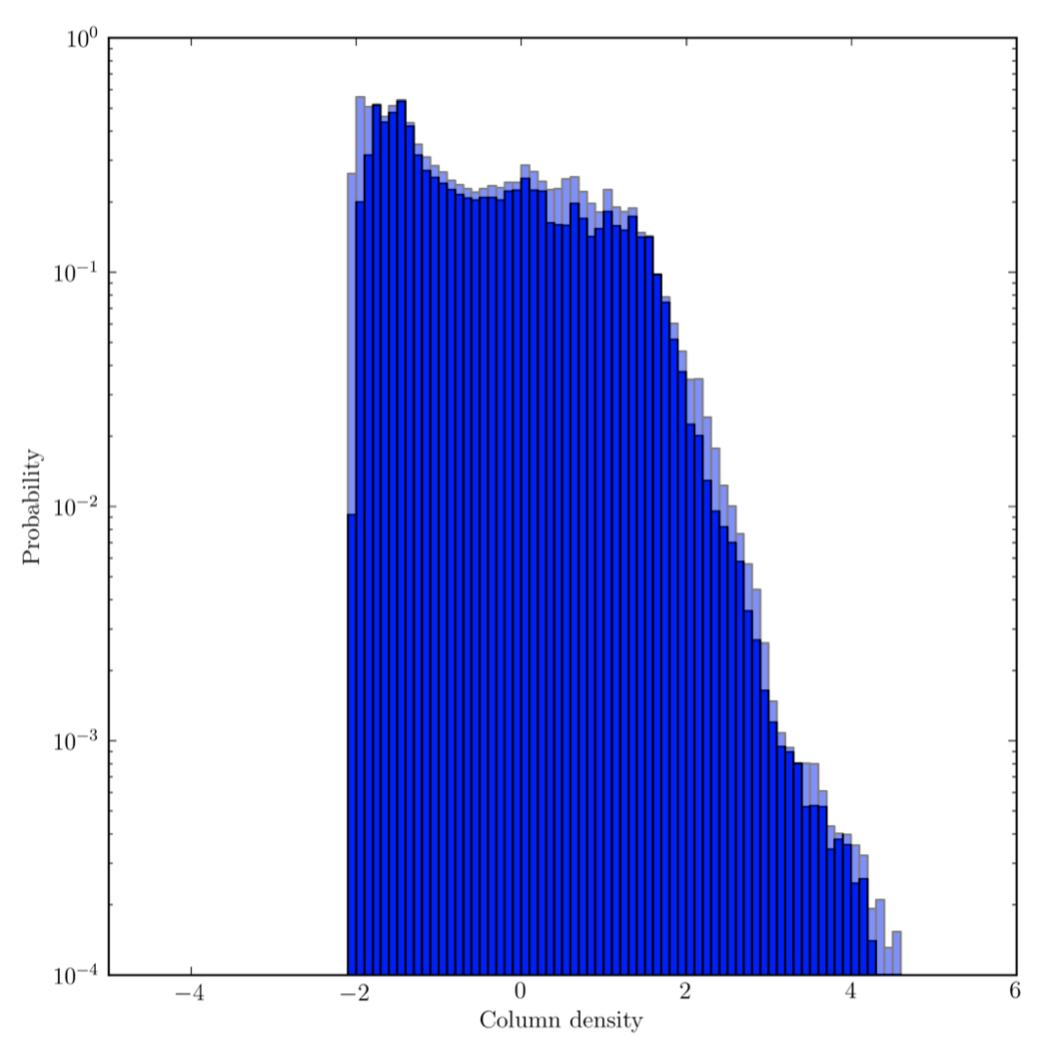}
	\includegraphics[width=0.33\textwidth]{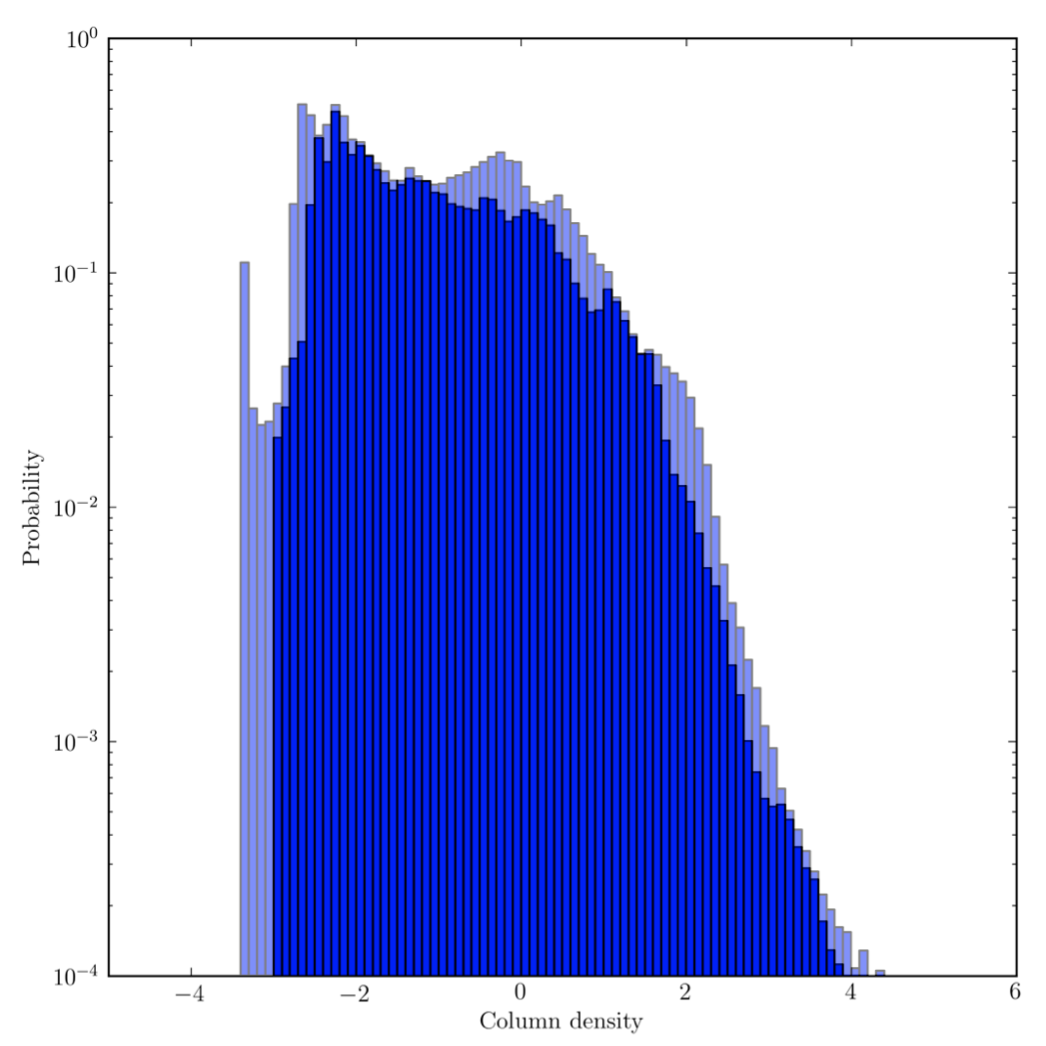}
	\includegraphics[width=0.33\textwidth]{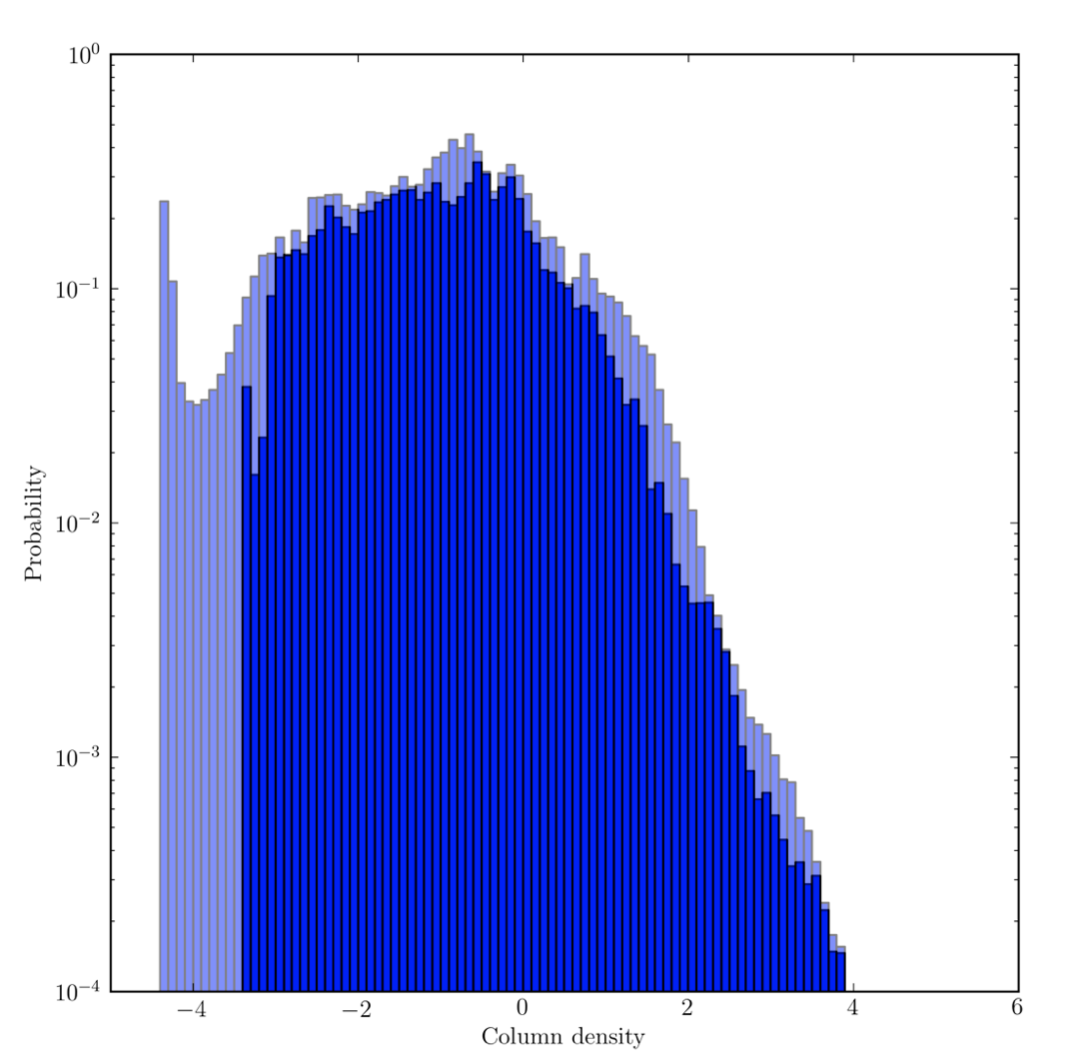}
    \caption{Column--density PDFs of the BB collision (half--tone) and control (full--tone) simulations at a time of 3.0,5.5 and 8.0\,Myr.}
    \label{fig:pdf}
\end{figure*}

Figure \ref{fig:sfe} presents the star formation efficiencies and rates for both collision and control simulations. The green lines and red dashed lines represent the collision and control runs respectively. In all the simulations star formation efficiency is calculated from\\
\begin{equation}
{\rm SFE} = \frac{\rm total\,\,stellar\,\,mass}{\rm total\,\,gas\,\,mass + \rm total\,\,stellar\,\,mass}
 \end{equation}
\\
\indent The first row in Figure \ref{fig:sfe} shows the star formation efficiencies and rates as a function of time of Run BB. The star formation efficiencies are almost identical at all times, but the control runs produces fractionally more stars over the $\approx$ 10 Myr span of the simulation. \\
\indent The second row in the figure shows that this pattern is repeated in Run UU, and the third row shows the star formation efficiency of Run UU is unchanged by the collision, but that fractionally \textit{fewer} stars are formed in the collision run. Like Run B the star formation efficiency remains the same for both collision and control runs. As expected, the star formation efficiencies are highest in the case where both clouds are initially bound and lowest when they are both unbound.\\
\indent These plots show that the effect of the collisions on the star formation process within the clouds is negligible. Star formation activity in these models is confined to the dense cores of the clouds, which are bound irrespective of the clouds' global virial ratios. Their density and boundedness makes the cloud cores immune to the effects of the collisions.\\
{\ \indent To further explore this point, we examined the column--density probability distribution function (PDFs) viewed along the $z$--axis (i.e. perpendicular to the collision axis), comparing the collision runs with the corresponding control runs at a given time. The high--density end of the column--density is a useful tool, particularly for observers, in examining the state of star formation. We find that the differences between corresponding PDFs of a pair of collision and control runs are slight, and certainly do not reveal any features that would allow one to identify a collision product from its PDF alone.\\
\indent We plot examples in Figure \ref{fig:pdf}, taken from Run BB at 3.0, 5.5 and 8.0\,Myr. At low densities, the collision simulation (shown in half--tone) shows more low--density material after the collision has occurred, as expected from gas which has been ejected from the clouds in the encounter. The collision run also shows more intermediate--density material, reflecting the shocking of the outer regions of the colliding clouds. However, the quantities of gas in the highest--density regions, in which star--formation is underway, are very similar, indicating that the high--density regions of the clouds are very little affected by the collision. This result is reflected in the similarity of star formation rates and efficiencies.}
\subsection{Angular momentum}
\begin{figure}
	\includegraphics[width=1.0\columnwidth]{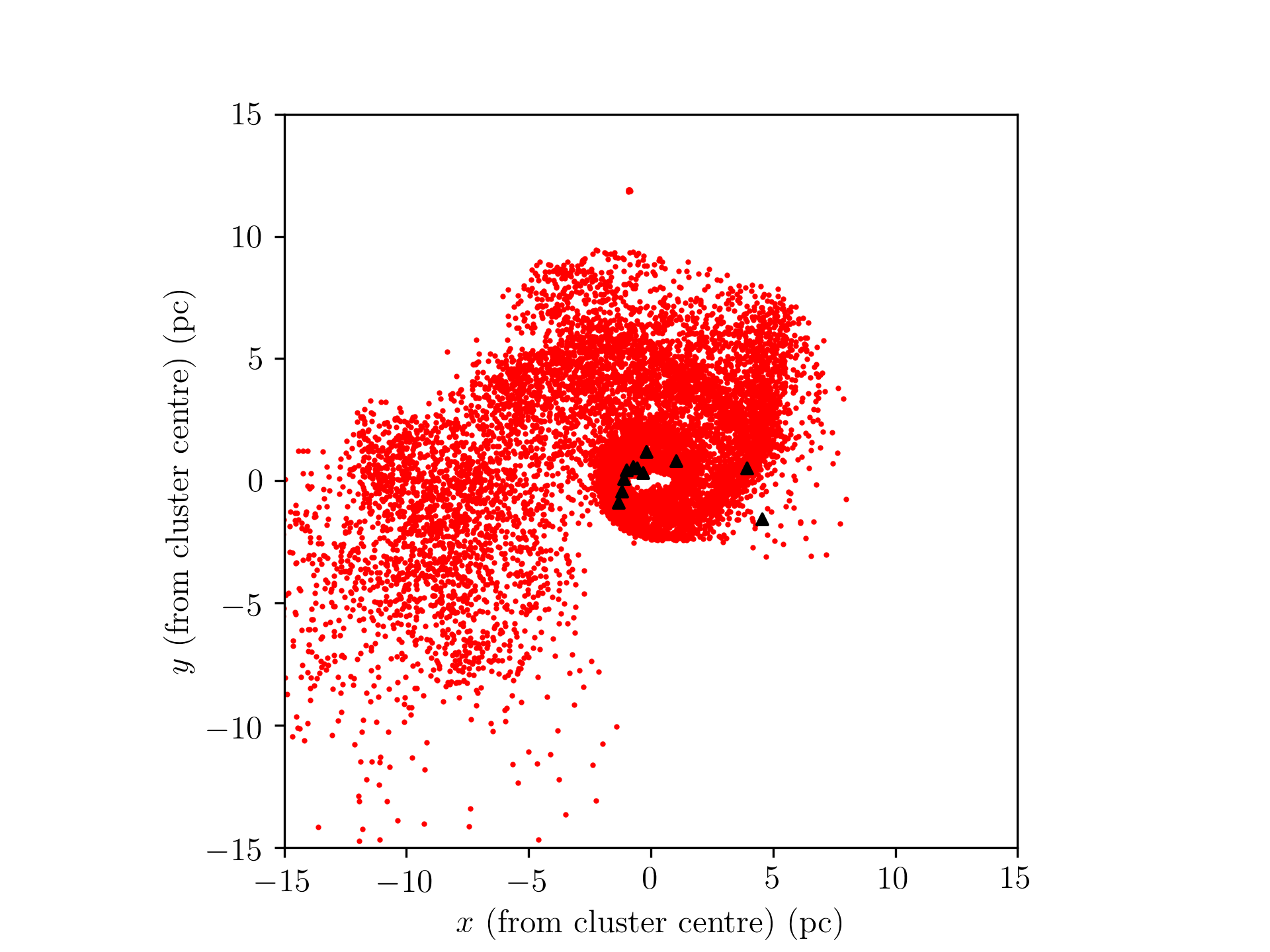}
    \caption{Particle plot of the rotating structure picked out by the procedure described in the text, around the right--hand cluster in the last output of Run BB. Red dots represent gas particles, and black triangles are sink particles.}
    \label{fig:run_A_rotating_struc}
\end{figure}

\indent Examination of the late--time column density images presented in Figures \ref{fig:runA}, \ref{fig:runB} and \ref{fig:runC} and the corresponding control simulations appears to show vortices of dense gas centred on the main concentrations of stars. Such a structure is particularly clear, for example, around the right--hand cluster in the $xy$ view of run BB at 9.50\,Myr in Figure \ref{fig:runA}. These structures suggest that considerable quantities of angular momentum are stored in the hydrodynamic flows near the clusters. Since the clusters are accreting from these flows, it is then natural to ask if the clusters themselves contain significant quantities of angular momentum.\\
\indent While there is considerable observational evidence of rotation in globular clusters \citep[e.g][]{1986AJ.....91..546P,2015A&A...573A.115L}, there is relatively little on the evidence for rotation in young or embedded systems \citep[e.g.][]{2012A&A...545L...1H,2015ApJ...807...27C}. Simulations, however, suggest that rotation in young clusters should be common \citep[e.g][]{2017MNRAS.467.3255M,2020MNRAS.496...49B}.\\
\indent To examine this issue, we first divide the stars in the final output dump from each of our simulations into `left' and `right' populations, based on whether their $x$--coordinates are positive or negative with respect to the centre of mass of the whole system. We then compute the centre--of--mass positions and velocities of the two populations of sinks,  ${\bf r}_{\rm left}$, ${\bf r}_{\rm right}$, ${\bf v}_{\rm left}$ and ${\bf v}_{\rm right}$ and transform the positions and velocities of all gas and sink particles at positive of negative $x$ (as appropriate) into this frame.\\
\indent We first examine the angular momentum of the gas. If the transformed position and velocity vectors of a given gas particle $i$ (of mass $m_{\rm gas}$ are ${\bf r}_{i}'$ and ${\bf v}_{i}'$ respectively, the angular momentum of the particle in the frame of the cluster of sinks is ${\bf J}_{i}=m_{\rm gas}{\bf r}_{i}'\times {\bf v}_{i}'$.\\
\indent We now wish to see if there is a preferred angular momentum direction in the gas. To do this, we choose a random sample of gas particles $i$ and for each $i$ we choose a random sample of gas particles $j$ and first compute $|{\bf r}'.{\bf v}'|/({\bf r}'||{\bf v}'|)$ for both particles. If this normalised dot product exceeds 0.9 for either particle, we consider that one or other has a trajectory too close to radial motion to make measuring its angular momentum legitimate and we discard the possible pairing.\\
\indent If neither particle is on a strongly radial path, we compute ${\bf J}_{i}.{\bf J}_{j}/(|{\bf J}_{i}||{\bf J}_{j})$ (note that it is too time--consuming to compute this for every possible pair of particles, hence the sampling). If this normalised dot product exceeds 0.7, we consider the particles to have similar angular momentum add the angular momentum of particles $i$ and $j$ (taking care not to double count) to a quantity ${\bf J}_{\rm corr}$. If there are a population of gas particles with similarly--oriented angular momentum vectors, their contributions will come to dominate ${\bf J}_{\rm corr}$.\\
\indent Once this is done, we go through \textit{all} the gas particles $i$ and compute ${\bf J}_{i}.{\bf J}_{\rm corr}/(|{\bf J}_{i}||{\bf J}_{\rm corr})$ and, if this normalised dot product exceeds 0.7, we take the gas particle to be a member of a coherent rotating structure. We also go through all the sink particles and use the same procedure to locate those whose angular momentum vectors are similar to those of the gas. In Figure \ref{fig:run_A_rotating_struc}, we show a particle plot centred on the right--hand cluster in the last output from run BB, with the selected gas particles shown in red and the selected sink particles shown in black.\\
\indent Two things are immediately apparent from Figure \ref{fig:run_A_rotating_struc}. Firstly, the procedure described above has picked out a structure in the gas resembling a vortex centred on the cluster. However, it appears that most of the sink particles have angular momentum vectors substantially misaligned with this structure, since they do not appear in the plot. We discuss this issue in more detail below.\\
\indent We plot on a logarithmic scale the modulus of the angular momentum vector against the radius from the cluster centre for the rotating structure shown in Figure \ref{fig:run_A_rotating_struc}. The colours represent the numbers of particles at that location in $r,J$--space.\\
\begin{figure}
	\includegraphics[width=1.0\columnwidth]{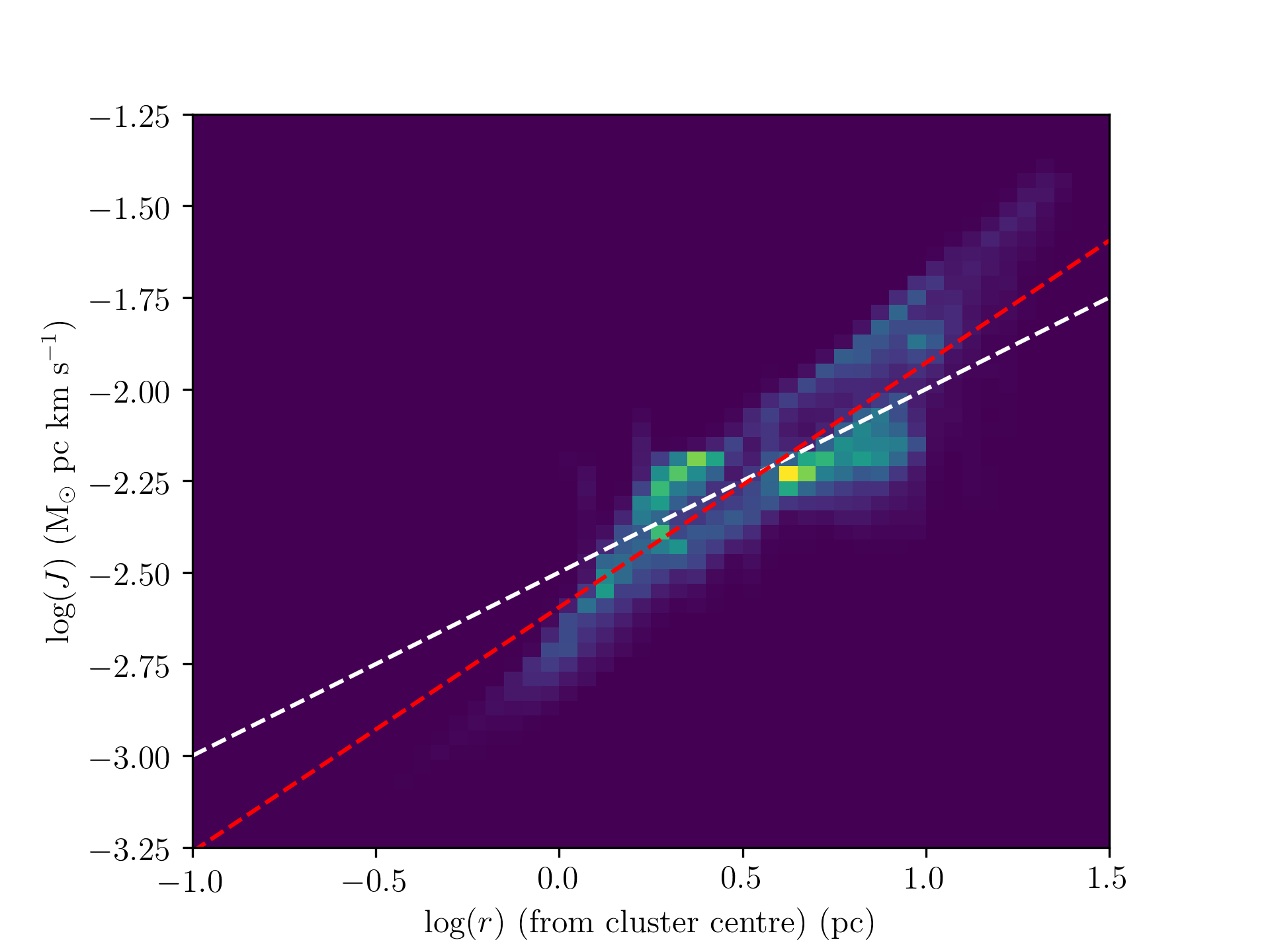}
    \caption{Particle plot of the logarithm of the tangential velocity against the logarithm of distance from the cluster centre for the gas particles in the rotating structure shown in Figure\ref{fig:run_A_rotating_struc}. The white dashed line shows the Keplerian scaling $J\propto r^{\frac{1}{2}}$ and the red dashed line is a fit to the simulated profile.}
    \label{fig:run_A_rotating_struc_J_r}
\end{figure}
\indent We compare approximately the run of angular momentum against radius in Figure \ref{fig:run_A_rotating_struc_J_r} with a Keplerian velocity field, in which $J\propto r^{\frac{1}{2}}$, shown by the white dashed line and a fit to the data as a red dashed line.\\
\indent In Figure \ref{fig:run_A_rotating_struc_J_r}, we plot $|{\bf J}_{i}|$ against radius, along with the specific angular momentum profile one would expect for Keplerian rotation (black dashed line). The angular momentum profile of the gas is steeper than the Keplerian profile, indicating that the gas at small radii has proportionately less angular momentum than that at large radii. This is likely an indication that gas at small radii is losing angular momentum by artificial viscosity and is being accreted by the cluster stars.\\
\indent We have shown that the gas near the main concentrations of stars exists in coherent rotating structures containing significant quantities of angular momentum. It is obvious to inquire if the star clusters at the centres of these flows also have net angular momentum.\\
\indent To assess this possibility, we compute the angular momentum of each star  ${\bf J}_{*}=m_{\rm *}{\bf r}_{*}'\times {\bf v}_{*}'$. We again eliminate stars on nearly radial orbits and compute for all pairs of stars (since there are relatively few) ${\bf J}_{*,i}.{\bf J}_{*,j}/(|{\bf J}_{*,i}||{\bf J}_{*,j})$ and plot histograms of the result. An example, for the same cluster as above, is shown in Figure \ref{fig:run_a_rotating_stars}.\\
\begin{figure}
	\includegraphics[width=1.0\columnwidth]{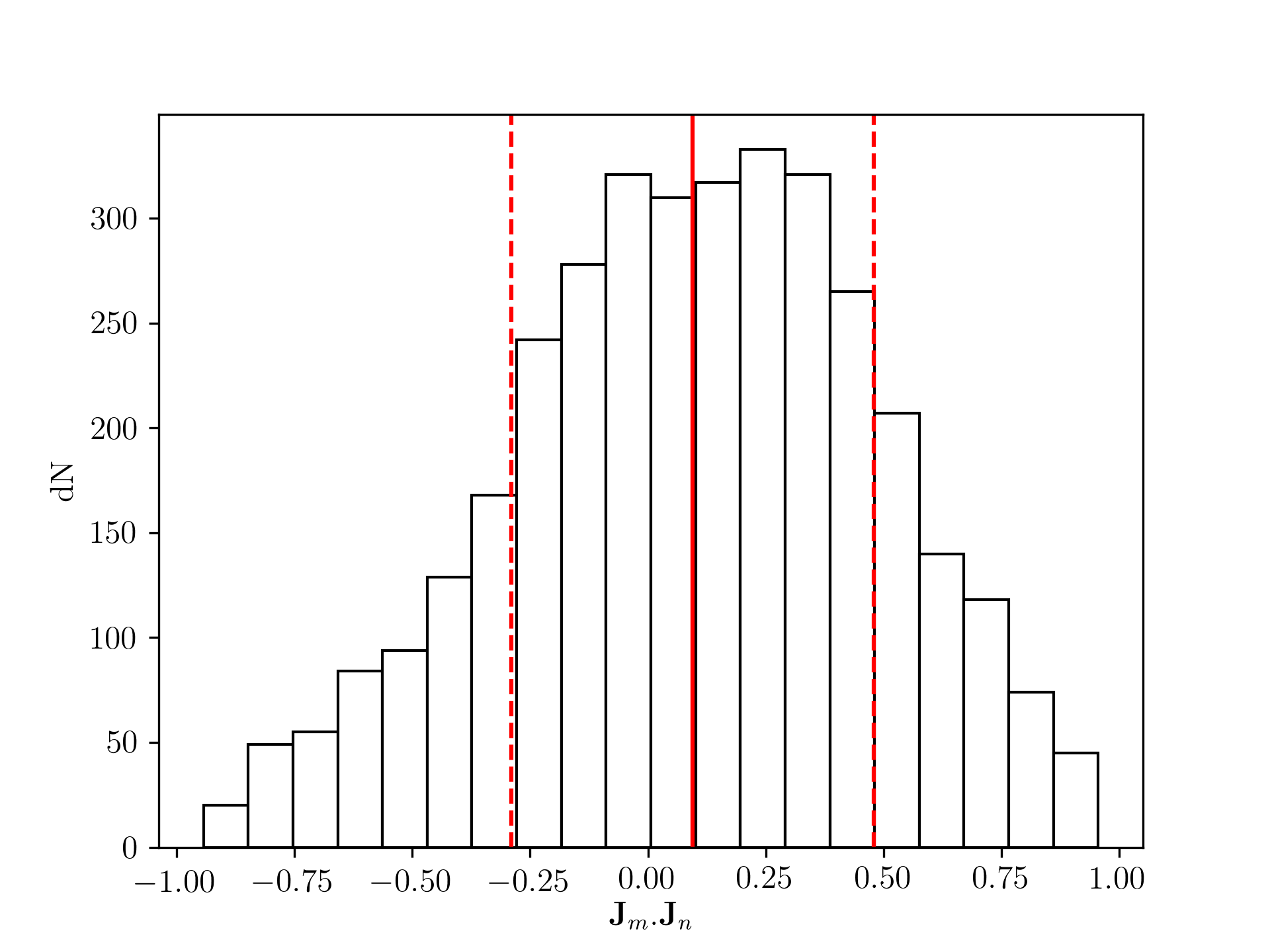}
    \caption{Histogram of dot products of pairs of normalised stellar angular momentum vectors in the right--hand cluster in Run BB at the last timestep (black lines), with the mean and mean $\pm$ one standard deviation shown as solid and dashed red lines.}
    \label{fig:run_a_rotating_stars}
\end{figure}
\begin{figure}
	\includegraphics[width=1.0\columnwidth]{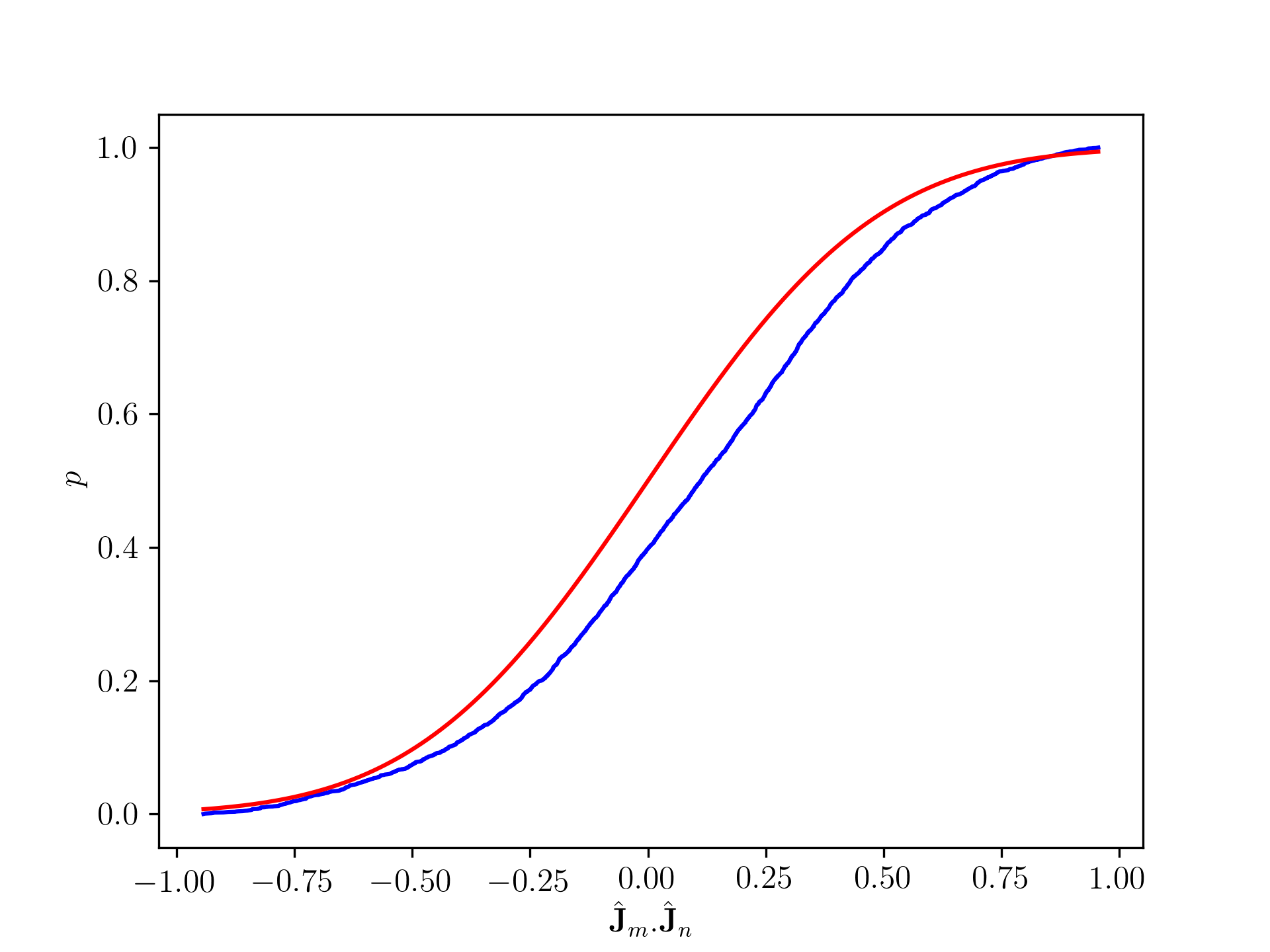}
    \caption{Normalised cumulative distribution of dot products of pairs of normalised stellar angular momentum vectors in the right--hand cluster in Run BB at the last timestep (blue line), compared with a normalised Gaussian with a mean of zero and the same standard deviation (red line).}
    \label{fig:run_a_rotating_stars_cu}
\end{figure}
\indent If the cluster had a preferred rotation, the distribution of dot products should be strongly skewed towards values close to unity. Figure \ref{fig:run_a_rotating_stars} instead shows a distribution which closely resembles a Gaussian, with a mean close to zero. There is a degree of asymmetry to the distribution, and its skewness is -1.38. In Figure \ref{fig:run_a_rotating_stars_cu}, we instead plot the unbinned cumulative distribution of the same dot products in blue. We overlay in red the cumulative distribution of a Gaussian with a mean of zero and same standard deviation as the data. The data are clearly skewed somewhat to larger values of the dot product, and a KS test comparing the Gaussian and the data returns a p--value of $\sim10^{-8}$. The distribution evidently has a bias towards alignment, but it is not particularly strong. Very similar results apply to the other cluster in Run BB, and those in runs BU and UU.\\
\indent It is tempting to suppose that these rotating structures are caused by the collisions of the clouds, since even an encounter with an impact parameter of zero may possess a net angular momentum if the clouds are inhomogenous. However, since we also observe such rotating structures in the control simulations, this is evidently not the case. Instead, it is due to the steep density profile of the clouds leading to star formation being confined to small volumes near their centres. The clusters so formed then accrete material from larger radii, which inevitably is carrying angular momentum. These structures are then a generic feature of the contraction of a strongly centrally--condensed turbulent cloud onto a compact star--forming core, and not anything particularly related to cloud--cloud collisions.\\ 
\subsection{Boundedness of collision products}
One of the most basic questions that should be addressed in a simulation modelling the collision of two GMCs is to what extent the result of the encounter is a single bound object. Collisions which convert one cloud into two result in coagulation, which may affect the overall cloud mass function.\\
 \begin{figure}
	\includegraphics[width=1.0\columnwidth]{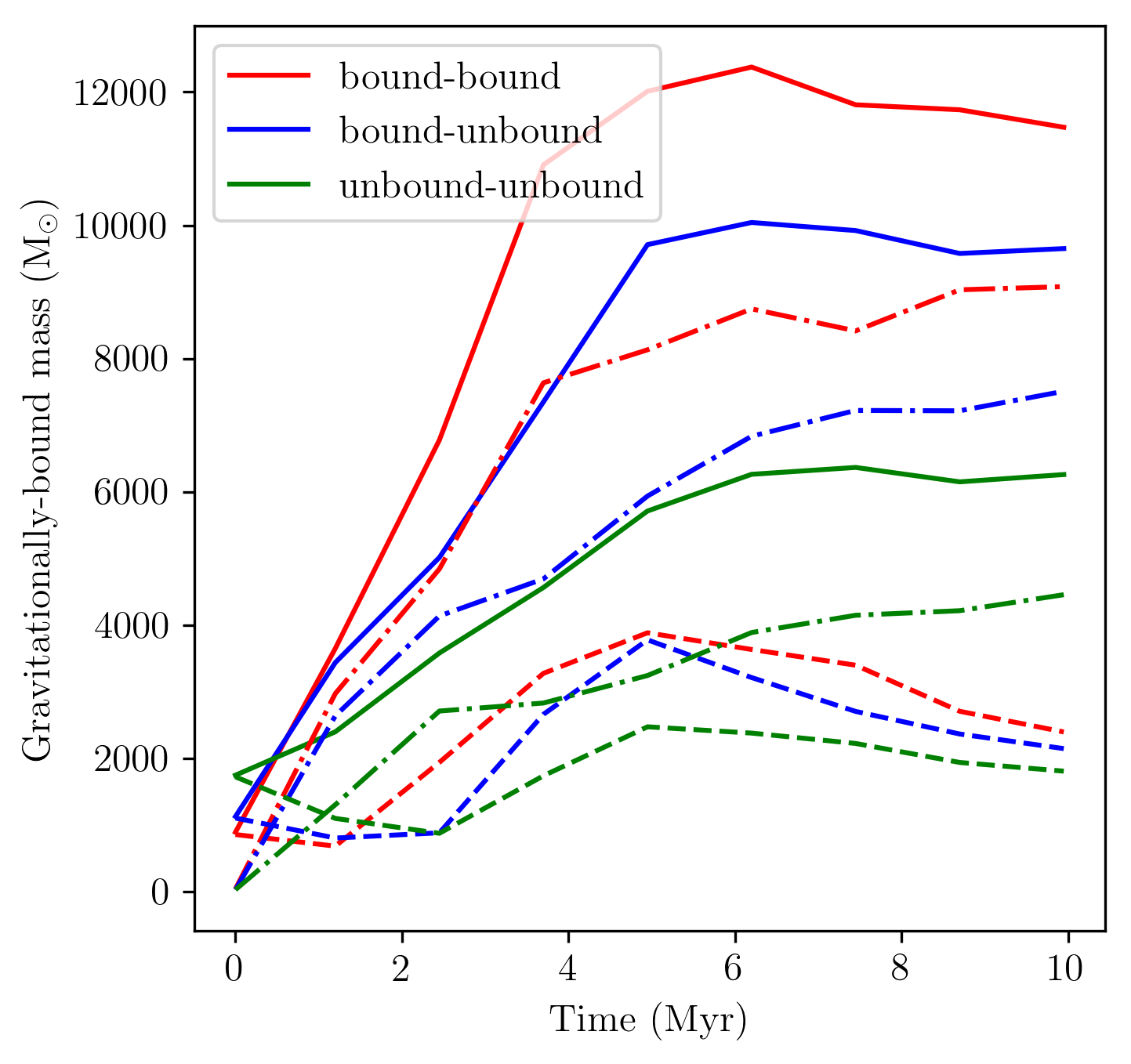}
    \caption{Evolution of quantities of total (solid lines), gas (dash-dotted lines) and stellar (dashed lines) mas in the bound--bound (red), bound--unbound (blue) and unbound--unbound (green) models.}
    \label{fig:boundness}
\end{figure}
\indent The relative velocities and masses of the clouds in these simulations are such that almost all the gas is initially \textit{unbound} in the COM frame. In this section we investigate whether the distribution of gas and stars resulting from the simulation will produce a single object which is bound by its own self gravity in the COM frame of reference. We compute for each gas and sink particle the kinetic energy in the COM frame, the gravitational potential energy due to all other particles, and the sum of these. Gas or sink particles for which this sum is negative are regarded as being bound and as members of a single object at rest in the COM frame.\\
\begin{figure}
	\includegraphics[width=0.8\columnwidth]{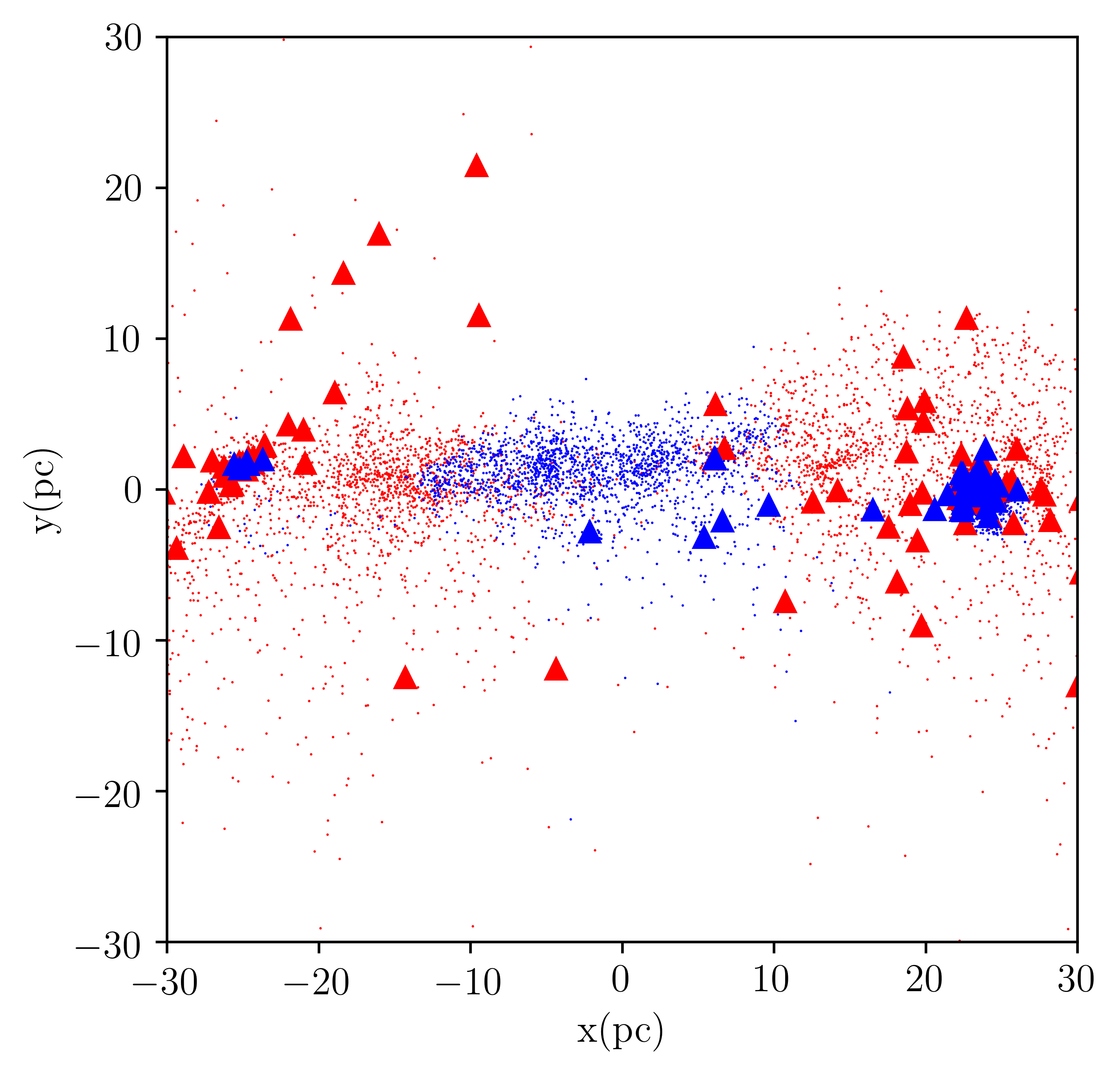}
		\includegraphics[width=0.8\columnwidth]{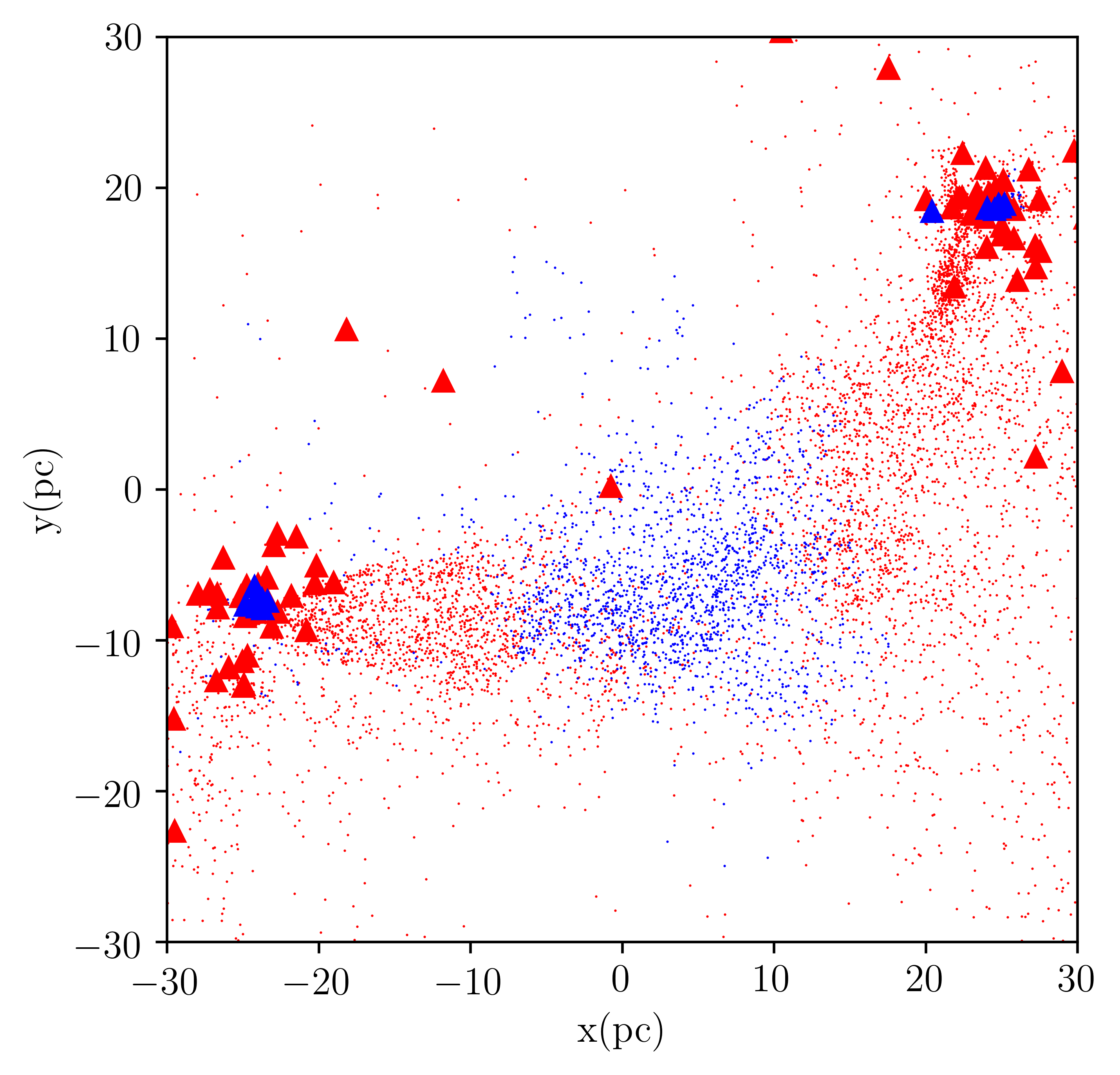}
		\includegraphics[width=0.8\columnwidth]{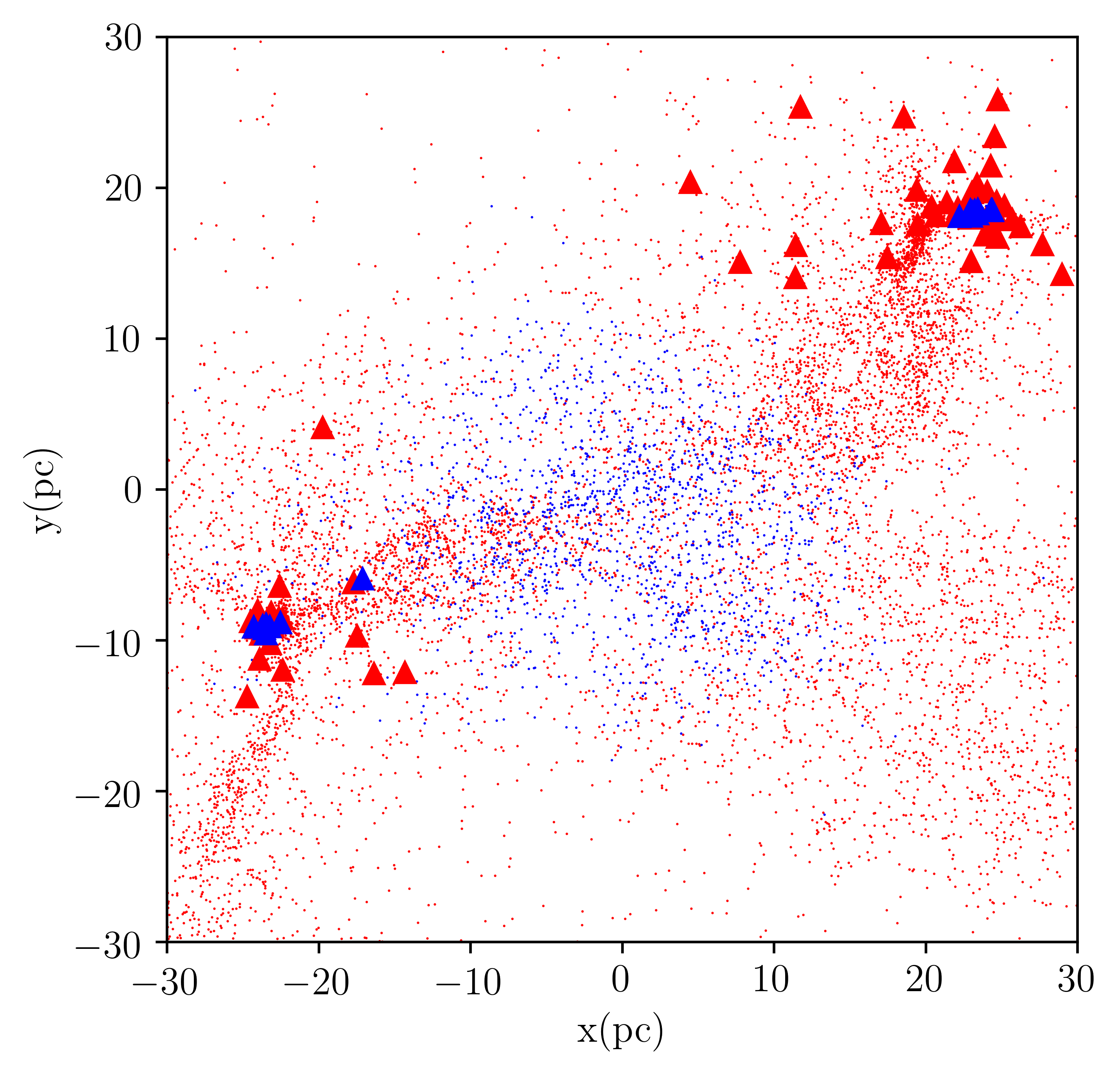}
    \caption{Particle and stellar plots of Run BB(first panel), BU (second panel), UU (third panel). Blue dots and triangle represent bound particles and stars (with respect to the COM frame) while the red dots and triangles represent the unbound particles and stars (with respect to the COM frame.)}
    \label{fig:particle}
\end{figure}
\indent Figure \ref{fig:particle} shows particle plots of the BB BU and UU simulations at 9.5 Myr. Blue and red dots represent, respectively, bound and unbound gas particles (only one tenth of the gas particles are shown for clarity), and blue and red triangles are respectively bound and unbound sinks.\\
\indent Towards the centre of each image there is a quantity of bound gas, most in the BB model, where the bound gas forms a likely long--lived filament,  and least in the UU model. This material corresponds approximately to the broad--bridge features seen in the PV images of the simulations, and consists of material which has been brought almost to a halt in the COM frame by shocks. The self--gravity of this gas does not need to be strong, since its velocity is so low. Much of the gas evidently remains unbound in the COM frame.\\
\indent Turning to the sink particles, most of the sinks also appear to be unbound except, in the main, those close to the centres of the clusters. However, a little thought shows that these sinks are not in fact part of a single bound structure in the COM frame. They are instead in the very deep potential wells of the clusters, so they are strongly bound to the clusters, but the clusters are \textit{not bound to each other}.\\
\indent We verified that this is the case by treating each simulation at 9.5 Myr as a three--body problem consisting of the left-- and right--hand clusters of masses $M_{\rm left}$ and $M_{\rm right}$, locations ${\bf r}_{\rm left}$ and ${\bf r}_{\rm right}$ and velocities ${\bf v}_{\rm left}$ and ${\bf v}_{\rm right}$, and the bound gas of mass $M_{\rm bnd}$ located at the origin with zero velocity. We can then compute the total kinetic energy as
\begin{equation}
KE_{3-body}=\frac{1}{2}M_{\rm left}{\bf v}_{\rm left}^{2}+\frac{1}{2}M_{\rm right}{\bf v}_{\rm right}^{2}
\end{equation}
and the potential energy as
\begin{equation}
PE_{3-body}=-\frac{GM_{\rm left}M_{\rm right}}{|{\bf r}_{\rm lr}|}-\frac{GM_{\rm left}M_{\rm bnd}}{|{\bf r}_{\rm left}|}-\frac{GM_{\rm right}M_{\rm bnd}}{|{\bf r}_{\rm right}|}
\end{equation}
where ${\bf r}_{\rm lr}={\bf r}_{\rm left}-{\bf r}_{\rm right}$.\\
\indent We computed these energies for all the final states of all three collision simulations and found that the kinetic energy exceeded the modulus of the potential energy by factors of $\approx 3$, indicating that the clusters pairs are not bound in the centre of mass frame. The dense cores of the clouds, where most of the stars and most of the star formation activity are to be found, are little affected by the collision.\\ 
\indent In Figure \ref{fig:boundness} we plot the quantities of bound stellar (dashed lines), gas (dash--dotted lines) and total (solid lines) mass as a function of time for all three simulations. Because of the issue discussed above in which the deep potential wells of the clusters imply sinks that sinks are bound when they are actually not, these quantities are strictly upper limits.\\
\indent As we observed in paper-I, the initial quantities of bound material are small, but there are sharp increases in bound mass before and during the collision. As the clouds approach, they generate a deeper potential well, and material is decelerated by shocks. This occurs fastest in Run BB and slowest in Run UU. This indicates that matter is likely to become more bound in the COM frame if the initial virial state of the clouds were more strongly bound. The quantities of the bound matter comes to a steady state around 4--6 Myr.\\ 
\indent The total mass of all models is $2\times10^{4}$\,M$_{\odot}$, so we see that, in the bound--bound and bound--unbound simulations, approximately half of the total mass is left bound, and most of the bound mass is gas. In the unbound--unbound model, approximately one third of the total mass is left bound, and approximately half of the bound mass consists of gas.\\
\indent It can be concluded that the collision between bound clouds will result in a large fraction of the total mass remaining bound in the COM frame. However in the case of collision between unbound clouds, only a small fraction (approx 30 percent) of the total mass becomes bound in the COM frame. More significantly, most of the stars and potentially star--forming gas do not become bound in the COM frame.\\
\section{Discussion and Conclusions}
\begin{figure}
	\includegraphics[width=0.8\columnwidth]{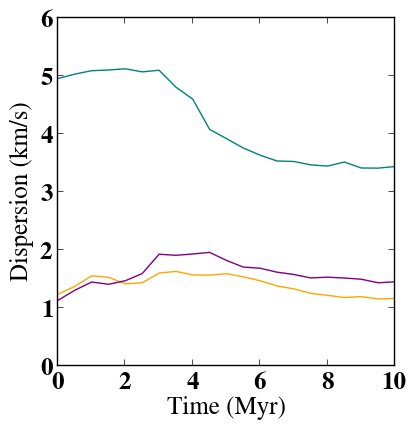}
		\includegraphics[width=0.8\columnwidth]{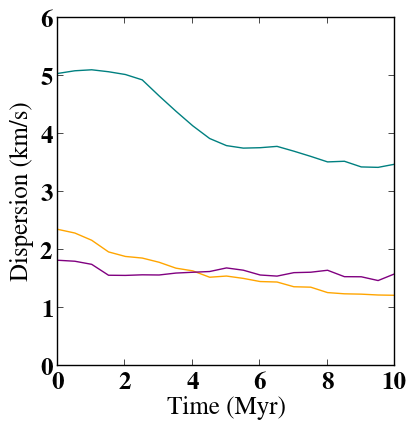}
		\includegraphics[width=0.8\columnwidth]{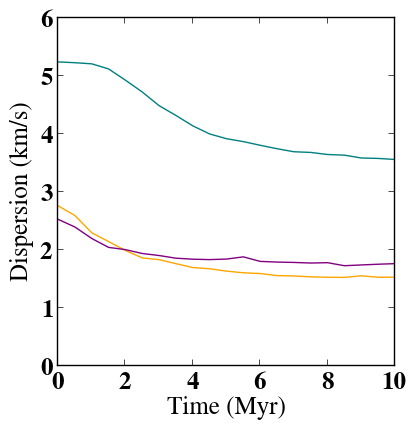}
    \caption{Velocity dispersion in the x (teal lines), y(orange lines) and z(purple lines) directions of Run BB(first panel), BU(second panel) and UU (third panel) simulations. The velocity dispersion in the x direction includes the clouds' bulk velocities.}
   \label{fig:vel}
\end{figure}

\indent In this paper we have investigated the impact of the initial density profile on collisions between molecular clouds. As in paper-I we have chosen to collide the clouds at a speed of 10\,km\,s$^{-1}$. This velocity is higher than the escape velocity of the clouds, so that most of the mass is unbound at the beginning of the simulation.\\
\indent {\bf We find that the density structure of the clouds dominates their evolution. The difference between the evolution of the collision and control simulations in terms of star formation rates and efficiencies and angular momentum of the stellar material, are rather minor. The dense cores of the clouds essentially detach themselves dynamically from the outer regions, and are sufficiently dense and gravitationally bound that the collisions have little effect on them.}\\
\indent For consistency with paper-I the simulations in this study have been run to $\approx$ 10 Myr, although we see that most of the quantities of interest, such as star--formation efficiency and fraction of bound material have settled down after 4--6 Myr in these models. We have once again not included the effects of feedback. In these models, the massive stars are all found deep in potential wells containing large quantities of dense gas, and which are continually accreting further gas. The timescale on which feedback might be expected to disperse the clouds -- if indeed it can -- is unclear. A separate paper will contain the effects of feedback and the simulations presented here will serve as a baseline to understand the effect of feedback.\\
\indent In this paper the model clouds have an initial density profile of $\rm \rho \propto R^{-2}$. If both the clouds are bound, then the collision results in the formation of a bound filamentary structure visible in both position-position and position-velocity plots. Dense cores from the original clouds survive the collision and moves towards the ends of the filament, and are the sole sites of star formation. If only one or neither clouds is bound, the collision does not create strong filamentary structures and the clusters are almost entirely unconnected to each other.\\
\indent These simulations exhibit higher star formation rates and efficiencies compared to the simulations involving uniform clouds presented in Paper I. Contrary to what was found in Paper I (see Figure 11 in that paper) in this study the collisions have almost no effect on the overall star formation efficiency, compared to the relevant control runs. There are likely several reasons behind this. In Paper I and here, the cloud collision time is $\approx$2 Myr. In paper I the shocking of the clouds does indeed generate high density materials which leads to a modest increase in the star formation efficiency. Due to their initial density profiles, the clouds in this study already have cores much denser than the clouds in Paper I, which begin forming stars very soon after the simulation starts and long before the collision. This is true even for the unbound clouds, since the cores of these clouds have strongly bound gas. Star formation remains heavily concentrated in the core regions of the clouds throughout the simulation.\\
\indent The shocking of the clouds, which affects first their low--density outer layers, has little effect on the cores and does not increase the quantity of gas at the highest densities. The shocks do not appear to affect the denser parts of the clouds or have any impact on the structures there. Infall motions onto the the clusters are neither accelerated nor retarded, therefore.\\
\indent Like Paper I the collision between the two clouds does not result in the formation of a singular object. Runs BB, BU, UU all generate appreciable quantities of bound material, consisting almost entirely of gas, but the clusters and dense cores of the clouds are not substantially decelerated by the collision and remain unbound approximately of unbound mass.\\
 \indent Figure \ref{fig:vel} shows the velocity dispersion in the gas along the coordinate axes.
There is a steady decline in velocity dispersion in the y and z directions as the turbulent velocity dissipates slowly. In Run A this is regulated by the contraction of the bound clouds.\\
\indent Along the collision axis the velocity dispersion however remains steady until the clouds come into contact at $\approx$2 Myr. A steady decline then lasts for another $\approx$2 Myr or 2 cloud crushing time. In this time period the clouds progressively enter the shocked region. The decline in kinetic energy in the x--direction is then the result of loss of clouds' bulk kinetic energy due to the isothermal shocks generated by the collision. Since the turbulence in the clouds was already creating substructure before the clouds came into contact, the clouds are strongly non--uniform and this process is not very efficient. A closer look at Figure \ref{fig:vel} shows that the velocity dispersion in the x--direction remains large enough to render some of the collision product unbound. This is because the velocity dispersion is deduced from the surviving gas from the clouds (i.e. material that did not form stars) and this is the material most likely moving at higher speed, and therefore most likely to be unbound. In none of the simulations was enough bulk kinetic energy lost to render the collision product completely bound. These plots are very similar to the plots presented in \cite{2020MNRAS.494..246T}.\\
\indent Once again magnetic fields and feedback from the massive stars have been neglected in these simulations, although both will be likely to influence the overall evolution of a cloud-collision product in several ways. Uniform magnetic fields oriented perpendicular to the collision axis may serve to cushion the collision, while a field orientated along the collision axis would suppress to some extent the tendency of bound clouds to contract into quasi--filamentary structures with this orientation. However, the effect of a turbulent field, as would inevitably be generated by our turbulent initial conditions, may simply act to slow the collapse of the clouds somewhat, and may have no discernible overall morphological influence.\\
\indent If it were the case that the magnetic field within the clouds were strong enough to support them against gravitational collapse, so that the star formation timescale became comparable to or longer than the collision timescale, then the collision might be expected to have a much stronger effect and may succeed in initiating the clouds' global collapse. We intend to investigate the effects of magnetic fields in subsequent papers.\\
\indent The effects of feedback would likely be strongly constrained by the fact that the massive stars in these calculations are in very deep potential wells, into which dense gas is constantly infalling. These would both serve to stifle the expansion of HII regions or wind bubbles. We aim to answer these questions in subsequent papers.
%\section*{Acknowledgements}
\section*{Data Availability}
upon request the simulation data can be shared with the corresponding author. 
%We once again thank the Dark Lord Sauron for his eternal blessing on our project. 

%%%%%%%%%%%%%%%%%%%%%%%%%%%%%%%%%%%%%%%%%%%%%%%%%%



\bibliographystyle{mnras}
\bibliography{example} % if your bibtex file is called example.bib

\begin{thebibliography}{}
\makeatletter
\relax
\def\mn@urlcharsother{\let\do\@makeother \do\$\do\&\do\#\do\^\do\_\do\%\do\~}
\def\mn@doi{\begingroup\mn@urlcharsother \@ifnextchar [ {\mn@doi@}
  {\mn@doi@[]}}
\def\mn@doi@[#1]#2{\def\@tempa{#1}\ifx\@tempa\@empty \href
  {http://dx.doi.org/#2} {doi:#2}\else \href {http://dx.doi.org/#2} {#1}\fi
  \endgroup}
\def\mn@eprint#1#2{\mn@eprint@#1:#2::\@nil}
\def\mn@eprint@arXiv#1{\href {http://arxiv.org/abs/#1} {{\tt arXiv:#1}}}
\def\mn@eprint@dblp#1{\href {http://dblp.uni-trier.de/rec/bibtex/#1.xml}
  {dblp:#1}}
\def\mn@eprint@#1:#2:#3:#4\@nil{\def\@tempa {#1}\def\@tempb {#2}\def\@tempc
  {#3}\ifx \@tempc \@empty \let \@tempc \@tempb \let \@tempb \@tempa \fi \ifx
  \@tempb \@empty \def\@tempb {arXiv}\fi \@ifundefined
  {mn@eprint@\@tempb}{\@tempb:\@tempc}{\expandafter \expandafter \csname
  mn@eprint@\@tempb\endcsname \expandafter{\@tempc}}}

\bibitem[\protect\citeauthoryear{{Balfour}, {Whitworth}, {Hubber}  \&
  {Jaffa}}{{Balfour} et~al.}{2015}]{2015MNRAS.453.2471B}
{Balfour} S.~K.,  {Whitworth} A.~P.,  {Hubber} D.~A.,   {Jaffa} S.~E.,  2015,
  \mn@doi [\mnras] {10.1093/mnras/stv1772}, \href
  {https://ui.adsabs.harvard.edu/abs/2015MNRAS.453.2471B} {453, 2471}

\bibitem[\protect\citeauthoryear{{Balfour}, {Whitworth}  \& {Hubber}}{{Balfour}
  et~al.}{2017}]{2017MNRAS.465.3483B}
{Balfour} S.~K.,  {Whitworth} A.~P.,   {Hubber} D.~A.,  2017, \mn@doi [\mnras]
  {10.1093/mnras/stw2956}, \href
  {https://ui.adsabs.harvard.edu/abs/2017MNRAS.465.3483B} {465, 3483}

\bibitem[\protect\citeauthoryear{{Ballone}, {Mapelli}, {Di Carlo},
  {Torniamenti}, {Spera}  \& {Rastello}}{{Ballone}
  et~al.}{2020}]{2020MNRAS.496...49B}
{Ballone} A.,  {Mapelli} M.,  {Di Carlo} U.~N.,  {Torniamenti} S.,  {Spera} M.,
    {Rastello} S.,  2020, \mn@doi [\mnras] {10.1093/mnras/staa1383}, \href
  {https://ui.adsabs.harvard.edu/abs/2020MNRAS.496...49B} {496, 49}

\bibitem[\protect\citeauthoryear{{Bhattal}, {Francis}, {Watkins}  \&
  {Whitworth}}{{Bhattal} et~al.}{1998}]{1998MNRAS.297..435B}
{Bhattal} A.~S.,  {Francis} N.,  {Watkins} S.~J.,   {Whitworth} A.~P.,  1998,
  \mn@doi [\mnras] {10.1046/j.1365-8711.1998.01503.x}, \href
  {https://ui.adsabs.harvard.edu/abs/1998MNRAS.297..435B} {297, 435}

\bibitem[\protect\citeauthoryear{{Cottaar} et~al.,}{{Cottaar}
  et~al.}{2015}]{2015ApJ...807...27C}
{Cottaar} M.,  et~al., 2015, \mn@doi [\apj] {10.1088/0004-637X/807/1/27}, \href
  {https://ui.adsabs.harvard.edu/abs/2015ApJ...807...27C} {807, 27}

\bibitem[\protect\citeauthoryear{{Dobbs}, {Pringle}  \&
  {Duarte-Cabral}}{{Dobbs} et~al.}{2015}]{2015MNRAS.446.3608D}
{Dobbs} C.~L.,  {Pringle} J.~E.,   {Duarte-Cabral} A.,  2015, \mn@doi [\mnras]
  {10.1093/mnras/stu2319}, \href
  {http://adsabs.harvard.edu/abs/2015MNRAS.446.3608D} {446, 3608}

\bibitem[\protect\citeauthoryear{{Fukui} et~al.,}{{Fukui}
  et~al.}{2014}]{2014ApJ...780...36F}
{Fukui} Y.,  et~al., 2014, \mn@doi [\apj] {10.1088/0004-637X/780/1/36}, \href
  {https://ui.adsabs.harvard.edu/abs/2014ApJ...780...36F} {780, 36}

\bibitem[\protect\citeauthoryear{{Fukui} et~al.,}{{Fukui}
  et~al.}{2016}]{2016ApJ...820...26F}
{Fukui} Y.,  et~al., 2016, \mn@doi [\apj] {10.3847/0004-637X/820/1/26}, \href
  {https://ui.adsabs.harvard.edu/abs/2016ApJ...820...26F} {820, 26}

\bibitem[\protect\citeauthoryear{{Furukawa}, {Dawson}, {Ohama}, {Kawamura},
  {Mizuno}, {Onishi}  \& {Fukui}}{{Furukawa}
  et~al.}{2009}]{2009ApJ...696L.115F}
{Furukawa} N.,  {Dawson} J.~R.,  {Ohama} A.,  {Kawamura} A.,  {Mizuno} N.,
  {Onishi} T.,   {Fukui} Y.,  2009, \mn@doi [\apj]
  {10.1088/0004-637X/696/2/L115}, \href
  {https://ui.adsabs.harvard.edu/abs/2009ApJ...696L.115F} {696, L115}

\bibitem[\protect\citeauthoryear{{Girichidis}, {Federrath}, {Banerjee}  \&
  {Klessen}}{{Girichidis} et~al.}{2011}]{2011MNRAS.413.2741G}
{Girichidis} P.,  {Federrath} C.,  {Banerjee} R.,   {Klessen} R.~S.,  2011,
  \mn@doi [\mnras] {10.1111/j.1365-2966.2011.18348.x}, \href
  {https://ui.adsabs.harvard.edu/abs/2011MNRAS.413.2741G} {413, 2741}

\bibitem[\protect\citeauthoryear{{Haworth} et~al.,}{{Haworth}
  et~al.}{2015a}]{2015MNRAS.450...10H}
{Haworth} T.~J.,  et~al., 2015a, \mn@doi [\mnras] {10.1093/mnras/stv639}, \href
  {http://adsabs.harvard.edu/abs/2015MNRAS.450...10H} {450, 10}

\bibitem[\protect\citeauthoryear{{Haworth}, {Shima}, {Tasker}, {Fukui},
  {Torii}, {Dale}, {Takahira}  \& {Habe}}{{Haworth}
  et~al.}{2015b}]{2015MNRAS.454.1634H}
{Haworth} T.~J.,  {Shima} K.,  {Tasker} E.~J.,  {Fukui} Y.,  {Torii} K.,
  {Dale} J.~E.,  {Takahira} K.,   {Habe} A.,  2015b, \mn@doi [\mnras]
  {10.1093/mnras/stv2068}, \href
  {http://adsabs.harvard.edu/abs/2015MNRAS.454.1634H} {454, 1634}

\bibitem[\protect\citeauthoryear{{H{\'e}nault-Brunet}
  et~al.,}{{H{\'e}nault-Brunet} et~al.}{2012}]{2012A&A...545L...1H}
{H{\'e}nault-Brunet} V.,  et~al., 2012, \mn@doi [\aap]
  {10.1051/0004-6361/201219472}, \href
  {https://ui.adsabs.harvard.edu/abs/2012A&A...545L...1H} {545, L1}

\bibitem[\protect\citeauthoryear{{Hubber}, {Batty}, {McLeod}  \&
  {Whitworth}}{{Hubber} et~al.}{2011}]{2011A&A...529A..27H}
{Hubber} D.~A.,  {Batty} C.~P.,  {McLeod} A.,   {Whitworth} A.~P.,  2011,
  \mn@doi [\aap] {10.1051/0004-6361/201014949}, \href
  {http://adsabs.harvard.edu/abs/2011A%26A...529A..27H} {529, A27}

\bibitem[\protect\citeauthoryear{{Hubber}, {Rosotti}  \& {Booth}}{{Hubber}
  et~al.}{2018}]{2018MNRAS.473.1603H}
{Hubber} D.~A.,  {Rosotti} G.~P.,   {Booth} R.~A.,  2018, \mn@doi [\mnras]
  {10.1093/mnras/stx2405}, \href
  {https://ui.adsabs.harvard.edu/abs/2018MNRAS.473.1603H} {473, 1603}

\bibitem[\protect\citeauthoryear{{Lardo} et~al.,}{{Lardo}
  et~al.}{2015}]{2015A&A...573A.115L}
{Lardo} C.,  et~al., 2015, \mn@doi [\aap] {10.1051/0004-6361/201425036}, \href
  {https://ui.adsabs.harvard.edu/abs/2015A&A...573A.115L} {573, A115}

\bibitem[\protect\citeauthoryear{{Mapelli}}{{Mapelli}}{2017}]{2017MNRAS.467.3255M}
{Mapelli} M.,  2017, \mn@doi [\mnras] {10.1093/mnras/stx304}, \href
  {https://ui.adsabs.harvard.edu/abs/2017MNRAS.467.3255M} {467, 3255}

\bibitem[\protect\citeauthoryear{{Monaghan}}{{Monaghan}}{1997}]{1997JCoPh.136..298M}
{Monaghan} J.~J.,  1997, \mn@doi [Journal of Computational Physics]
  {10.1006/jcph.1997.5732}, \href
  {http://adsabs.harvard.edu/abs/1997JCoPh.136..298M} {136, 298}

\bibitem[\protect\citeauthoryear{{Ohama} et~al.,}{{Ohama}
  et~al.}{2010}]{2010ApJ...709..975O}
{Ohama} A.,  et~al., 2010, \mn@doi [\apj] {10.1088/0004-637X/709/2/975}, \href
  {https://ui.adsabs.harvard.edu/abs/2010ApJ...709..975O} {709, 975}

\bibitem[\protect\citeauthoryear{{Price} \& {Monaghan}}{{Price} \&
  {Monaghan}}{2004}]{2004MNRAS.348..139P}
{Price} D.~J.,  {Monaghan} J.~J.,  2004, \mn@doi [\mnras]
  {10.1111/j.1365-2966.2004.07346.x}, \href
  {https://ui.adsabs.harvard.edu/abs/2004MNRAS.348..139P} {348, 139}

\bibitem[\protect\citeauthoryear{{Pryor}, {McClure}, {Fletcher}, {Hartwick}  \&
  {Kormendy}}{{Pryor} et~al.}{1986}]{1986AJ.....91..546P}
{Pryor} C.,  {McClure} R.~D.,  {Fletcher} J.~M.,  {Hartwick} F.~D.~A.,
  {Kormendy} J.,  1986, \mn@doi [\aj] {10.1086/114035}, \href
  {https://ui.adsabs.harvard.edu/abs/1986AJ.....91..546P} {91, 546}

\bibitem[\protect\citeauthoryear{{Shima}, {Tasker}, {Federrath}  \&
  {Habe}}{{Shima} et~al.}{2018}]{2018PASJ...70S..54S}
{Shima} K.,  {Tasker} E.~J.,  {Federrath} C.,   {Habe} A.,  2018, \mn@doi
  [\pasj] {10.1093/pasj/psx124}, \href
  {https://ui.adsabs.harvard.edu/abs/2018PASJ...70S..54S} {70, S54}

\bibitem[\protect\citeauthoryear{{Shu}}{{Shu}}{1977}]{1977ApJ...214..488S}
{Shu} F.~H.,  1977, \mn@doi [\apj] {10.1086/155274}, \href
  {https://ui.adsabs.harvard.edu/abs/1977ApJ...214..488S} {214, 488}

\bibitem[\protect\citeauthoryear{{Springel} \& {Hernquist}}{{Springel} \&
  {Hernquist}}{2002}]{2002MNRAS.333..649S}
{Springel} V.,  {Hernquist} L.,  2002, \mn@doi [\mnras]
  {10.1046/j.1365-8711.2002.05445.x}, \href
  {http://adsabs.harvard.edu/abs/2002MNRAS.333..649S} {333, 649}

\bibitem[\protect\citeauthoryear{{Takahira}, {Tasker}  \& {Habe}}{{Takahira}
  et~al.}{2014}]{2014ApJ...792...63T}
{Takahira} K.,  {Tasker} E.~J.,   {Habe} A.,  2014, \mn@doi [\apj]
  {10.1088/0004-637X/792/1/63}, \href
  {https://ui.adsabs.harvard.edu/abs/2014ApJ...792...63T} {792, 63}

\bibitem[\protect\citeauthoryear{{Tanvir} \& {Dale}}{{Tanvir} \&
  {Dale}}{2020}]{2020MNRAS.494..246T}
{Tanvir} T.~S.,  {Dale} J.~E.,  2020, \mn@doi [\mnras] {10.1093/mnras/staa665},
  \href {https://ui.adsabs.harvard.edu/abs/2020MNRAS.494..246T} {494, 246}

\bibitem[\protect\citeauthoryear{{Tasker} \& {Tan}}{{Tasker} \&
  {Tan}}{2009}]{2009ApJ...700..358T}
{Tasker} E.~J.,  {Tan} J.~C.,  2009, \mn@doi [\apj]
  {10.1088/0004-637X/700/1/358}, \href
  {https://ui.adsabs.harvard.edu/abs/2009ApJ...700..358T} {700, 358}

\bibitem[\protect\citeauthoryear{{Torii} et~al.,}{{Torii}
  et~al.}{2011}]{2011ApJ...738...46T}
{Torii} K.,  et~al., 2011, \mn@doi [\apj] {10.1088/0004-637X/738/1/46}, \href
  {https://ui.adsabs.harvard.edu/abs/2011ApJ...738...46T} {738, 46}

\bibitem[\protect\citeauthoryear{{Torii} et~al.,}{{Torii}
  et~al.}{2015}]{2015ApJ...806....7T}
{Torii} K.,  et~al., 2015, \mn@doi [\apj] {10.1088/0004-637X/806/1/7}, \href
  {https://ui.adsabs.harvard.edu/abs/2015ApJ...806....7T} {806, 7}

\bibitem[\protect\citeauthoryear{{Torii} et~al.,}{{Torii}
  et~al.}{2017}]{2017ApJ...835..142T}
{Torii} K.,  et~al., 2017, \mn@doi [\apj] {10.3847/1538-4357/835/2/142}, \href
  {https://ui.adsabs.harvard.edu/abs/2017ApJ...835..142T} {835, 142}

\bibitem[\protect\citeauthoryear{{Whitworth}, {Bhattal}, {Chapman}, {Disney}
  \& {Turner}}{{Whitworth} et~al.}{1994a}]{1994MNRAS.268..291W}
{Whitworth} A.~P.,  {Bhattal} A.~S.,  {Chapman} S.~J.,  {Disney} M.~J.,
  {Turner} J.~A.,  1994a, \mn@doi [\mnras] {10.1093/mnras/268.1.291}, \href
  {https://ui.adsabs.harvard.edu/abs/1994MNRAS.268..291W} {268, 291}

\bibitem[\protect\citeauthoryear{{Whitworth}, {Bhattal}, {Chapman}, {Disney}
  \& {Turner}}{{Whitworth} et~al.}{1994b}]{1994A&A...290..421W}
{Whitworth} A.~P.,  {Bhattal} A.~S.,  {Chapman} S.~J.,  {Disney} M.~J.,
  {Turner} J.~A.,  1994b, \aap, \href
  {http://adsabs.harvard.edu/abs/1994A%26A...290..421W} {290, 421}

\makeatother
\end{thebibliography}

%\begin{thebibliography}{99}
%\end{thebibliography}

%%%%%%%%%%%%%%%%%%%%%%%%%%%%%%%%%%%%%%%%%%%%%%%%%%

%%%%%%%%%%%%%%%%% APPENDICES %%%%%%%%%%%%%%%%%%%%%

%%%%%%%%%%%%%%%%%%%%%%%%%%%%%%%%%%%%%%%%%%%%%%%%%%

% Don't change these lines
\bsp	% typesetting comment
\label{lastpage}
\end{document}